%% file: R2.tex
\documentclass[a4paper,fleqn,usenatbib]{mnras}

\usepackage{newtxtext,newtxmath}
\usepackage[T1]{fontenc}
\usepackage{ae,aecompl}

\usepackage{graphicx}	
\usepackage{amsmath}	

\usepackage{hyperref}
\usepackage[nolist,nohyperlinks]{acronym}
\usepackage{supertabular}
\usepackage{pdflscape}
\usepackage{float}
\usepackage{epstopdf}
\usepackage{multicol}
\usepackage{graphicx}
\usepackage{subcaption}
\usepackage{geometry}
\usepackage{changepage}
\usepackage{siunitx}
\usepackage{dcolumn}
\usepackage{longtable}
\usepackage{afterpage}
\usepackage{array}
\usepackage{varwidth}

\input{Files/catalogue_pars}

\usepackage{xcolor}

\def\HII{\hbox{H\,{\sc ii}}}

\newcommand{\D}{$^\circ$}

\def\arcmin{\hbox{$^\prime$}}
\def\arcsec{\hbox{$^{\prime\prime}$}}

\title[Radio Continuum Sources behind the LMC]{Radio Continuum Sources behind the Large Magellanic Cloud}

\author[M. D. Filipovi\'c et al.]{M. D. Filipovi\'c,$^{1}$\thanks{E-mail: m.filipovic@westernsydney.edu.au} 
I. S. Boji\v ci\'c,$^{1}$
K. R. Grieve,$^{1}$
R. P. Norris,$^{1,2}$ 
N. F. H. Tothill,$^{1}$  
\newauthor
D. Shobhana,$^{1}$ 
L. Rudnick,$^{3}$ 
I. Prandoni,$^{4}$
H. Andernach,$^{5}$
N. Hurley-Walker,$^{6}$
\newauthor
R. Z. E. Alsaberi,$^{1}$ 
C. S. Anderson,$^{2}$
J. D. Collier,$^{1,7}$
E. J. Crawford,$^{1}$
B.-Q. For,$^{7,8}$ 
\newauthor
T. J. Galvin,$^{6}$
F. Haberl,$^{10}$
A. M. Hopkins,$^{11,1}$
A. Ingallinera$^{12}$
P.~J.~Kavanagh,$^{13}$
\newauthor
B. S. Koribalski,$^{1,2}$
R. Kothes,$^{14}$
D. Leahy,$^{15}$
H. Leverenz,$^{1}$
P. Maggi,$^{16}$
C. Maitra,$^{10}$
\newauthor
J. Marvil,$^{17}$ 
T. G. Pannuti,$^{18}$
L. A. F. Park,$^{1}$
J. L.~Payne,$^{1}$
C. M. Pennock,$^{19}$
S. Riggi,$^{12}$
\newauthor
G. Rowell,$^{20}$
H. Sano,$^{21}$
M. Sasaki,$^{22}$ 
L. Staveley-Smith,$^{7,8}$
V. Trigilio,$^{12}$
G. Umana,$^{12}$
\newauthor
D. Uro\v sevi\' c,$^{23,24}$
J. Th. van Loon,$^{19}$
E. Vardoulaki,$^{25}$
\\
$^{1}$Western Sydney University, Locked Bag 1797, Penrith South DC, NSW 2751, Australia \\
$^{2}$CSIRO Astronomy and Space Science, PO Box 76, Epping, NSW 1710, Australia \\
$^{3}$Minnesota Institute for Astrophysics, School of Physics and Astronomy, University of Minnesota, 116 Church Street SE, Minneapolis, MN 55455, USA\\
$^{4}$INAF -- Istituto di Radioastronomia, Via P. Gobetti 101, 40129, Bologna, Italy\\
$^{5}$Depto. de Astronom\'ia, DCNE, Universidad de Guanajuato, Cj\'on. de Jalisco s/n, Col. Valenciana, Guanajuato, CP 36023, Gto., Mexico\\
$^{6}$International Centre for Radio Astronomy Research, Curtin University, Bentley, WA 6102, Australia\\
$^{7}$The Inter-University Institute for Data Intensive Astronomy (IDIA), Department of Astronomy, University of Cape Town, Rondebosch 7701, South Africa \\
$^{8}$ARC Centre of Excellence for All Sky Astrophysics in 3 Dimensions (ASTRO 3D), Australia \\ 
$^{9}$International Centre for Radio Astronomy Research, University of Western Australia, 35 Stirling Hwy, Crawley, WA 6009, Australia \\ 
$^{10}$Max-Planck-Institut f\"{u}r extraterrestrische Physik, Gie{\ss}enbachstra{\ss}e 1, D-85748 Garching, Germany \\
$^{11}$Australian Astronomical Optics, AAO-Macquarie, Faculty of Science and Engineering, Macquarie University, \\
105 Delhi Rd, North Ryde, NSW 2113, Australia \\
$^{12}$INAF -- Osservatorio Astrofisico di Catania, via Santa Sofia 78, I-95123 Catania, Italia \\
$^{13}$School of Cosmic Physics, Dublin Institute for Advanced Studies, 31 Fitzwilliam Place, Dublin 2, Ireland \\
$^{14}$Dominion Radio Astrophysical Observatory, Herzberg Astronomy and Astrophysics, National Research Council Canada, \\
PO Box 248, Penticton BC V2A 6J9, Canada \\
$^{15}$Department of Physics and Astronomy, University of Calgary, University of Calgary, Calgary, Alberta, T2N 1N4, Canada\\
$^{16}$Observatoire Astronomique de Strasbourg, Universit\'e de Strasbourg, CNRS, 11 rue de l'Universit\'e, F-67000 Strasbourg, France \\
$^{17}$National Radio Astronomy Observatory, 1003 Lopezville Road, Socorro, NM 87801, USA \\
$^{18}$Department of Physics, Earth Science and Space Systems Engineering, Morehead State University, 235 Martindale Drive, Morehead, KY 40351, USA\\
$^{19}$Lennard-Jones Laboratories, Keele University, Staffordshire ST5 5BG, UK\\
$^{20}$School of Physical Sciences, The University of Adelaide, Adelaide 5005, Australia\\
$^{21}$National Astronomical Observatory of Japan, Mitaka, Tokyo 181-8588, Japan\\
$^{22}$Remeis Observatory and ECAP, Universit\"{a}t Erlangen-N\"{u}rnberg, Sternwartstra{\ss}e 7, D-96049 Bamberg, Germany \\
$^{23}$Department of Astronomy, Faculty of Mathematics, University of Belgrade, Studentski trg 16, 11000 Belgrade, Serbia\\
$^{24}$Isaac Newton Institute of Chile, Yugoslavia Branch\\
$^{25}$Th\"{u}ringer Landessternwarte, Sternwarte 5, 07778 Tautenburg, Germany\\
}

\date{Accepted 2021 July 22. Received 2021 July 22; in original form 2021 March 18}

\pubyear{2021}

\begin{document}
\label{firstpage}
\pagerange{\pageref{firstpage}--\pageref{lastpage}}
\maketitle


\begin{abstract}
We present a comprehensive multi-frequency catalogue of radio sources behind the \ac{LMC} between 0.2 and 20~GHz, gathered from a combination of new and legacy radio continuum surveys. This catalogue covers an area of $\sim$144~deg$^2$ at angular resolutions from 45~arcsec to $\sim$3~arcmin. We find \SourcesTotalUnique\ discrete radio sources in total, of which \SpecIndexSourceCount\ are detected at two or more radio frequencies. We estimate the median spectral index ($\alpha$; where $S_{v}\sim\nu^\alpha$) of $\alpha$~=~\SpecIndexLowestTwoFreq\ and mean of \SpecIndexLowestTwoFreqMean\ $\pm$\SpecIndexLowestTwoFreqSD\ for \SpecIndexLowestTwoFreqSourceCount\ sources detected exclusively at two frequencies (0.843 and 1.384~GHz) with similar resolution (\ac{FWHM} $\sim$40--45~arcsec). The large frequency range of the surveys makes it an effective tool to investigate \ac{GPS}, \ac{CSS} and Infrared Faint Radio sources populations within our sample. We find \SourcesGPS\ \ac{GPS} candidates with peak frequencies near 5~GHz, from which we estimate their linear size. \SourcesCSS\ sources from our catalogue are \ac{CSS} candidates with $\alpha~<-0.8$. We found six candidates for \ac{HFP} sources, whose radio fluxes peak above 5~GHz and no sources with unconstrained peaks and $\alpha~>0.5$. We found optical counterparts for \OpticalMatchesUnique\ of the radio continuum sources, of which \OpticalMatchesUniqueWithZ\ have a redshift measurement. Finally, we investigate the population of 123 \ac{IFRSs} found in this study.
\end{abstract}


\begin{acronym}[AWGN]
\acro{2MASS}{Two Micron All Sky Survey}
\acro{2MASX}{Two Micron All Sky Survey Extended Source Catalogue}
\acro{30Dor}[30 Dor]{30~Doradus}
\acro{AGN}{active galactic nuclei}
\acro{ATCA}{Australia Telescope Compact Array}
\acro{ATESP}{Australia Telescope ESO Slice Project}
\acro{ATNF}{Australia Telescope National Facility}
\acro{ATOA}{Australia Telescope Online Archive}
\acro{AT20G}{Australia Telescope 20 GHz Survey}
\acro{ASKAP}{Australian Square Kilometre Array Pathfinder}
\acro{BETA}{Boolardy Engineering Test Array}
\acro{BL}[BL Lac]{BL Lacertae Objects}
\acro{CASS}{CSIRO Astronomy and Space Science}
\acro{CABB}{Compact Array Broadband Back-end}
\acro{CHII}[{\sc CHii}]{Compact \textsc{Hii}}
\acro{CMB}{Cosmic Microwave Background}
\acro{CSIRO}{Australian Commonwealth Scientific and Industrial Research Organisation}
\acro{CSS}{Compact Steep Spectrum}
\acro{DS9}[\textsc{DS9}]{\textsc{SAOImage DS9}}
\acro{DSS}{Digital Sky Survey}
\acro{EM}{Electromagnetic}
\acro{EMU}{Evolutionary Map of the Universe}
\acro{ev}[eV]{electronvolt\acroextra{: 1 eV $\approx 1.6 \times 10^{-19}$ J}}
\acro{FITS}[\textsc{Fits}]{Flexible Image Transport System}
\acro{FRB}{Fast Radio Bursts}
\acro{FSRQ}{Flat Spectrum Radio Quasars}
\acro{FWHM}{Full Width at Half-Maximum}
\acro{GLEAM}{GaLactic Extragalactic All-sky MWA}
\acro{GPS}{Gigahertz Peak Spectrum}
\acro{HEMT}{High Electron Mobility Transistor}
\acro{HFP}[HFP]{High Frequency Peaker}
\acro{HPBW}{Half Power Beam Width} 
\acro{HzRGs}{High Redshift Radio Galaxies}
\acro{pHFP}[pHFP]{Potential High Frequency Peaker}
\acro{HI}[H{\sc i}]{Neutral Atomic Hydrogen} 
\acro{HST}{\textit{Hubble Space Telescope}}
\acro{IAU}{International Astronomical Union}
\acro{IFRSs}{Infrared Faint Radio Sources}
\acro{IFRS}{Infrared Faint Radio Source}
\acro{IR}{Infrared}
\acro{JY}[Jy]{Jansky\acroextra{, 1 Jy = $10^{-26} \times \mathrm{W~ m}^{-2}~\mathrm{Hz}^{-1}$}} 
\acro{LLS}{Largest Linear Size}
\acro{LFAA}{Low-Frequency Aperture Array}
\acro{LMC}{Large Magellanic Cloud}
\acro{LSO}[LSOs]{Large Scale Objects}
\acro{MACHO}{Massive Astrophysical Compact Halo Objects}
\acro{MC}[MCs]{Magellanic Clouds}
\acro{mc}[MC]{Magellanic Cloud}    
\acro{MCELS}{Magellanic Cloud Emission Line Survey}
\acro{MW}{Milky Way}
\acro{MIRIAD}[\textsc{Miriad}]{Multichannel Image Reconstruction, Image Analysis and Display}
\acro{MIT}{Massachusetts Institute of Technology}
\acro{MOST}{Molonglo Observatory Synthesis Telescope}
\acro{MQS}{Magellanic Quasars Survey}
\acro{MRC}{Molonglo Reference Catalogue of Radio Sources}
\acro{MWA}{Murchison Widefield Array}
\acro{NRAO}{National Radio Astronomy Observatory}
\acro{NVSS}{NRAO VLA Sky Survey}
\acro{OPAL}{Online Proposal Applications \& Links}
\acro{OVV}{Optically Violent Variable Quasars}            
\acro{PAF}{Phased Array Feed}
\acro{pc}{parsec\acroextra{: 1 pc $\simeq 3.09 \times 10^{16}$ m}}
\acro{PMN}{Parkes-MIT-NRAO}
\acro{PNe}{Planetary Nebulae}
\acro{QSO}{Quasi-Stellar Object}
\acro{RA}{Right Ascension}
\acro{RFI}{Radio-Frequency Interference}
\acro{RMS}[rms]{Root Mean Squared}
\acro{SDSS}{Sloan Digital Sky Survey}
\acro{SED}{spectral energy distribution}
\acro{SI}[$\alpha$]{Spectral Index\acroextra{, $S \propto \nu^\alpha$}}
\acro{SKA}{Square Kilometre Array}
\acro{SMB}{Super Massive Blackholes}
\acro{SMC}{Small Magellanic Cloud}
\acro{SN}{Supernova}
\acro{SNR}{Supernova Remnant}
\acro{SNRs}{Supernova Remnants}
\acro{SUMSS}{Sydney University Molonglo Sky Survey}
\acro{TOPCAT}[\textsc{Topcat}]{Tool for OPerations on Catalogues And Tables}
\acro{USS}{Ultra Steep Spectrum}
\acro{WBAC}{Wide-Band Analogue Correlator}
\acro{WIFES}[WiFeS]{Wide-Field Spectrograph}
\acro{WISE}{Wide-Field Infrared Survey Explorer}
\acro{VLBI}{Very Long Baseline Interferometry} 
\acro{VLSR}[\textbf{$v_{lsr}$}]{Velocity in the Line of Sight}
\end{acronym}




\section{Introduction}

Various peoples of the southern hemisphere, including Australian Aborigines, incorporated the \ac{MC} as an important part of their culture. These obvious astronomical objects of the southern sky were seen by Portuguese explorer Ferdinand Magellan in the 16$^\mathrm{th}$ century. He described the \ac{MC} as the most vivid celestial objects of the southern sky seen during his mesmerising expeditions \citep[for more details on historical observations of \ac{MC} see:][]{1996ASPC..112...91F}. 

The \ac{LMC} is thought to be a barred spiral galaxy, dynamically disrupted due to its proximity to the \ac{MW} and other interactions with the \ac{SMC} \citep{1974ApJ...190..291M}. The \ac{LMC} is currently $\sim$50~kpc \citep{2013Natur.495...76P,2019Natur.567..200P,Riess_2019} from the centre of the \ac{MW}, while the \ac{SMC} is further away at a distance of $\sim$60~kpc \citep{2005MNRAS.357..304H}. In addition, this well-known distance and motion of the \ac{LMC} allows for accurate calculations of the total energy that has been released from \ac{SNRs} \citep{2016A&A...585A.162M,2017ApJS..230....2B,2019A&A...631A.127M,2019AJ....158..149L,2021MNRAS.500.2336Y}. A detailed optical study of 715 \ac{LMC} \ac{PNe} is presented by \citet{2013MNRAS.436..604R} and \citet{2014MNRAS.438.2642R}. To date, 31 of these objects have been detected at radio wavelengths by \citet{2008SerAJ.177...53P}, \citet{2009MNRAS.399..769F}, \citet{2009A&A...503..855V} and \cite{2017MNRAS.468.1794L} with 28 having a complete radio-surface-brightness-to-diameter ($\sum$ -- $D$) relation derived from 6~cm surveys. Each of these objects have been observed at multiple radio wavelengths from 3 to 20~cm. Since the \ac{LMC} is seen almost face-on from the \ac{MW}, it is a perfect laboratory for studying complete populations with accurate energy measurements \citep{Riess_2019}. 

The last few decades have seen extensive multi-frequency surveys of the sky in the direction of these two galaxies, primarily  investigating objects within the \ac{MC} themselves. These various surveys present a prime opportunity to study not just intrinsic \ac{mc} objects but also a large population of background \ac{AGN} and quasars located in the sky area around the \ac{LMC}. This massive collection of data allows for a high level analysis of the \ac{AGN} population with reasonable spectral resolution, sensitivity and spatial resolution.

Large scale radio continuum surveys by \cite{1996A&AS..120...77F,1998PASA...15..128F,1998A&AS..130..421F} allowed for a detailed overview of \HII\ and star forming regions intrinsic to the \ac{MC}. 469 discrete sources in the direction of the \ac{LMC} were detected using the Parkes telescope between 1.4 and 8.55~GHz \citep{1995A&AS..111..311F}. Of these, 209 have been classified as \HII\ regions or \ac{SNRs}. This analysis has allowed for a detailed understanding of the overall structure of the \ac{LMC} and \ac{SMC}. Additionally, the detection of these objects allowed cross identification with infrared surveys to investigate the radio--to--infrared comparison \citep{1998A&AS..130..441F}. Most recently, \cite{2018MNRAS.480.2743F}, presented a survey of the \ac{MC} using a low-frequency wideband telescope, the Murchison Widefield Array \citep{2015PASA...32...25W}.

Large background source catalogues have a twofold significance: \textbf{1)} they enable population studies to understand the evolution of radio objects and by -- extension -- the evolution of the Universe, and \textbf{2)} they allow for the study of foreground media, through Faraday rotation imparted on polarised sources. These Faraday rotation measurements further improve the understanding of the magnetic field strength and direction as measured by \cite{1991A&A...252..475H}. Building on this, \cite{2005Sci...307.1610G,2012ApJ...759...25M,2017MNRAS.467.1776K} utilised these background catalogues to more accurately determine the magnetic field strength within the \ac{LMC}.  

From this wealth of data, combined with new surveys and observations, we can further understand the evolution and composition of the \ac{AGN} population. In this paper, we describe the process of source finding and cross-matching these various source lists to construct a large catalogue of radio sources in the direction of the \ac{LMC}. We examine each source to determine its radios properties. 
Our new catalogue presented here is best compared and similar to the Australia Telescope Large Area Survey \citep[ATLAS; at 1400~MHz][]{2006AJ....132.2409N,2008AJ....135.1276M} that covered a total of $\sim$6~deg$^2$ on the sky, down to a \ac{RMS} noise level of $<$30~$\mu$Jy. The ATLAS survey uncovered over 3\,000 distinct radio sources out to a redshift of 2. Our similar resolution and increased sky coverage will serve to significantly increase our knowledge of these distant objects.


\section{Observational Data}
 \label{sec:LMC_Surveys}
We have used a number of radio continuum images from archival data to detect discrete sources in the direction of the \ac{LMC}. These data have been taken with the \ac{MWA}, \ac{ASKAP}, the \ac{MOST} and the \ac{ATCA} telescopes (details are shown in Table~\ref{tab:Surveys}). These observations were originally used to detect and categorise sources and regions intrinsic to the \ac{LMC}. Below, we describe each of the data sets that we used in this study.

\subsection{MWA Images}
 \label{sec:MWA}
Low-frequency surveys are becoming increasingly important in the understanding of objects such as young \ac{SNRs}, \ac{CSS} and \ac{GPS} sources. In particular, this is important for \ac{GPS} sources as the turnover can be more accurately identified at this end of the radio frequency bandwidth. As part of the \ac{GLEAM} survey \citep{2015PASA...32...25W}, \citet{2018MNRAS.480.2743F} presented a series of selected images of the \ac{MC}. We used the 0.2~GHz (robust=$-$1) image, which has the lowest noise and best angular resolution for this study of individual sources (see Table~\ref{tab:Surveys}).

\subsection{\ac{ASKAP}-Beta Image}
 \label{sec:ASKAPImage}

In this paper, we present a new 0.843~GHz image from the Australian Square Kilometre Array Pathfinder (ASKAP) Telescope \ac{BETA} \citep{2014PASA...31...41H} that covers a $9.5\degr \times 9.8\degr$ area of the \ac{LMC} field. This observation was performed in 2015~May as an early \ac{ASKAP} testing field with a bandwidth of 1~MHz. At this time, \ac{ASKAP} was operated as a 6 antenna interferometer using first generation \ac{PAF} receivers. Each receiver produced nine independent beams which were arranged on the sky as a $3 \times 3$ grid with a grid spacing of 1.46~degrees. The extent of the \ac{LMC} was covered by mechanically re-pointing the antennas at four field centres to reposition the beam pattern on the sky. To increase the density of the mosaic pattern, an additional four field centres were observed, each of which was offset from the original field centre by 0.73 degrees in both RA and Dec (see \cite{2016MNRAS.457.4160H} for a more detailed description).  

The observations cycled through these 8 field centres approximately once per hour for 10 hours, producing the equivalent of a 72-pointing mosaic with a traditional, single-beam instrument. These data from each of the 72 beam/field combinations were processed independently using the \textsc{ASKAPSoft} software package and pipeline. The bandpass and flux density were calibrated using a separate observation of PKS~B1934$-$638. This source was observed at the centre of each of the 9 \ac{PAF} beams, and calibrated against the \cite{1994Reynolds} model. These data for each beam/field combination were then imaged and self-calibrated independently and the final images were merged into a linear mosaic. From this, we achieve an average \ac{RMS} noise level of 0.71~mJy~beam$^{-1}$ and a \ac{FWHM} of 61 $\times$ 53~arcsec.

Note that our \ac{ASKAP}-Beta image was made at a very early stage of the \ac{ASKAP} testing. As expected, a range of issues such as positional accuracy, beam forming accuracy, and calibration were discovered. We have made every effort to identify and correct these problems. These efforts have subsequently contributed to understanding and improving the performance of \ac{ASKAP} generally. As a result, a newer higher-sensitivity and resolution \ac{ASKAP} survey at 888~MHz was recently produced \citep[][]{2021arXiv210612013P}. This new survey is built upon results presented here.

\subsection{MOST 0.843~GHz Image}
To further enhance our frequency coverage of sources in the direction of the \ac{LMC}, we use the 0.843~GHz image from the \ac{MOST} using observations from 1981 that pre-dates the \ac{SUMSS} (Sect.~\ref{sec:SUMSS}) as described by \cite{1984IAUS..108..283M}. This image reaches a \ac{RMS} of between 0.3 and 0.6~mJy~beam$^{-1}$ with a resolution of 43~arcsec and a bandwidth of 3~MHz. This survey was initially constructed to maximise detection of \ac{SNRs} and \HII\ regions within the \ac{MC}. In this study, however, we decided to use only \ac{SUMSS} as a core catalogue (also see Sect.~\ref{sec:MOST}) to be used for the source spectral index investigation as we found the \ac{MOST} flux densities to be unreliable, especially for the fainter sources.

\subsection{\ac{ATCA} 1.384~GHz Image}
Radio continuum emission at $\sim$1.4~GHz is produced from a variety of sources on all scales. Typically, the use of only one telescope (or array configuration) is not sufficient to fully sample all spatial scales. To counteract this, we have chosen to use the 1.384~GHz images from \citet{2006MNRAS.370..363H,2007MNRAS.382..543H} and \citet{2009SerAJ.178...65P} who combine the \ac{ATCA} and Parkes telescopes \ac{LMC} surveys \citep{1991A&A...252..475H,1995A&AS..111..311F,1996A&AS..120...77F,1998A&AS..130..421F}. The \ac{ATCA} Telescope observations were undertaken with four different array configurations providing 40 unique baselines, with a maximum baseline of 750~m. This allows for high resolution observations to be combined with zero spacing (single-dish) data. Once combined, the final image has a \ac{FWHM} of 40~arcsec with a \ac{RMS} of $\sim$0.5~mJy~beam$^{-1}$. 
The data from these two telescopes were taken between 1994 and 1996, covering a $10.8\degr\times12.3\degr$ area. This field fully covers the \ac{LMC} with sufficient room beyond the extent of the diffuse emission of the \ac{LMC}, allowing for a detailed study of various source types to be undertaken.

\subsection{\ac{ATCA} 4.8 and 8.64~GHz Images}
\cite{2005AJ....129..790D} observed the \ac{LMC} using the \ac{ATCA}'s dual band observing mode; one 128~MHz band was placed at a central frequency of 8.64~GHz and the second at a central frequency of 4.8~GHz. The main science goal for this survey was to observe and identify \ac{SNRs} and \HII\ regions that reside within the \ac{LMC}. These data were combined with observations made with the Parkes telescope \citep{1991A&A...252..475H,1995A&AS..111..311F,1996A&AS..120...77F,1998A&AS..130..421F} to allow the broad, diffuse radio structure of the \ac{LMC} to be analysed. 7085 separate \ac{ATCA} pointing centres were used to cover the $6\degr\times6\degr$ area, with additional sampling around the \ac{30Dor} region. With angular resolutions of 33~arcsec (at 4.8~GHz) and 20~arcsec (at 8.64~GHz), images were created to be sensitive to the target objects and at the same time closely matching the then-existing X-ray, optical and \ac{IR} surveys. The resolution from the 352-m and 367-m arrays (configured by excluding the sixth antenna) presents an issue of poor resolution ($>$1~arcmin) so that sources could not easily be distinguished as background objects or objects within the \ac{LMC}. From this, a \ac{RMS} sensitivity of 0.28~mJy~beam$^{-1}$ and 0.5~mJy~beam$^{-1}$ was achieved for the 4.8 and 8.64~GHz images, respectively.

\subsection{Supplementary Surveys }
\label{sec:SupplementarySurveys}
We make use of an additional four radio surveys and five optical catalogues as summarised in Tables \ref{tab:Surveys} and \ref{tab:OpticalSurveys} respectively. These supplementary surveys allow for radio continuum spectra to be sampled over the range of 0.408 to 20~GHz, to detect sources with curved or peak radio spectra. Additionally, the inclusion of the \ac{SUMSS} catalogue provides 0.843~GHz data for the full \ac{ATCA} 1.384~GHz field. We further include optical surveys to obtain redshift measurements for a small subset of objects.

\subsubsection{Molonglo Reference Catalogue of Radio Sources}
 \label{sec:MRC}

The \ac{MRC} \citep{1981MNRAS.194..693L} was established using the \ac{MOST} telescope between 1968 and 1974, detecting 12\,141 sources brighter than 0.7~Jy at 408~MHz. This was achieved with a \ac{RMS} of 18~--~60~mJy~beam$^{-1}$. The total survey area covers 7.85~sr ($\sim$25770~degrees$^2$), with an overall source density of 1\,500~sr$^{-1}$ ($\sim$0.456 per 1~degree$^2$). The catalogue is reasonably complete above $\sim$1~Jy, however, the flux measurements between 0.7 and 1~Jy may be unreliable.

\subsubsection{The Sydney University Molonglo Sky Survey }
 \label{sec:SUMSS}
The Sydney University Molonglo Sky Survey (SUMSS) \citep{1999AJ....117.1578B, 2003MNRAS.342.1117M} was completed in 2007 using \ac{MOST} and covering the Southern sky south of $-30\degr$ with $|b| > \pm10\degr$. This survey was designed as an extension of the \ac{NVSS} on the same grid and at a similar resolution. However, the \ac{SUMSS} Survey observed at 0.843~GHz with a bandwidth of 3~MHz compared to the $\sim50$~MHz bandwidth of the 1.4~GHz \ac{NVSS} survey. These two surveys, \ac{SUMSS} and \ac{NVSS}, overlap in the region \mbox{--40$<$Dec$<-30\degr$} and catalogues have been published that contain 210\,412 and 1\,773\,484 sources, respectively \citep{2002yCat.8065....0C,2008yCat.8081....0M}. While the \ac{RMS} noise level of \ac{SUMSS} ($\sim$1~mJy~beam$^{-1}$) is about twice that of \ac{NVSS}, the $u-v$ coverage for \ac{SUMSS} is superior to \ac{NVSS}, allowing for extended objects with lower surface brightness to be detected.

\subsubsection{The \ac{PMN}}

The Parkes-\ac{MIT}-\ac{NRAO} Survey of the sky South of Dec=$+10\degr$ was performed in 1990 using the Parkes Radio Telescope outfitted with the \ac{NRAO} seven beam receiver at 4.850~GHz \citep{1993AJ....105.1666G} with \ac{FWHM} of 252~arcsec. 36\,640 sources across the Southern sky were detected down to a 5$\sigma$ level with a \ac{RMS} of $>$4.2~mJy~beam$^{-1}$. As for the \ac{SUMSS}, we make use of this catalogue but not of the images.

\subsubsection{The Australia Telescope 20 GHz Survey}

The \ac{AT20G} is a blind survey that looks for objects above a given \ac{RMS} anywhere in the given sky patch. It was conducted at 19.904~GHz using a bandwidth of 248~MHz that surveyed the entire Southern sky at Dec$<0\degr$. The primary objective was to categorise and analyse radio populations at high frequencies on an unprecedented scale. \cite{2010MNRAS.402.2403M} have presented a catalogue of 5\,890 sources detected at a flux density level of at least 40~mJy with a 91 per cent completeness above 100~mJy for declination south of $-15\degr$, and 79 per cent above 50~mJy~beam$^{-1}$ (some data were lost due to weather for declinations above $-15\degr$).

\subsubsection{6dF Galaxy Survey}

The 6dF Galaxy Survey \citep{2005PASA...22..277J,2005ASPC..329...11J} is an optical survey conducted with the 1.2~m UK Schmidt Telescope with the goal of obtaining spectroscopy for a magnitude limited sample of galaxies to (K, H, J, $r_{F}$, $b_{J}$) = (12.75, 13.00, 13.75, 15.60, 16.75). This was achieved with the 6 degree Field (6dF) instrument, using 150 spectroscopic fibres positioned by a robot on the focal plane. From this, 136\,304 redshifts of galaxies were measured for galactic latitudes $|b| > $10~deg, over an area of 17\,000~deg$^2$. Each source in the 6dF catalogue was selected from the \ac{2MASX} catalogue \citep[][]{2000AJ....119.2498J,2006AJ....131.1163S} (K$_{tot} < 12.75$). A near-infrared selection has a twofold effect: \textbf{1)} the target sample is less affected by dust extinction; \textbf{2)} typically, candidates selected solely through optical means will be dominated by young, blue stellar populations. However, candidates chosen through near-infrared criteria are dominated by their oldest population of stars, allowing for more robust galactic mass measurement, and are more likely to host an \ac{AGN}.

\subsubsection{Magellanic Quasars Survey (MQS)}

\cite{2009ApJ...701..508K,2011ApJS..194...22K,2012ApJ...746...27K,2013ApJ...775...92K} classified a total of 6\,103 optical sources in the direction of the \ac{LMC} as \ac{QSO}'s using Australian Astronomical Telescope (AAT). Of these sources, 962 have redshift measurements. Crucially, these sources were selected using multiple methods to confirm that they do not belong to the \ac{LMC}. Each source was inspected for its mid-IR colours compared against existing optical surveys, with the Spitzer SAGE survey \citep{2006AJ....132.2268M} and the OGLE-III survey \citep{1994AcA....44..165U}.  
The redshift measurements presented in these surveys are of great importance as they allow us to disentangle sources within the \ac{LMC} against the background \ac{AGN}. Additionally, these surveys allow for the ease of cross-matching radio sources to the CatWISE2020 \ac{IR} source catalogue (as shown in Sect.~\ref{sec:CatWISE2020}).

\subsubsection{MACHO Survey}

The MACHO survey, as presented by \cite{2012ApJ...747..107K}, provides measurements of 663 \ac{QSO} candidates in the direction of the \ac{LMC} with the intent of searching for \ac{MACHO}. These sources were cross-matched with additional surveys, such as Spitzer SAGE, \ac{2MASS}, \textit{Chandra}, the {\it XMM-Newton} and a \ac{LMC} UBVI catalogs. The authors applied machine learning techniques to these data, allowing for a high confidence classification of each source. This gives an estimated false positive rate of 1~per~cent for cross identification with other surveys. This allows us to identify sources in our radio catalogue with \ac{QSO} candidates from \cite{2012ApJ...747..107K}.

\subsubsection{CatWISE2020 Catalogue}
 \label{sec:CatWISE2020}

The \ac{WISE} space telescope has mapped the whole sky at wavelengths of 3.4, 4.6, 11.6 and \SI{22}{\micro\metre} (hereafter W1, W2, W3 and W4) as presented by \cite{2010AJ....140.1868W}. From this, the CatWISE2020 catalogue was created \citep{Marocco2021} where all existing data products from all phases were combined to make the most sensitive all-sky mid-IR survey to date. In total, two complete and 10 additional all sky passes were made (data collected from 2010 to 2018), allowing for identification and rejection of asteroids and other solar system objects\footnote{\url{http://wise2.ipac.caltech.edu/docs/release/neowise/neowise_2019_release_intro.html}}. This survey was conducted until the telescope ran out of liquid helium, pushing the \ac{RMS} significantly higher than initially designed. Recently, \citet{Marocco2021} produced a catalogue of over 1\,890\,715\,640 entries for the W1 and W2 bands.

\begin{table*} 
       \caption{Radio continuum surveys of the \ac{LMC} used in this work. 
       Surveys flagged with an asterisk were used to create discrete sources catalogued (Table~\ref{tab:Catalogue}) and to estimate spectral index. $\dag$ indicates that the field is restricted to the 1.384~GHz image size.        }
    \begin{tabular}{ l l c c c c c c l }
    	\hline
   		Survey & Telescope & Freq.          & Field Size & \ac{RMS} & Synthesised & Mean Maj. \& Min. & Num. of & Reference  \\
   		&      & (GHz)     & (deg $\times$ deg) & (mJy~beam$^{-1}$)     &  \ac{FWHM} (\arcsec)  & Source Size  (\arcsec)       & Sources  & \\
    	\hline
        \ac{GLEAM}$^{*}$ & \ac{MWA}  & 0.200  & $10.8 \times 12.3 \dag$  & 17 -- 20  & 162 $\times$ 139 & 171 $\times$ 146 & \SourcesMWAsingle   & This Work\\
		\ac{MRC}$^{*}$   & \ac{MOST} & 0.408  & $10.8 \times 12.3 \dag$  & 18 -- 60  & 156 $\times$ 172 & N/A              & \SourcesMRC       & \cite{1981MNRAS.194..693L}\\
		\ac{SUMSS}$^{*}$ & \ac{MOST} & 0.843  & $10.8 \times 12.3 \dag$  & $\sim$1.2 & 45 $\times$ 47   & 54.27 $\times$ 46.98 & \SourcesSUMSS & \cite{2003MNRAS.342.1117M} \\
		\ac{ASKAP}-Beta  & \ac{ASKAP}& 0.843  & $9.5 \times 9.8$         & 0.71      & 61 $\times$ 53 & 66.81 $\times$ 56.73 & \SourcesASKAPBETA & This Work \\
		\ac{MOST}        & \ac{MOST} & 0.843  &  $7.7 \times 7.9$        & 0.30 -- 0.40 & 45 $\times$ 47   & 48.90 $\times$ 44.19 & \SourcesMOST   & This Work \\
		\ac{ATCA}$^{*}$  & \ac{ATCA} & 1.384  & $10.8 \times 12.3$       & 0.50      & 40 & 46.98 $\times$ 41.26 & \SourcesTwentyCm  & This Work \\
		\ac{ATCA}$^{*}$  & \ac{ATCA} & 4.80   & $6.0 \times 6.3$         & 0.28      & 33 & 40.97 $\times$ 34.89 & \SourcesSixCm     & This Work \\
		\ac{PMN}         & Parkes    & 4.85   & $10.8 \times 12.3 \dag$  & $\sim$8   & 252 & 312 $\times$ 254 & \SourcesPMN       & \cite{1993AJ....105.1666G} \\
		\ac{ATCA}$^{*}$  & \ac{ATCA} & 8.64   & $6.0 \times 6.3$         & 0.50      & 20 & 25.64 $\times$ 21.70 & \SourcesThreeCm   & This Work  \\
		\ac{AT20G}$^{*}$ &\ac{ATCA}  & 19.904 & $10.8 \times 12.3 \dag$  & $\sim$10  & 13.9 & N/A & \SourcesAT        & \cite{2010MNRAS.402.2403M} \\
        \hline
         \label{tab:Surveys}
    \end{tabular}   
\end{table*}

\begin{table*} 
       \caption{Optical surveys of the \ac{LMC} used in this work. }
    \begin{tabular}{ l c c c c l }
    	\hline
   		Survey & Telescope & Unique & Matched & Matched & Reference  \\
   		& & Source & Sources & Sources & \\
   		& & Count &  & with z & \\
    	\hline
		6dF & 1.2-m UK Schmidt Telescope & 540 & \OpticalMatchesSixDf\ & \OpticalMatchesWithZSixDf\ & \cite{2009MNRAS.399..683J} \\		
		MACHO & 50" Mt. Stromlo Telescope & 3\,407 & \OpticalMatchesMACHO\ & \OpticalMatchesWithZMACHO\ & \cite{2012ApJ...747..107K} \\	 
		Magellanic Quasars Survey. I & Anglo-Australian Telescope & 4\,679 & \OpticalMatchesKozOne\ & \OpticalMatchesWithZKozOne\ & \cite{2009ApJ...701..508K} \\
		Magellanic Quasars Survey. II & Anglo-Australian Telescope & 677 & \OpticalMatchesKozTwo\ & \OpticalMatchesWithZKozTwo\ & \cite{2012ApJ...746...27K} \\
		Magellanic Quasars Survey. III & Anglo-Australian Telescope & 565 & \OpticalMatchesKozThree\ & \OpticalMatchesWithZKozThree\ &
		\cite{2013ApJ...775...92K} \\
        \hline
         \label{tab:OpticalSurveys}
    \end{tabular}   
\end{table*}


\section{Radio Source Catalogue}
 \label{sec:rsc}

We generated five new source lists of discrete radio-continuum sources from the 0.843 (\ac{ASKAP}-Beta), 0.843 (MOST), 1.384, 4.8 and 8.64~GHz \ac{ATCA} images. When combined with the existing \ac{SUMSS}, a total of \SourcesTotalUnique\ unique discrete radio sources have been detected in at least one of these catalogues. The source finding for each image was performed as a blind search without the use of prior knowledge from any existing catalogues.

Sources detected in these lists are a combination of sources belonging to the \ac{LMC} (such as \ac{SNRs}, \HII\ regions or \ac{PNe}) and background objects in the direction of the \ac{LMC}. By cross-matching (radius of 10~arcsec) with objects in \textsc{Simbad} Database \citep{2000A&AS..143....9W}, no known objects from our Galaxy were detected in any of the previous radio surveys of this area that we study here in this paper. Galactic radio sources are expected to have a low surface density in the direction of the \ac{LMC} due to its high Galactic latitude \citep{1998A&AS..130..421F}.

\subsection{Source Finding and Cross-matching}
\label{sec:SourceFinding} 

The \textsc{Aegean} Source Finding suite of tools \citep{2012ascl.soft12009H, 2012MNRAS.422.1812H,2018PASA...35...11H} have been used to process the core radio surveys, creating a catalogue of the sources within the field, including source positions, sizes and integrated flux densities. A two-step process was applied to the images for source detection, firstly forming a background and \ac{RMS} map, and subsequently running the source finding routine. The background and \ac{RMS} maps were created with The Background And Noise Estimation tool (\textsc{Bane}) using a moving box-car smoothing method. Using this \ac{RMS} map improves the integrity of the flood fill operation used in \textsc{Aegean} as it is able to more accurately determine the source extent, and thus the integrated flux density of the source. For both \textsc{Aegean} and \textsc{Bane} we set tuning parameters to their default values. The cross identification and visual analyses were performed with the \ac{TOPCAT} \citep{2005ASPC..347...29T,2011ascl.soft01010T} and \ac{DS9} applications \citep{2003ASPC..295..489J}.

\begin{figure}
		\centering
		\includegraphics[width=0.475\textwidth]{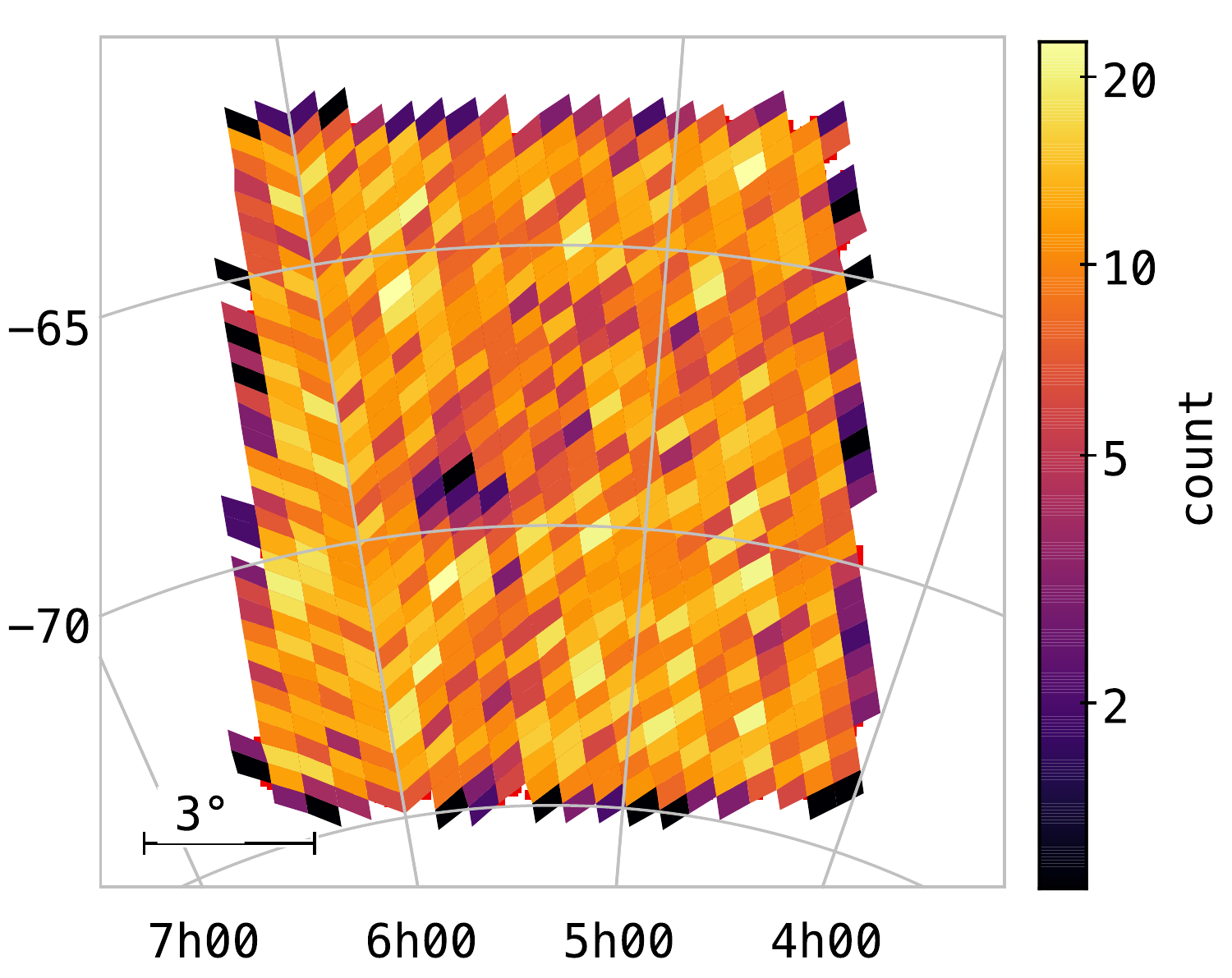} \\
		\caption{A map of the source counts in HEALPix pixel map in equatorial coordinate system (X axis is RA\,J2000 and Y axis is Dec\,J2000). The map is produced using \ac{TOPCAT} with HEALPix level 7 (pixel area is $\approx0.21$~deg$^2$). As scale bar shows, black coloured pixel indicate low density source population which is in agreement with intentionally avoided fields such as the \ac{30Dor} and outer edge of the survey where our coverage is limited.}
		\label{fig:sourcedensity}
\end{figure}

We used the following five steps to create the final source lists presented in this paper:
\begin{enumerate}
    \item Initially we began with a 5$\sigma$ detection level, creating preliminary source lists from the 0.843 (\ac{ASKAP} and \ac{MOST} images), 1.384, 4.8 and 8.64~GHz images. 
    \item 
    We manually examined sources that were close to S/N$ \sim$5 and $\theta_{major}/\theta_{minor}>2$ to determine if they were spurious. Detections that were not genuine (e.g. due to imaging artefacts, or an increased noise level towards the field edge as evident in Figure~\ref{fig:sourcedensity}) were flagged and removed.
    \item We visually inspected large, extended sources belonging to the \ac{LMC} using the \textsc{Simbad} Database and HASH PN database \citep[][hashpn.space]{Parker2016} to identify known \ac{SNRs}, \ac{PNe} and \HII\ regions \citep{1998PASA...15..128F}. The \ac{SNRs} and \ac{PNe} found here are addressed in \citet{2009MNRAS.399..769F} and \citet{2017ApJS..230....2B} and are excluded together with all known \HII\ regions from the catalogue presented in this paper. 
    \item The \ac{ATCA} 1.384~GHz source list was taken as the starting point which for the cross-match with the other source lists extracted from the above listed radio surveys. We use a nearest neighbour matching radius of 30~arcsec to determine whether there was a detection at any other frequencies. With a secondary internal match we were able to identify sources that have multiple components at the higher frequencies of 4.8 and 8.64~GHz, while being unresolved at the lower frequencies, 0.843 and 1.384~GHz. After each cross-match, a best match position was created for each source to aid in subsequent cross-matching. This best-match position is taken from the highest-resolution component for the source.
    \item A second and final, targeted source finding sweep was conducted on all images using prior knowledge from three \ac{ATCA} source lists. We also apply this prior knowledge where the source does not have a detection above 5$\sigma$ for the image in question. Consequently, this allows for a second source finding pass to detect sources down to 3$\sigma$ and cross-matching (and merging) these into the existing source lists. 
\end{enumerate}

We estimate a false cross-match rate of $\sim$1~per~cent by shifting each source list in RA by three times the cross-matching criterion. For example, the 4.8~GHz to 1.384~GHz cross-match radius is 30~arcsec, hence we shift the 4.8~GHz image northward by 90~arcsec and repeat the cross-matching procedure. This removes any real matches so that we are left with only random matches. 

Supplementary source lists have been added to more accurately constrain the spectral index of each source as described in Sect.~\ref{sec:SupplementarySurveys}. \ac{AT20G} was the first of these supplementary source lists to be cross-matched with  all of the preceding surveys. The addition of this source list reveals a limited number of sources (\SourcesAT) but provides a significant increase in the positional accuracy and a wider frequency range which is essential to a better constrained source \ac{SED}. The best-match position was then replaced with the position from the highest angular resolution survey. In line with this, the next supplementary source list to be added was the \ac{SUMSS} catalogue, as it had the next best resolution. Similarly, the best match position was recalculated before repeating this cycle with the \ac{MWA}, \ac{MRC} and \ac{PMN} source lists. The \ac{PMN} catalogue was cross-matched with a 252~arcsec search radius due to its poor resolution compared to our other surveys. Unlike for the four core radio surveys (\ac{SUMSS} at 0.843~GHz and \ac{ATCA} at 1.384, 4.8 and 8.64~GHz), a further internal cross-match was not performed on these supplementary surveys. 

Multiple cross-matches of a source may be present throughout each of the source lists that have been acquired from existing surveys or generated through \textsc{Aegean}. This is due to the differing resolutions of each image, allowing for one detection to be resolved into multiple detections at higher frequencies. These sources have been flagged in our source lists and catalogued as ``Confused Source''. Sources of this nature have been excluded from the following analysis, because of their complexity.

Care must be taken when cross-matching these detections due to the differing resolutions and positional accuracies, both of which create false matches. These false matches and erroneous data points can be potentially corrected by understanding trends in the cross-match between different pairs of catalogues as shown in Sections~\ref{sec:Completeness}, \ref{sec:SourceFinderErrors} and \ref{sec:RMSEffects}. Once identified, a manual approach (source-by-source visual inspection) is used to investigate the erroneous properties {bf such as flux density, spectral index and exact position}. This cycle of analysis is vital in creating a reliable catalogue.

Additionally, extensive cross checking with the \ac{MCELS} in H$\alpha$, [S{\sc ii}], and [O{\sc iii}] images presented by \cite{1998PASA...15..163S} has been performed to identify sources in both the optical and radio bands. These matches are most likely intrinsic to the \ac{LMC} \citep[see for example][]{2021MNRAS.500.2336Y}. Radio sources with optical counterparts as detected in \ac{MCELS} have been excluded from further analysis in this work. We note that the \ac{MCELS} images have been continuum subtracted and no compact stellar optical sources can be detected. Hence, these objects detected in the \ac{MCELS} images are likely to have an \ac{LMC} origin and will be studied in our subsequent papers. Also, we make use of other existing optical catalogues that have known background sources, as determined from their redshifts (discussed in Sect.~\ref{sec:results}). 98 such radio sources from the \ac{MCELS} have extended emission (optically diffuse objects) while 129 sources have a compact optical object at or near the centre of the radio source. An additional cross check was made with known objects, such as 80+ \ac{SNRs}, 20+ \ac{PNe} and 100+ \HII\ regions. Variable sources may be present but are hard to detect as this catalogue has been compiled from datasets with sparse coverage over a long time-scale ($\sim$31~years).

A sample of the final catalogue is shown in Table~\ref{tab:Catalogue}. This table includes our combined selected radio source lists from all surveys apart from two 0.843~GHz lists from ASKAP-Beta and \ac{MOST} surveys as discussed in Sects.~\ref{sec:ASKAPflux} and \ref{sec:MOST}. The first column of Table~\ref{tab:Catalogue} is using the standard \ac{IAU} source naming format of Jhhmmss--ddmmss \citep{2000BaltA...9..564D}. This name has been derived from the position of the highest resolution detection for each source from all input source lists. Columns 2 and 3 give the source position from the highest-resolution detection of the source. The full catalogue is available through the VizieR Catalogue Service \citep{2000A&AS..143...23O}. For each survey in the final combined catalogue, we provide a source position, size, integrated flux density, peak flux density and a local \ac{RMS} measure. From this, we calculated a spectral index ($\alpha$) where possible which is defined as $S_{\nu}$~$\propto$~$\nu^\alpha$, where $S_{\nu}$ is flux density at frequency $\nu$.

\subsection{Selection Effects and Completeness}
\label{sec:Completeness}
Our catalogue of discrete radio sources presented here is not complete because of various problems including (but not limited to) detection limits, diffuse emission from the \ac{LMC}, imaging artifacts, variable and changing sensitivity (across the field of a given survey) as well as the resolution limitations for both foreground and background objects. This is most notable in the \ac{30Dor} region where large side-lobes are produced by strong and diffuse emission, especially in our \ac{ATCA} images. This area has been excluded from the analysis to prevent contamination from this emission and the spurious clean components created during the imaging phase (see Figure~\ref{fig:sourcedensity}). We define this exclusion area around \ac{30Dor} after inspecting the 1.384~GHz and 0.843~GHz images and then we applied this exclusion to all subsequent source finding catalogues. This is the area between RA(J2000) of 5$^h$30\arcmin and 5$^h$47$^m$ and Dec(J2000) from --68\D50$^m$ to --70\D. The \ac{RMS} estimates presented in this work are the best case measurement, i.e. taken from source free regions on the outskirts of the \ac{LMC}. As mentioned above (Sect.~\ref{sec:SourceFinding}), we have excluded all known \ac{SNRs} and \HII\ regions listed by \cite{2017ApJS..230....2B} and various other surveys covering sources that are a part of the \ac{LMC}. This has allowed us to have a less contaminated population of background (such as Quasars and \ac{AGN}) and foreground objects (mainly Galactic radio stars).

\subsection{Discrete Source Finding Errors}
\label{sec:SourceFinderErrors}

It is widely accepted that all source finders have an error rate\footnote{Includes the number of false positives and the missing real sources.} of at least a few per cent. However, this depends a lot on the threshold settings. The compromise is always between completeness (of real sources) and reliability. \textsc{Aegean} is no exception to this, as shown in \citet{2015PASA...32...37H}. The catalogue from each image was carefully visually inspected for spurious sources due to image artifacts and regions of strong emission. For example, detections were made along the edge of each image which were either incomplete or false detections, requiring a manual pass to remove these sources. This manual pass was extended to the catalogue as a whole, with a focus on regions of strong emission (either sidelobes or extended emission) and sources with unrealistic properties (such as extremely large gaussians, strong sources and source positions beyond the limits of the observed area). 
 
\subsection{Discrete Source Position and Integrated Flux Density Uncertainties}
 \label{sec:errors}

The richness of data available for this field allows us to compare and, if necessary, post-calibrate our new source lists. This was particularly useful for the \ac{ASKAP}-Beta data, which was re-calibrated against the more reliable \ac{SUMSS} catalogue.

\input{Files/main_fluxes}


\subsubsection{ASKAP-Beta and MOST Survey Positional Correction}
 \label{sec:ASKAPpos}
\begin{figure*}
		\centering
		\begin{tabular}{@{}cc@{}}
		    \includegraphics[width=.45\textwidth]{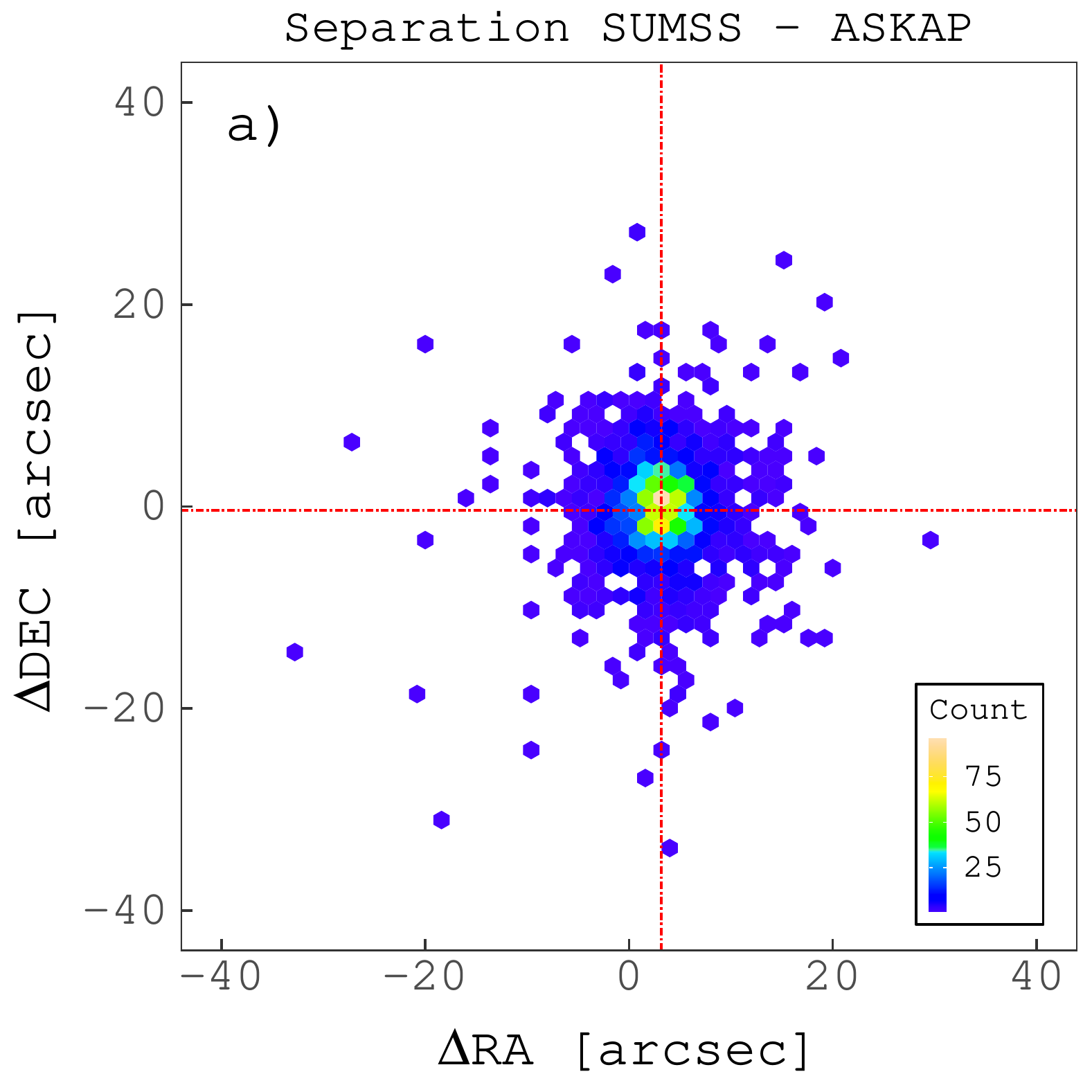} &
		    \includegraphics[width=.45\textwidth]{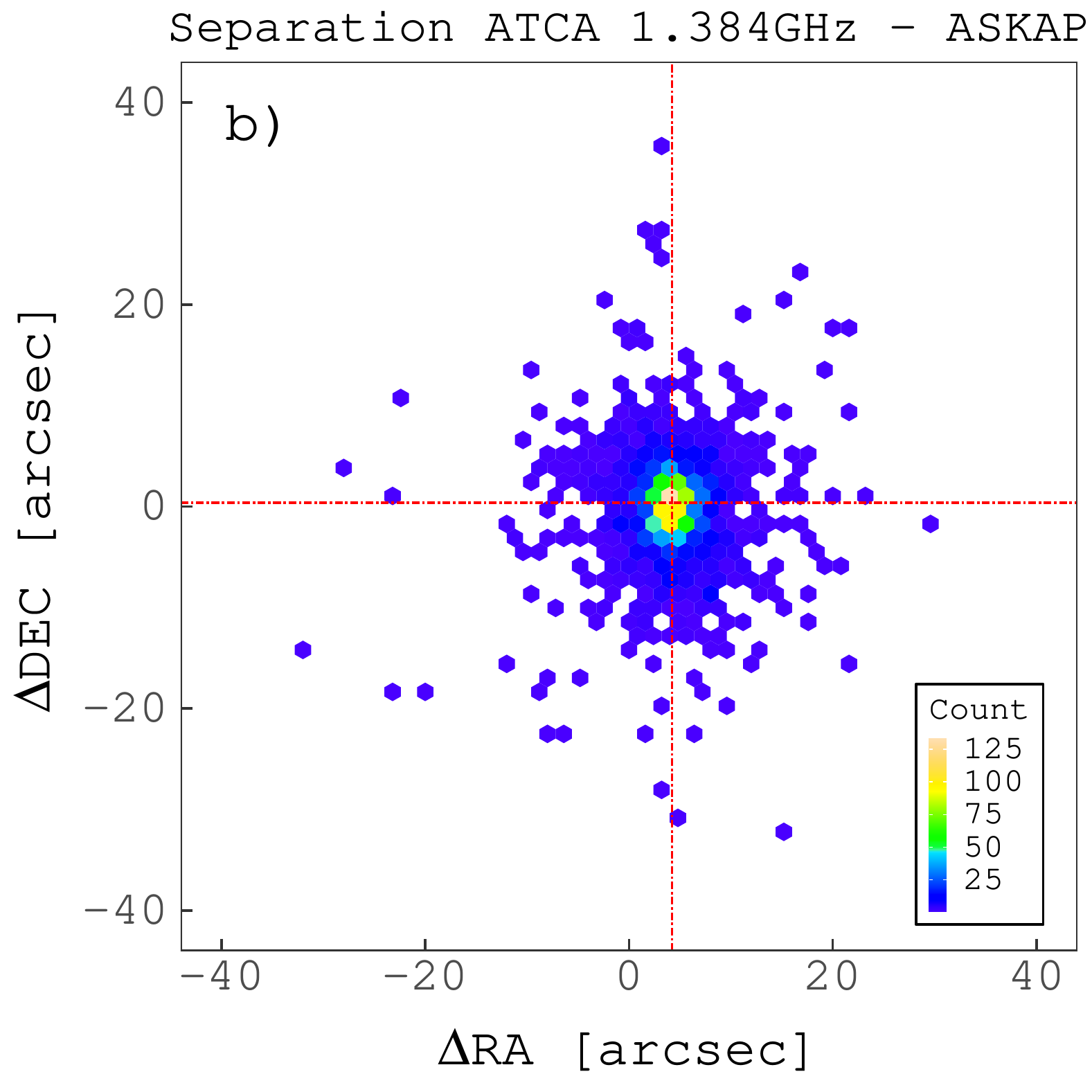} \\
		    \includegraphics[width=.45\textwidth]{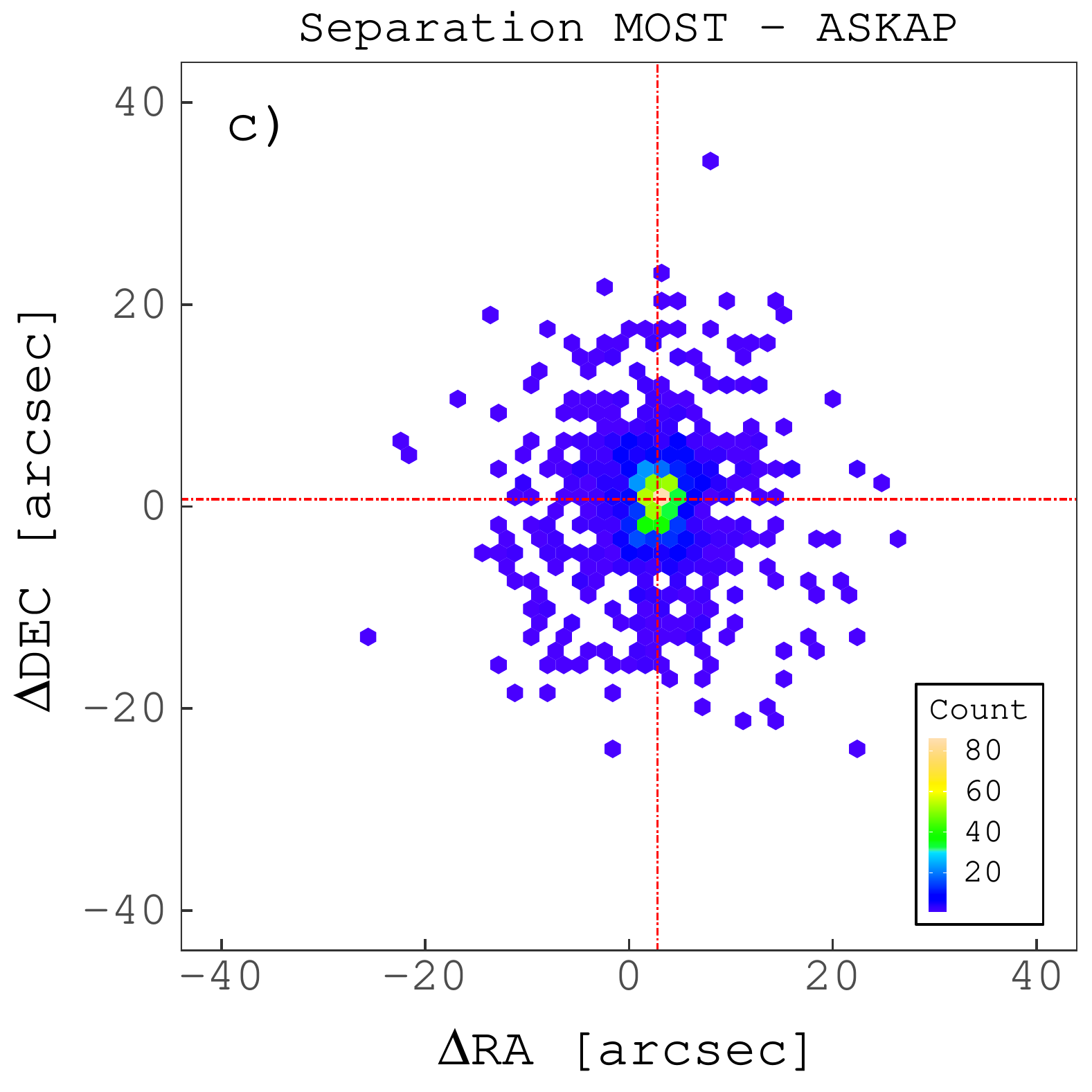} &
		    \includegraphics[width=.45\textwidth]{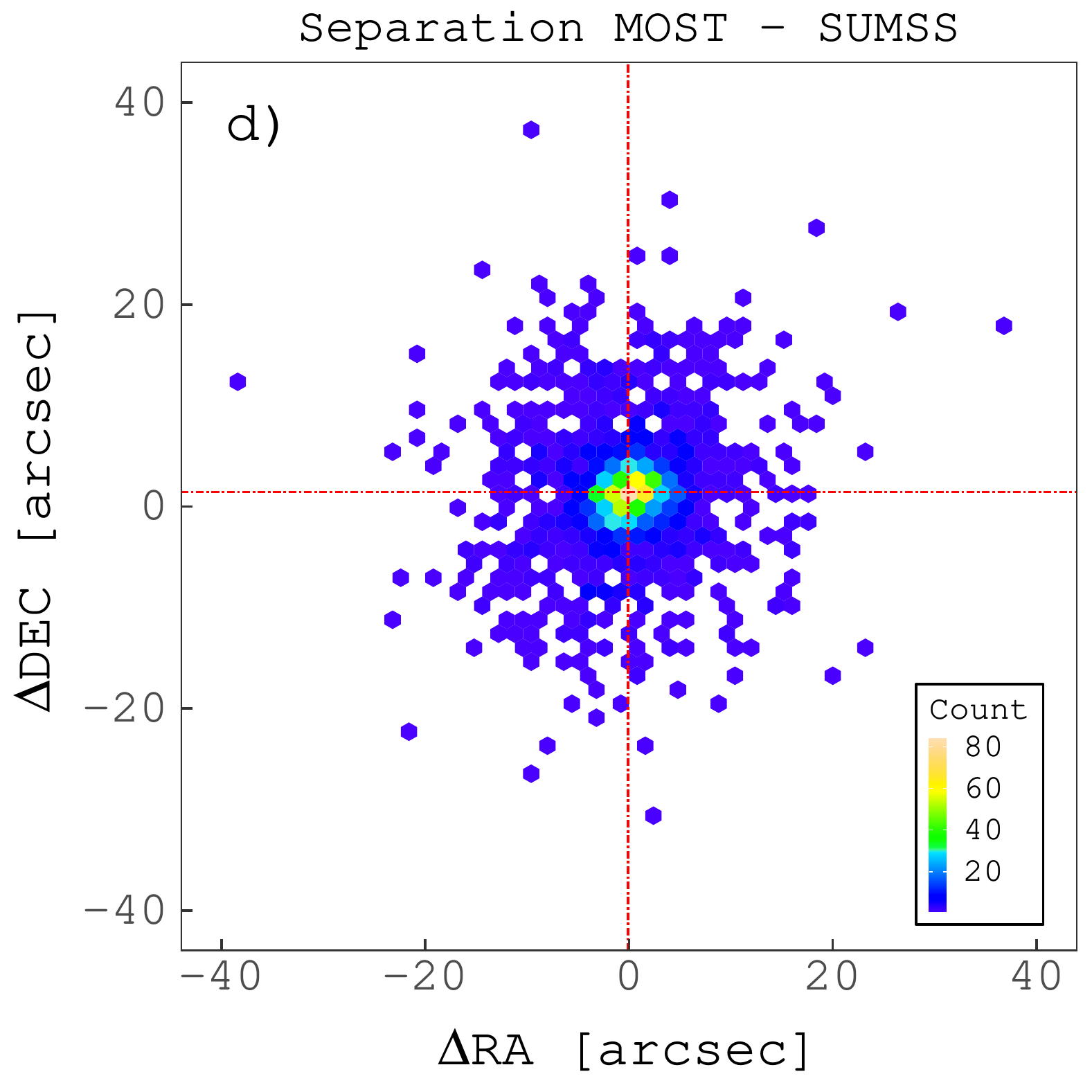} \\
		\end{tabular}
		\caption{Positional comparisons between \textbf{a)} \ac{SUMSS} -- \ac{ASKAP}-Beta, \textbf{b)} \ac{ATCA} 1.384~GHz -- \ac{ASKAP}-Beta, \textbf{c)} \ac{MOST} -- \ac{ASKAP}-Beta and \textbf{d)} 
		\ac{MOST} -- \ac{SUMSS} catalogues without applying  ASKAP-Beta positional corrections. The median RA and Dec offsets are shown with vertical and horizontal red dot-dashed line, respectively. The median positional offsets in RA and Dec, respectively, are: \textbf{a)} \OffsetSumssVSaskapRA\ and \OffsetSumssVSaskapDEC~arcsec, \textbf{b)} \OffsetAtcaTwcmVSaskapRA\ and \OffsetAtcaTwcmVSaskapDEC~arcsec, \textbf{c)} \OffsetSumssVSaskapRA\ and \OffsetSumssVSaskapDEC~arcsec and \textbf{d)} \OffsetMostVSsumssRA\ and 
		\OffsetMostVSsumssDEC~arcsec.}
		\label{fig:Separations}
\end{figure*}

To determine the positional accuracy of the ASKAP-Beta data set, we have made comparisons with the \ac{SUMSS} 0.843~GHz source list. From this, we find that the ASKAP-Beta dataset requires a $\sim$3~arcsec correction of its source position to be consistent with existing catalogues. Larger offsets are found when comparing the \ac{ATCA} 1.384~GHz catalogue with the ASKAP-Beta catalogue as the median positional offset between the two catalogues is \OffsetAtcaTwcmVSaskapRA\ and \OffsetAtcaTwcmVSaskapDEC~arcsec in RA and Dec respectively. A similar discrepancy is identified when positional comparisons are made with the 0.843~GHz \ac{MOST} source list. The median positional offsets are  \OffsetMostVSaskapRA\ and \OffsetMostVSaskapDEC~arcsec in RA and Dec, respectively, as shown in Figure~\ref{fig:Separations}. The cross-matching process was performed after correcting the positional error. A comparison of positional offsets between pairs of catalogues is shown in Figure~\ref{fig:Separations}.

\subsubsection{ASKAP-Beta Flux Density Post-Calibration}
\label{sec:ASKAPflux}

We make use of the \ac{SUMSS} survey as an established, external catalogue in order to determine the completeness and quality of our newly produced source lists of \ac{ASKAP}-Beta and \ac{MOST} surveys. 

It is now understood that the \ac{ASKAP}-Beta system and, by extension, the observations made during that time had an inaccurate amplitude calibration on the order of $\sim$10~--~15~per~cent. This affected the flux density measurement and caused an additional positional offset (Sect.~\ref{sec:ASKAPpos}). 

A systematic correction was derived from the comparison with the \ac{SUMSS} survey as no frequency interpolation is required to compare fluxes. Additionally, we have used the \ac{SUMSS} catalogue to measure the completeness and reliability of the 0.843~GHz MOST source list (as shown in Sect.~\ref{sec:Completeness}). Using these surveys, we perform a least squares linear fit to determine the systematic positional offset present within the \ac{ASKAP}-Beta dataset. Before correcting these data, the ratio of $S_{\mathrm{ASKAP-Beta}}/S_{\mathrm{MOST}}$ for the integrated flux was 0.880; similarly the ratio of $S_{\mathrm{ASKAP-Beta}}/S_{\mathrm{SUMSS}}$ was 0.893. From these results, we have applied a 12.75 per cent systematic increase to the \ac{ASKAP}-Beta fluxes. 

Even after applying the above corrections however, at lower flux densities ($<15\sigma$) our \ac{ASKAP}-Beta flux densities still show a large variance. In Figure~\ref{fig:IngtegratedFlux}a we show a direct comparison between the estimated flux densities and histograms of the flux density differences normalised to the weighted root sum square of the corresponding uncertainties (see section~4 of \cite{Gregory1991}). The plots clearly show that the deviation from the expected value is significant and it appears to be a non-linear systematic function of the flux density (i.e. the lower \ac{ASKAP} fluxes appear to be systematically underestimated). Not just flux scale, but poorly known \ac{ASKAP} beam shapes are a huge problem. The primary beam correction factors were poorly known and variable at the time of testing.

The resulting source list is provided in Table~\ref{tab:askaptable}. While we found that this list is useful for the confirmation (and detection) of faint sources, we decided not to use the \ac{ASKAP} data in the following analysis and in our combined catalogue shown in Table~\ref{tab:Catalogue}. we also note the problem that S$_{int}$/S$_{peak}$ is systematically higher than one, by typically 20 per cent.

\subsubsection{MOST Survey Flux Density Post-Calibration}
\label{sec:MOST}
The \ac{SUMSS} survey reaches a \ac{RMS} of $\sim$1~mJy~beam$^{-1}$ allowing for $\sim$78 per cent (\CommonMostVsSumss\ sources) of sources in the 0.843~GHz MOST source list to be cross-matched and compared against the sources within \ac{SUMSS}. From this, we find that all \ac{SUMSS} sources have an 0.843~GHz \ac{MOST} counterpart (within the overlap region). However, not all 0.843~GHz MOST sources within the same limits (\ac{RMS} and spatial) have a \ac{SUMSS} detection. As expected, we detect an additional \ExtraMostSources\ sources in the 0.843 GHz \ac{MOST} image that are undetected in \ac{SUMSS}.

\begin{figure*}
		\centering
		\begin{tabular}{@{}cc@{}}
			\includegraphics[width=.33\textheight]{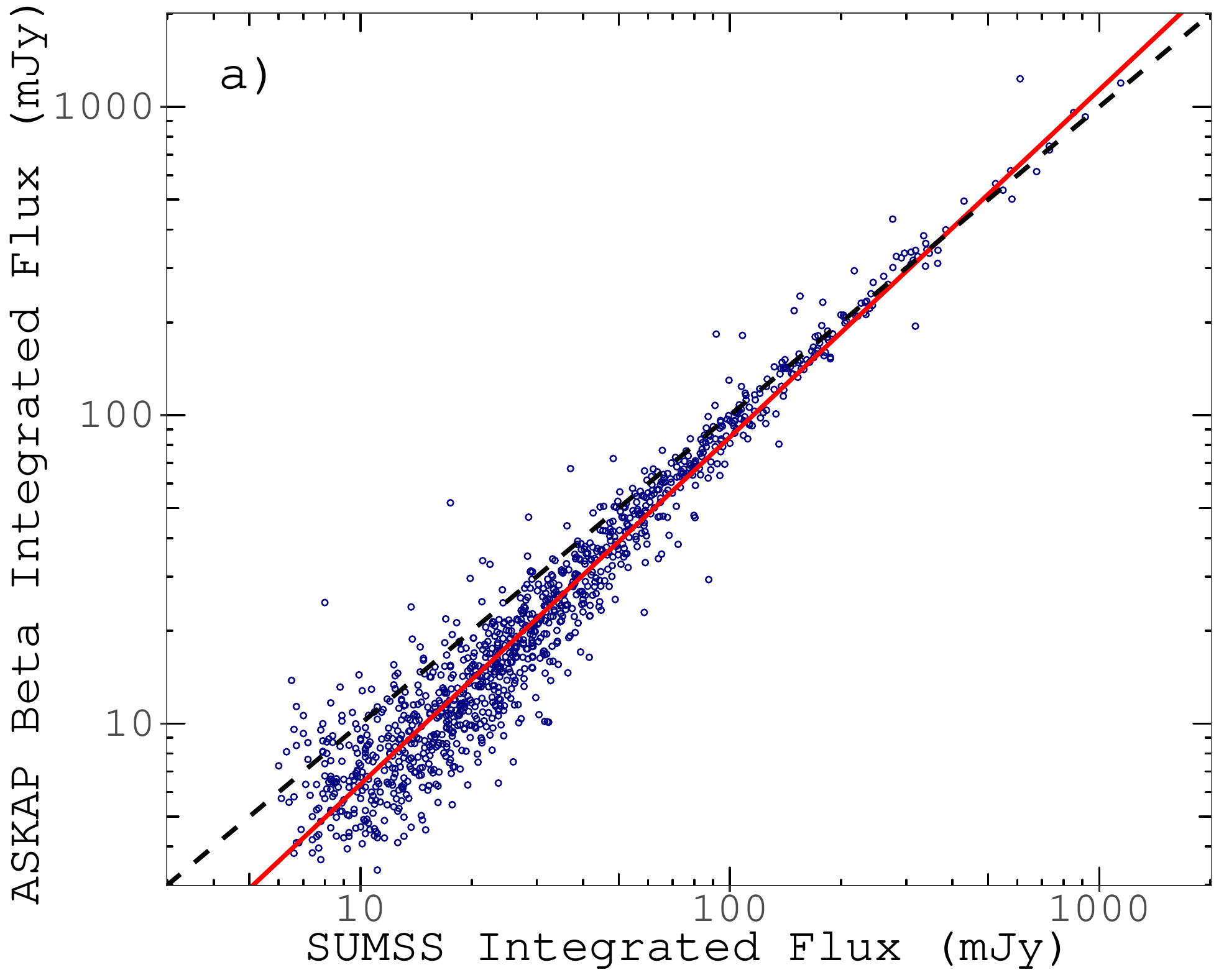} & 
			\includegraphics[width=.33\textheight]{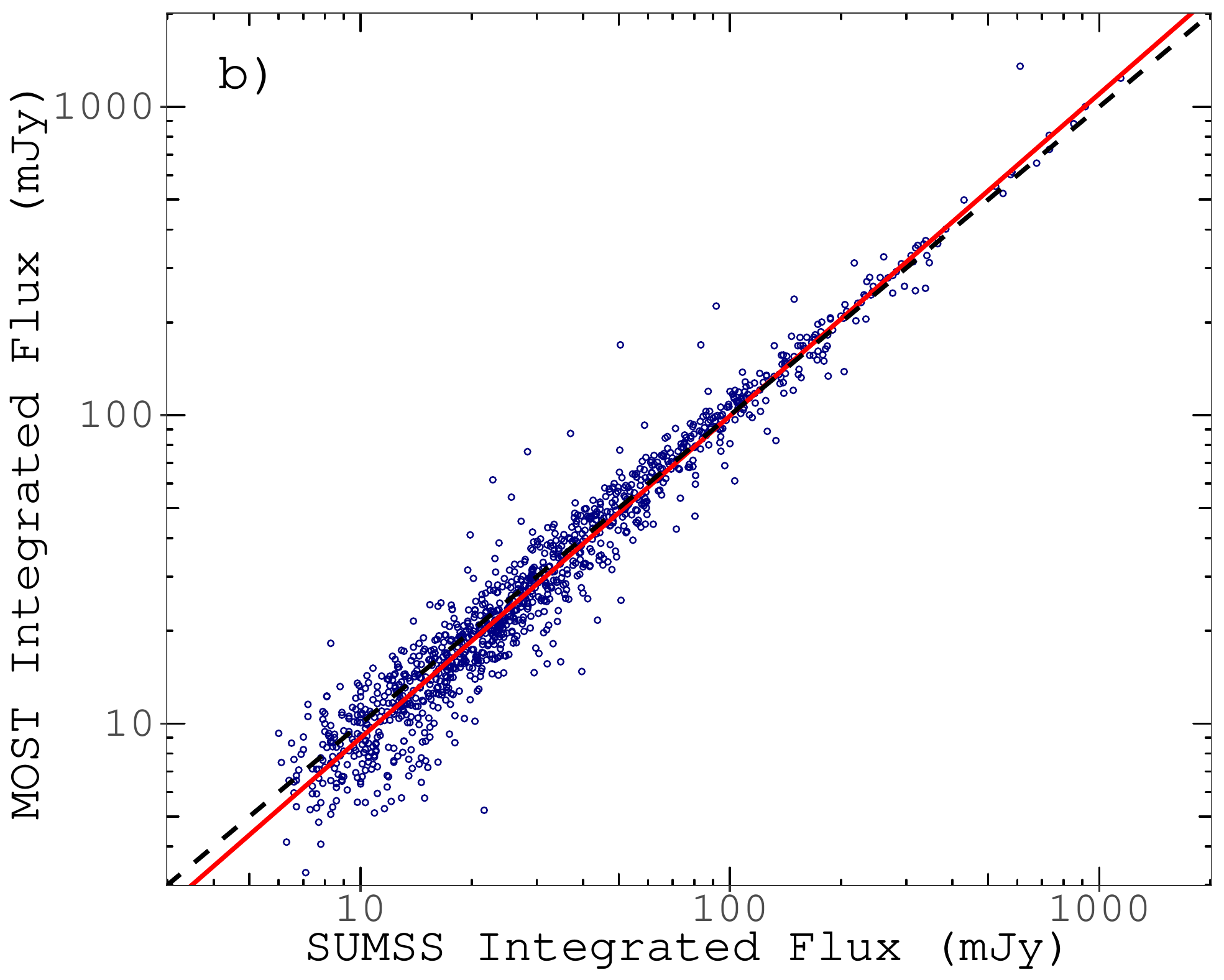}\\
			\includegraphics[width=.33\textheight]{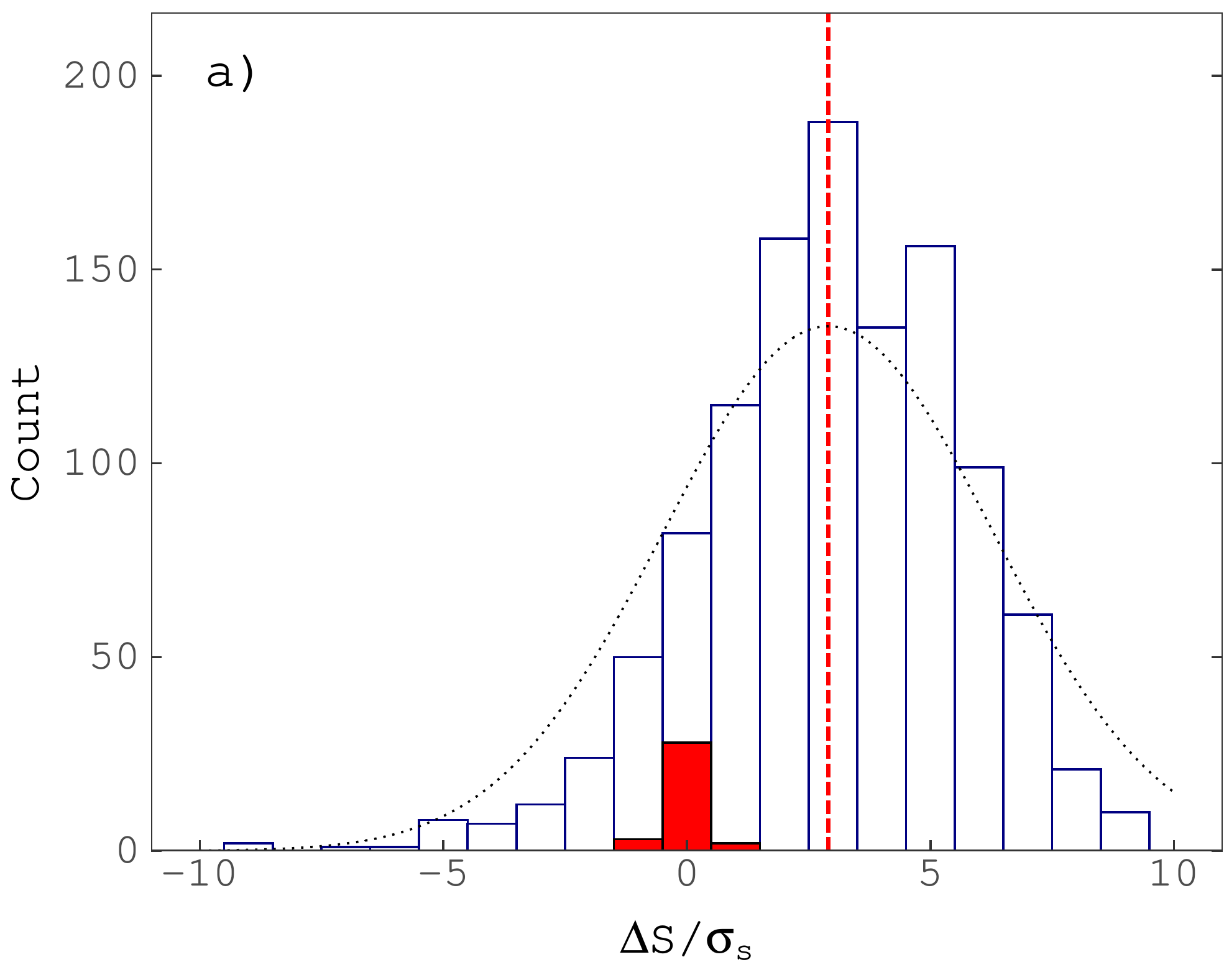} & 
			\includegraphics[width=.33\textheight]{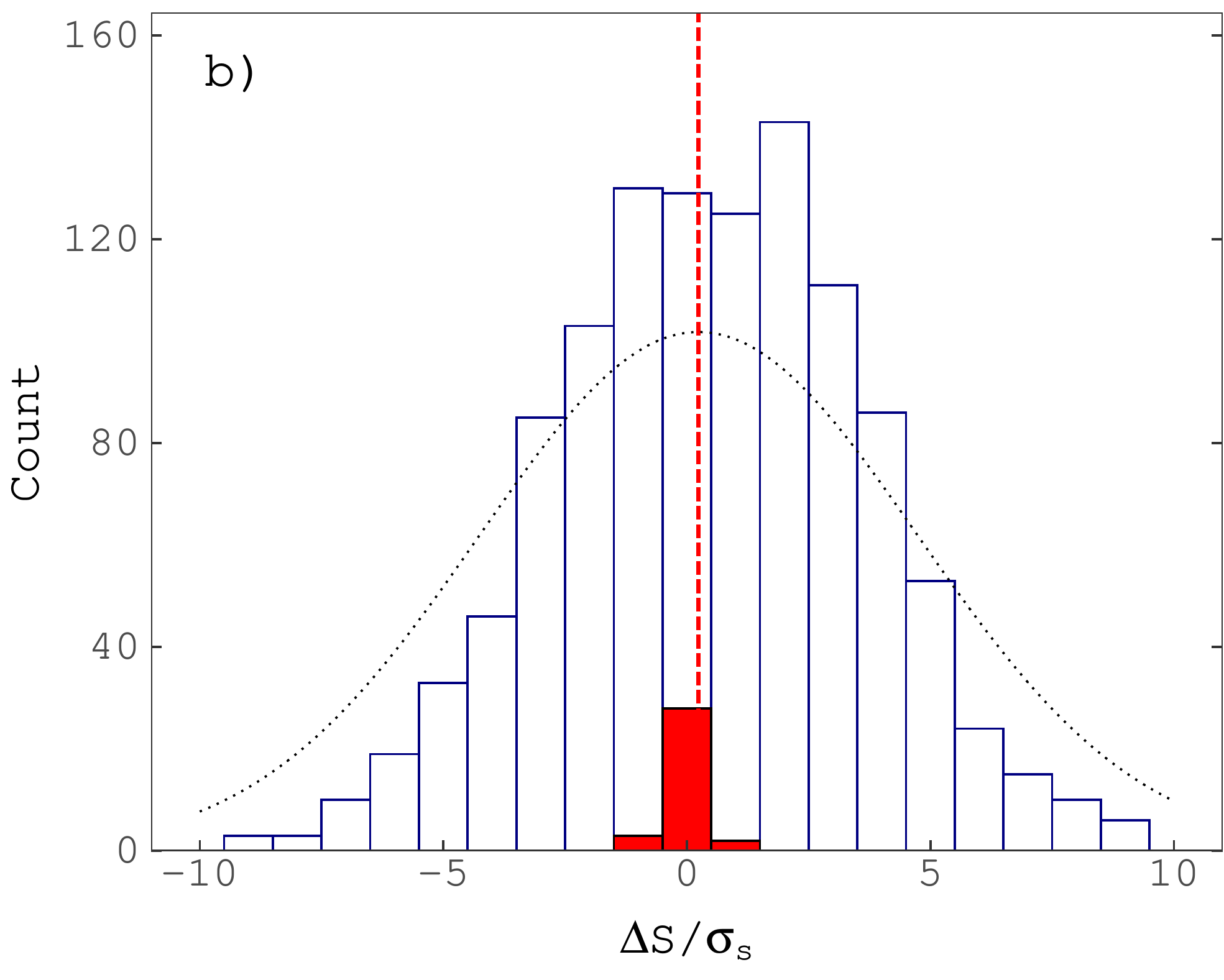}\\ 
	\end{tabular}
		\caption{{\bf Top row:} An integrated flux density comparison of the \ac{ASKAP}-Beta (a) and \ac{MOST} (b) 0.843~GHz source lists, with the \ac{SUMSS} (see Sects.~\ref{sec:ASKAPflux} and \ref{sec:MOST} for more details). In each figure, we represent a ratio of 1 with the black dashed line and the best fit line is in red. {\bf Bottom row:} Histogram of flux density differences weighted with the root sum of squares of corresponding uncertainties. Histogram has a bin size of 1 normalised flux difference. Vertical red dashed line represents the mean of the distribution. Dotted line is a Gaussian distribution  with the same mean and standard deviation as in the histogram. Red-filled histogram show the distribution of $\Delta S/\sigma_s$ between matched sources from \ac{PMN} and \ac{ATCA} \citep{2005AJ....129..790D}, respectively, at 4.8~GHz. 
		}
		\label{fig:IngtegratedFlux}
\end{figure*}

These \CommonMostVsSumss\ sources common to both lists, allow us to determine the flux density accuracy of our 0.843~GHz \ac{MOST} sources (as shown in Figure~\ref{fig:IngtegratedFlux}b). The ratio of the integrated flux densities for these two surveys ($S_{\mathrm{SUMMS}}/S_{\mathrm{MOST}}$) is 1.02$\pm$0.01. This ratio does not show any systematic offset between two source lists. In Figure~\ref{fig:IngtegratedFlux}b (lower right panel) we can see that the distribution of normalised flux density differences are much wider than acceptable. 

As for the \ac{ASKAP}-Beta flux density analysis, we can use the \ac{MOST} detections for cross-checking of faint source detections. However, we are not using these flux densities in source spectral index estimates and in constructing our main catalogue Table~\ref{tab:Catalogue}. We present the MOST 0.843~GHz source list with its positions and all measured flux densities as a separate Table~\ref{tab:mosttable}.

\input{Files/askap_fluxes}

\input{Files/most_fluxes}

\subsubsection{Other Flux Density Errors and Uncertainties}

Uncertainties in flux density estimates are mainly a product of gain calibration uncertainties ($\sigma_g$; usually given as a percentage of the measured flux density $S$) and the uncertainty ($\sigma_n$) caused by fitting in the presence of the local noise. The total uncertainty ($\sigma_{tot}$) can be then calculated as:
\begin{equation}
\sigma_{tot}^2 = \sigma_n^2 + (S\cdot\sigma_g)^2
\end{equation}

We applied a 10~per~cent uncertainty in the gain ($\sigma_g$) to our new \ac{ATCA} flux estimates at 1.384, 4.8 and 8.64~GHz, as recommended in \cite{harvey-smith2018}.

\begin{figure*}
	\centering
	\begin{tabular}{@{}cc@{}}
	\includegraphics[width=.35\textheight]{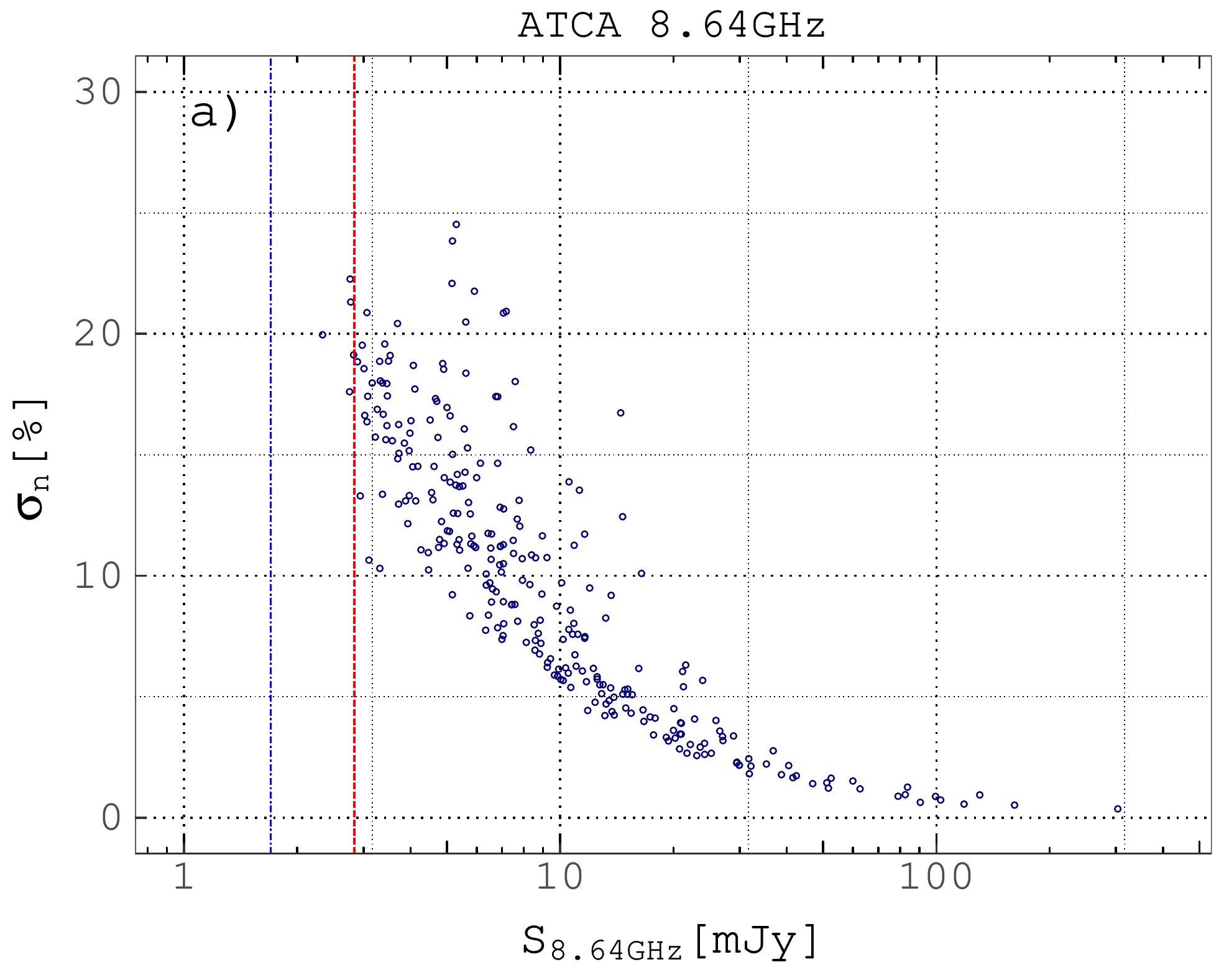} &
	\includegraphics[width=.35\textheight]{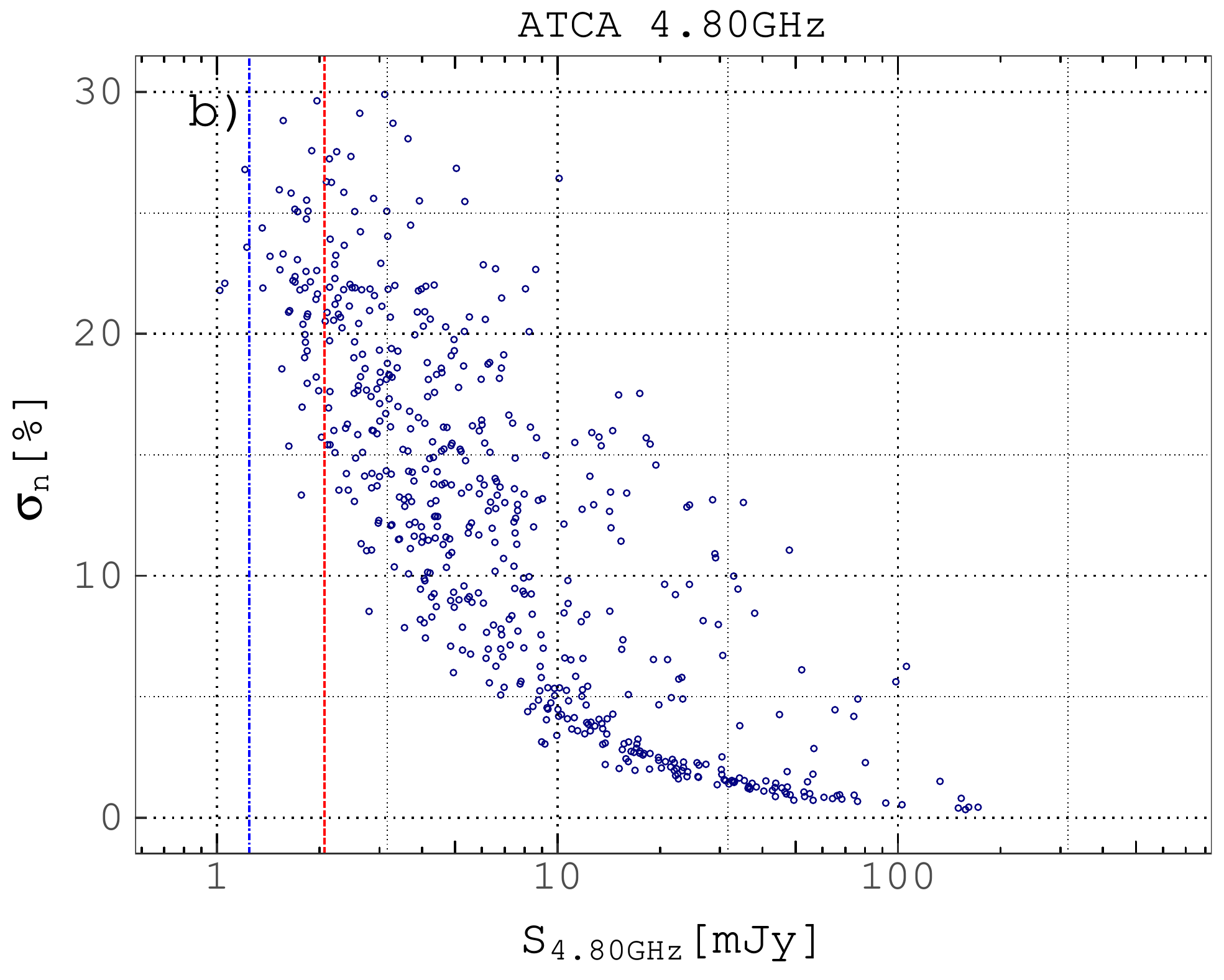} \\
	\includegraphics[width=.35\textheight]{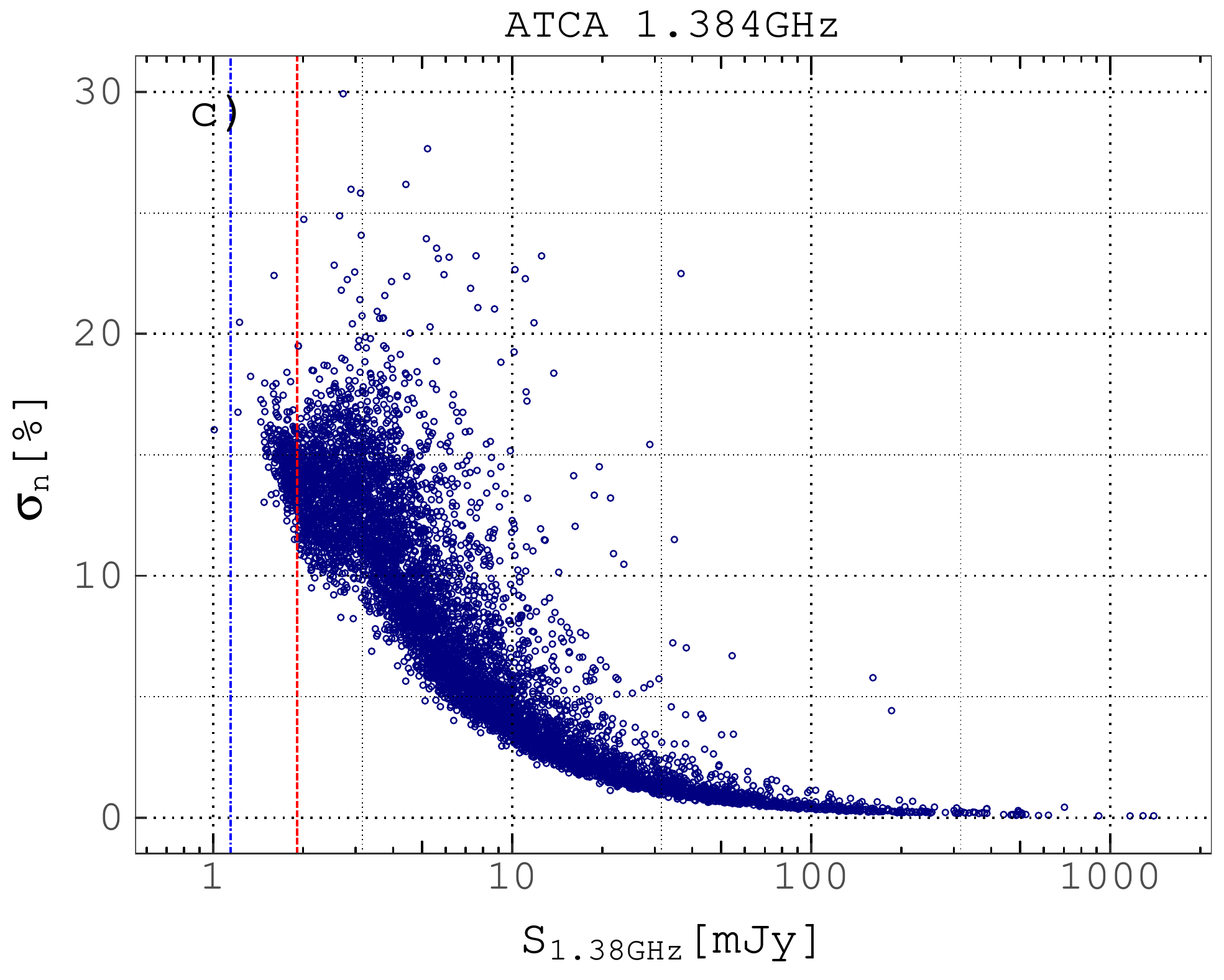} &
	\includegraphics[width=.35\textheight]{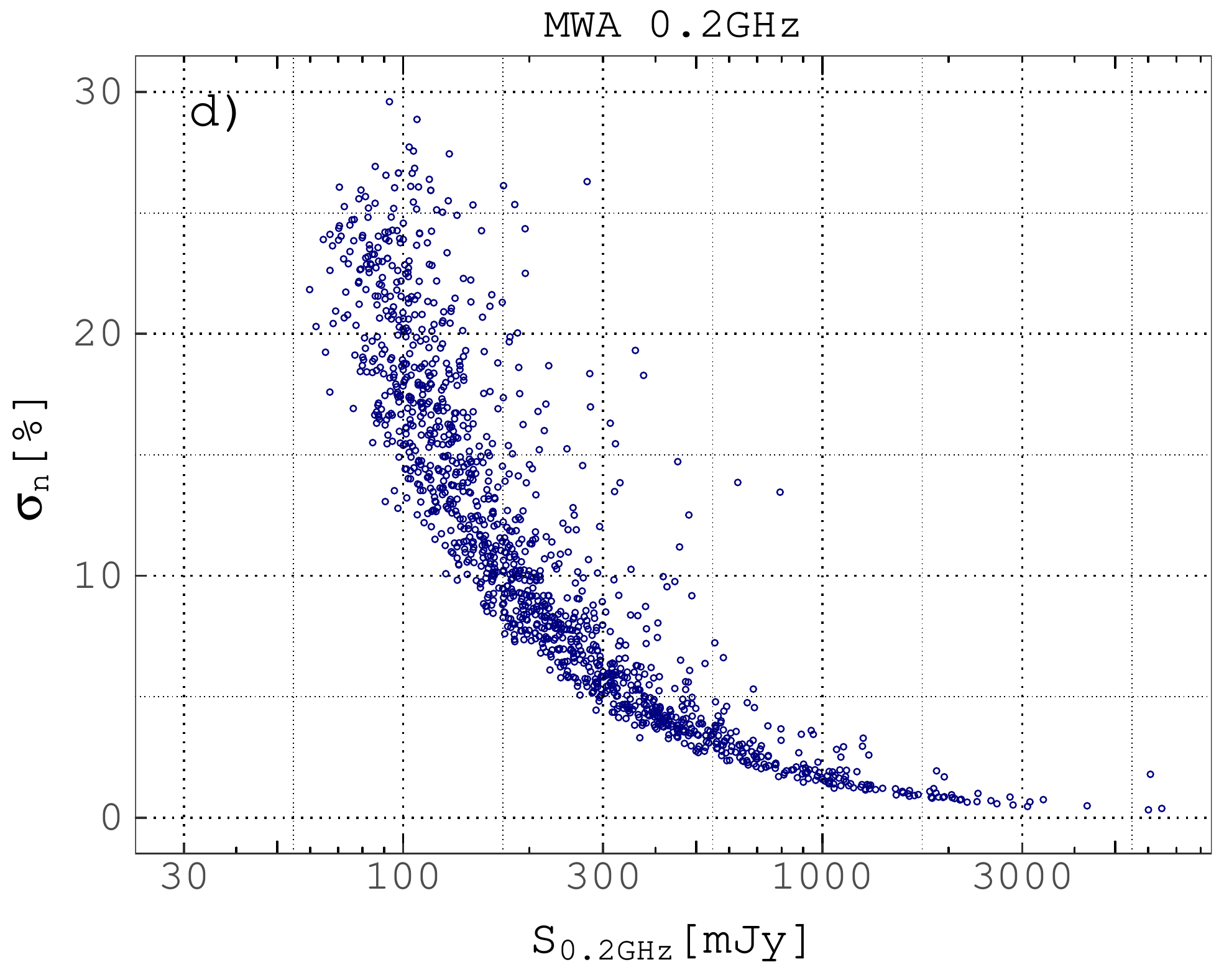} \\
	\end{tabular}
	\caption{The fitting flux density error ($\sigma_{n}$) for each source as a per cent of the total integrated flux density. The blue dot-dashed line represents the 3$\sigma$ \ac{RMS} limit of each catalogue, and the 5$\sigma$ cutoff is shown with a red dotted line.}
	\label{fig:IngtegratedFluxErr}
\end{figure*}

All source finding tools have a degree of error in their flux density estimation (see also Sect.~\ref{sec:SourceFinderErrors}). The main source of this inaccuracy is due to the noise present within any image, causing the sources at or near the detection threshold to have large flux density errors as shown in Figure~\ref{fig:IngtegratedFluxErr}. In addition to this effect, source finding tools can incorrectly determine the extent of a source, and in most cases will overestimate source flux density \citep{2015PASA...32...37H}.

To determine this, we measure the error of the integrated flux density as a function of fitting error ($\sigma_n$) to find outlier sources that have been fitted poorly, as shown in Figure~\ref{fig:IngtegratedFluxErr}. As expected, the relative error increases with weaker sources. Statistically, $\sigma_n$ for the majority of our sources is less than 15~per~cent across all five main catalogues. However, for some smaller portion of sources it can reach up to 30~per~cent.

\subsubsection{ \ac{MWA} Survey Flux Density Post-Calibration}

The radio images presented by \cite{2018MNRAS.480.2743F} were processed by \cite{2017MNRAS.464.1146H} as part of a large analysis of the GLEAM survey to produce an extragalactic catalogue. During the flux density calibration of the final mosaics of the GLEAM survey, a correction was made for the primary beam of the instrument, which was poorly modelled at the time of processing. This was performed by measuring the ratio of flux density of sources between GLEAM and ancillary catalogues in the Northern sky, then assuming symmetry and applying this correction to the Southern sky. At declinations far from the local zenith ($\delta\sim-27^\circ$) the correction was less accurate, resulting in an estimated flux density calibration accuracy of 8~per~cent for $-72^\circ<\delta<18^\circ$ and 13~per~cent outside of this range, escalating to $\approx90$~per~cent at the celestial pole.

The GLEAM data comprise a ``wideband'' 170--231~MHz image, typically used for source detection, and $20\times7.68$~MHz ``sub-band'' images, covering 72--231~MHz. In the catalogue formed by \cite{2017MNRAS.464.1146H}, sources detected in the wideband image at peak flux density $>5\sigma$, where $\sigma$ is the local \ac{RMS} noise, are measured in the sub-band images using the \textit{priorised} fit function of \textsc{Aegean} \citep{2018PASA...35...11H}. We used an identical procedure to \cite{2017MNRAS.464.1146H} to produce a catalogue for this region of sky, including accurate \ac{RMS}, background, and PSF measurement. Note that from these multi-frequency data, SEDs can be measured for brighter sources, enabling a more accurate flux density measurement, but for fainter sources the spectra are biased towards being flat by the uncertainties, and the wideband images provide the best measurement of the integrated source flux densities, which we here label as $S_\mathrm{200MHz,GLEAM}$.

The high-frequency data presented in this paper enable us to test the flux density scale accuracy and recalibrate the GLEAM data. We first selected sources which would not be confused in the \ac{MWA} data by excluding any sources within 2~arcmin of any other source, yielding a catalogue of 5\,419~sources. For each source we fitted a weighted linear power-law SED using the \textit{scipy.stats.leastsq} implementation of the Levenberg-Marquardt algorithm to all of the high-frequency data. From these fitted spectra we predicted the 200-MHz integrated flux densities, $S_\mathrm{200MHz,predicted}$. We crossmatched these sources with the results from the GLEAM source-finding, yielding 1\,178 matches within 2~arcmin positional threshold. From these we excluded the sources with GLEAM $S_\mathrm{200MHz,GLEAM}>1$~Jy, leaving 43~sources with high signal-to-noise and clean SEDs. We compared the predicted flux densities to those measured by GLEAM in the wideband image, finding that $S_\mathrm{200MHz,GLEAM} = (0.80\pm0.13)~S_\mathrm{200MHz,predicted}$. We applied this factor to the final cross-matched ensemble of GLEAM integrated flux densities, yielding the 200-MHz measurements referred to as ``GLEAM'' in Table~\ref{tab:Surveys}.

We used the Positional Update and Matching Algorithm \citep[PUMA; https://github.com/JLBLine/PUMA;][]{line2017}) to establish the final cross-match between GLEAM and our main catalogue. PUMA is an open source software capable of cross-matching between radio surveys with different spectral (frequency) and resolving capabilities. The GLEAM catalogue was used as the base catalogue to match sources from our \ac{ATCA} 1.384~GHz catalogue within 2~arcmin and in combination with objects catalogued in the \ac{SUMSS} 843~MHz survey. Since PUMA relies on spectral characteristics of cross-matched sources, as well as on their positions, we used the newly re-calibrated GLEAM flux densities as described above. We adopted as positive matches all sources flagged by the PUMA algorithm as (a) isolated, i.e. only single cross-match is found within the search region, or (b) dominant, i.e. multiple cross-matches are found but one matched source fits a power law significantly better than the others (see the PUMA manual\footnote{https://github.com/JLBLine/PUMA} for more details). We found a set of \SourcesMWAmultiple\ sources flagged as {\it multiple}, consisting of multiple higher resolution (see Sect.~\ref{sec:3.5}) sources blended into a single GLEAM source. For these objects we have used GLEAM flux densities as upper limits for cross-matched, high resolution, components. 

Finally, sources fainter than the source detection limit of $5\sigma$ are considered undetected, and their flux density in our catalogue is given as an upper limit as $5\sigma$, with the rescaling factor applied. We provide an additional column in the electronic catalogue with flag ($<$) designating an upper limit for the flux density.

\subsection{Spectral Index Uncertainties }
\label{sec:3.5}

Spectral index ($\alpha$) is defined in Sect.~\ref{sec:SourceFinding}. We use \textsf{R}'s package {\sc metafor} \citep{metafor} to estimate $\alpha$ using an unweighted mixed-effects linear regression model. The model takes into account uncertainties in the flux measurements to estimate $\alpha$ uncertainties. If no uncertainty is reported in the original measurement we assumed 10~per~cent of the flux value. Large uncertainties in spectral index ($\Delta\alpha >0.5$) are mainly the result of only two frequencies available.


\section{Results and Discussion}
 \label{sec:results}


 \subsection{Optical and Radio Counterparts}
 \label{sec:Optical}
 
We make use of multiple optical catalogues (as described in Table~\ref{tab:OpticalSurveys}) and we find \OpticalSourcesTotalUnique\ unique optical sources that are thought to be sources behind the \ac{LMC}. We apply a 5~arcsec search radius when matching optical counterparts to the highest frequency radio component. From this, we find \OpticalMatchesUnique\ unique optical sources (Table~\ref{tab:optdet}) of which \OpticalMatchesUniqueWithZ\ have a spectroscopic redshifts. For these \OpticalMatchesUniqueWithZ\ sources, we calculate a mean redshift of 0.39. The average positional offset between the radio continuum and optical sources is $1.59$ and $1.72$~arcsec in RA and Dec respectively.

\input{Files/optical_detections}

\subsection{AGN Populations}
\cite{2003MNRAS.346.1055K} used the Sloan Digital Sky Survey to argue that \ac{AGN} of all types and luminosities are typically found in the most massive galaxies. They also concluded that the optically luminous \ac{AGN} reside in galaxies with younger stellar populations with strong indications of recent starburst activity. 

A subset of this population are the radio-loud \ac{AGN} which form highly collimated, fast jets of material that drive through the surrounding galaxy into the intergalactic medium, at close to the speed of light. This class of \ac{AGN} can be further divided into Quasars, \ac{BL} or \ac{FSRQ}, \ac{OVV} and Radio Galaxies. Even though these jets are seen to primarily emit synchrotron radiation, their composition and production mechanism are still widely debated. These different classifications are thought to describe similar host galaxies (\ac{AGN}) seen at different orientation angles with respect to our line of sight, creating differences both in the observed radio morphology as well as the optical activity type \citep{1993ARA&A..31..473A,1995PASP..107..803U,2003ASPC..290....3U,2012AstRv...7c..92K}.

We use the definition from \cite{2006MNRAS.371..898S} to categorise the emission mechanisms of a source by its spectral index for centimetre wavelengths, i.e.: flat-spectrum sources with an $\alpha>-0.5$ are thought to be dominated by self absorbed, compact objects, while sources with $\alpha<-0.5$ are dominated by optically thin synchrotron emission \citep{2018MNRAS.477..578C,2018MNRAS.474..779G}.

\subsubsection{GPS and CSS Sources}
 \label{sec:gps_css}

\ac{GPS} and \ac{CSS} sources are thought to contribute $\sim$30--40 per cent of radio sources \citep{0004-637X-760-1-77,2016AN....337..194L}. However, their true nature is still under discussion. The debate about the absorption mechanism that contributes to the unique spectral features of this population has continued for over thirty years. The most extensive study of \ac{GPS} and \ac{CSS} sources is presented by \cite{1998PASP..110..493O} and more recently \citet{2021A&ARv..29....3O} who hypothesised regarding their origin. They additionally use these sources as constraints on \ac{AGN} physical evolution as it is believed that these sources represent the early stage of the \ac{AGN} phase \citep{1995A&A...302..317F, 2003PASA...20...69P, 2009AN....330..120F, 2011MNRAS.416.1135R,2017ApJ...836..174C,2018MNRAS.477..578C}.  

\ac{GPS} and \ac{CSS} sources are usually small (point) radio sources without jets and have similar characteristics in that they are both compact with steep spectra ($\alpha < -0.8$). \ac{GPS} sources have a much higher turnover frequency apart from their steep spectrum beyond the turnover frequency. This observable property allows their physical size to be estimated using the relation shown by \citet{doi:10.1093/mnras/stt2217} who agree with previous work by \citet[][Eq.~4]{1997AJ....113..148O}. They define the relation (fit) as:
\begin{equation}
\mathrm{log}~\nu_m = (-0.21\pm0.05) - (0.65\pm0.05)~\mathrm{log}~\mathrm{LLS}
\label{eq:turnover}
\end{equation} 

\noindent where $\nu_m$ is the turnover frequency (in GHz) and LLS is the largest linear size (in kpc). This suggests a simple physical process common to both \ac{GPS} and \ac{CSS} populations that have been separately defined due to an arbitrary selection of turnover frequencies. Generally, it is believed that these objects are young, whose growth was frustrated by dense gas and dust or are simply young radio sources moving along  their evolutionary path. \cite{1995A&A...302..317F} presented the ``young source'' hypothesis, postulating that these sources typically have ages of $10^6$ years and tend to decrease their radio luminosity with time and grow in size from a few kpc to a few hundred kpc.

\ac{CSS} sources are classified as such if they show an $\alpha$ of $< -0.8$, while \ac{GPS} sources have an average $\alpha$ of $+0.56$ below the turnover frequency and $-0.77$ above the turnover. In this study we detected \SourcesCSS\ \ac{CSS} candidate sources with an $\alpha$ of $< -0.8$, indicating candidates for further analysis as discussed in Sect.~\ref{sec:SpectralIndicies}.

\subsubsection{High Frequency Peakers (HFP)}

We follow the convention given by \cite{2000A&A...363..887D} in naming radio continuum sources with spectra that peak above 5~GHz, \ac{HFP}s. These high-frequency peaking sources are thought to be very young and compact, following the anti-correlation shown in Equation \ref{eq:turnover}. Additionally, these sources are thought to be rapidly changing in radio flux density, size and flux density peak frequency on time scales of a few decades. These sources are thought to then evolve into \ac{GPS} sources, then become \ac{CSS} sources and finally evolve into \ac{LSO} with a significantly broader (flatter; less peaked) spectrum \citep{1995A&A...302..317F,1996ApJ...460..634R,2000MNRAS.319....8A,2003PASA...20...46M,2008A&A...477..807O,0004-637X-760-1-77}.

\cite{2000A&A...363..887D} presented a catalogue of 55 \ac{HFP} sources above 300~mJy near the peak frequency of 4.9~GHz, with the majority of these being unresolved sources at small angular scales. Additionally, they defined a sample of true \ac{HFP}s as sources with an $\alpha < -0.56$ above the turnover frequency. Understanding and modelling these sources is of vital importance for three main reasons: \textbf{1)} their evolutionary path is currently not very well understood; \textbf{2)} these sources are highly variable on short time scales, with this change likely driven by significant events in their host galaxy; \textbf{3)} due to the high frequency of their emission peak, these sources can strongly contribute to measurements of the \ac{CMB} radiation as observed with the WMAP and PLANCK telescopes and as such these sources need to be accurately subtracted \citep{2000A&A...354..467D}. \cite{2016AN....337..120D} show that a common behaviour of \ac{HFP} sources is a spectral profile change (a change in time of the shape of the radio spectrum) correlated with an increase in optically thick emission and a decrease in optically thin emission.

We find \SourcesHFP\ potential \ac{HFP} characterised by an inverted spectrum (i.e. $\alpha > 0.5$) and a possible turnover in their radio spectrum above 5~GHz, as discussed in Section~\ref{sec:RadioTwoColour}.

\subsubsection{\ac{FSRQ} and \ac{BL} Candidates}

We investigated the sample of blazar candidates behind the \ac{LMC} as presented by \citet{2018ApJ...867..131Z}. They identified 37 such objects (their Tables~2 and 3) of which 23 are classified as \ac{FSRQ} candidates and 14 are \ac{BL} candidates. Using a 30~arcsec search radius, we find 17 out of 23 \ac{FSRQ} have a radio counterpart in common with the catalogue presented here (Table~\ref{tab:Catalogue}), and 9 out of 14 \ac{BL} candidates in common. We note that all 11 objects (6 \ac{FSRQ} and 5 \ac{BL} candidates) that are not detected in our surveys are listed in \citet{2018ApJ...867..131Z} as ``dubious objects''.

We denote these sources in our radio catalogue as the \ac{FSRQ} or \ac{BL} accordingly (Table~\ref{tab:Catalogue}; Col.~13). The \ac{FSRQ} candidate sample has a mean $\alpha$ of --0.43 $\pm$0.12 (compared to \cite{2018ApJ...867..131Z} findings of --0.40 $\pm$0.09) while the \ac{BL} sample has a somewhat steeper mean $\alpha$ of --0.72 $\pm$0.13, and a combined $\alpha$ of --0.53 $\pm$0.20. This is expected to be flatter than the average \ac{AGN} population ($\alpha$ = \SpecIndexMean; Table~\ref{tab:SpecIndexSubset}) as discussed in Sect.~\ref{sec:SpectralIndicies}. We note that \ac{BL} objects are known to be highly variable and as such their spectral indices should be treated with caution as our radio continuum data were not obtained simultaneously. Additionally, we estimate the average optical redshift of \ac{FSRQ} sample studied here is 1.37$\pm$0.17 based on various $z$ measurements described in Sect.~\ref{sec:SupplementarySurveys}.

\subsubsection{\ac{IFRSs} and HzRG Candidates}
 \label{sec:HzRG}
 
Finding \ac{HzRGs} is important in understanding the formation and evolution of galaxies at higher redshifts and in dense environments. Traditionally, radio sources with an \ac{USS} ($\alpha < -1.3$) have been the most efficient tracers of \ac{HzRGs} at $z>$2. This technique is based on the observed steepening of the radio spectrum with both redshift and frequency. Almost all known \ac{HzRGs} including the most distant one at $z=5.72$ \citep{2018MNRAS.480.2733S} have been identified using the \ac{USS} selection method only. However, several studies \citet{yamashita2020wide, jarvis2009discovery, waddington1999nicmos} demonstrated that the \ac{USS} selection criterion is not fully efficient as they discovered radio galaxies with flatter spectral indices. Finding high redshift radio-\ac{AGN}s without relying on \ac{USS} selection technique is now possible with the discovery of Infrared Faint Radio Sources (\ac{IFRSs}) from the deep ATLAS radio continuum survey a decade ago.

\ac{IFRSs} are a class of high redshift radio loud \ac{AGN} originally identified as bright radio sources with no co-spatial IR counterpart when cross-matching the 1.4~GHz ATLAS radio survey with the deep SWIRE IR survey between 3.6~$\mu$m and 24~$\mu$m \citep{2006AJ....132.2409N}. As radio sources, the nature of \ac{IFRSs} was unknown at the time. It was believed that all radio sources whether it is star forming galaxies or \ac{AGN} would produce IR emission. Numerous follow up studies have been conducted since their discovery to select more \ac{IFRSs} and to understand what they are. Since the original selection criterion from \citet{2006AJ....132.2409N} was survey specific, \citet{2011A&A...531A..14Z} redefined it and proposed a set of survey independent criteria, which even enable the selection of a brighter population of \ac{IFRSs} with very faint IR counterpart. They have radio-to-IR flux density ratios above 500 and secondly, 3.6~$\mu$m flux density less than 30~$\mu$Jy. To date, a total of $\sim$1400 \ac{IFRSs} have been identified from various radio surveys using the criteria of \cite{2011A&A...531A..14Z}. All studies so far suggested that most of the known \ac{IFRSs}, if not all, are radio-\ac{AGN}s at $z>2$. \cite{orenstein2019redshift} further confirmed the high redshift nature of IFRSs by presenting the largest sample of 108 \ac{IFRSs} with median spectroscopic redshift of 2.68. Additionally, some studies presented the evidence that IFRS population not only encompasses \ac{USS} sources but also flat spectrum sources too \textbf{\citep[][]{2014MNRAS.439..545C}}. Thus, selecting \ac{IFRSs} serves as a valuable technique to discover all sorts of high redshift radio-\ac{AGN}s.

We search for high redshift radio \ac{AGN}s behind the \ac{LMC} by selecting the potential IFRS candidates following the procedures from \cite{2014MNRAS.439..545C} and \citet{orenstein2019redshift} utilising the IR photometry from the CatWISE2020 \citep{Marocco2021} and SAGE \citep{2006AJ....132.2268M} surveys. We cross-matched the radio sources in the \ac{LMC} field with these two Mid-IR source catalogues, using a match radius of 1~arcsec. Because of the large radio beam sizes ($\sim$40~arcsec resolution), we chose 1~arcsec as the matching radius in order to minimise the false identifications. We acknowledge that by increasing the search radius we would certainly find many more true IFRS candidates but then the confidence level on these matches would be significantly lower.

\begin{figure}
    \centering
   \includegraphics[width=1.0\columnwidth]{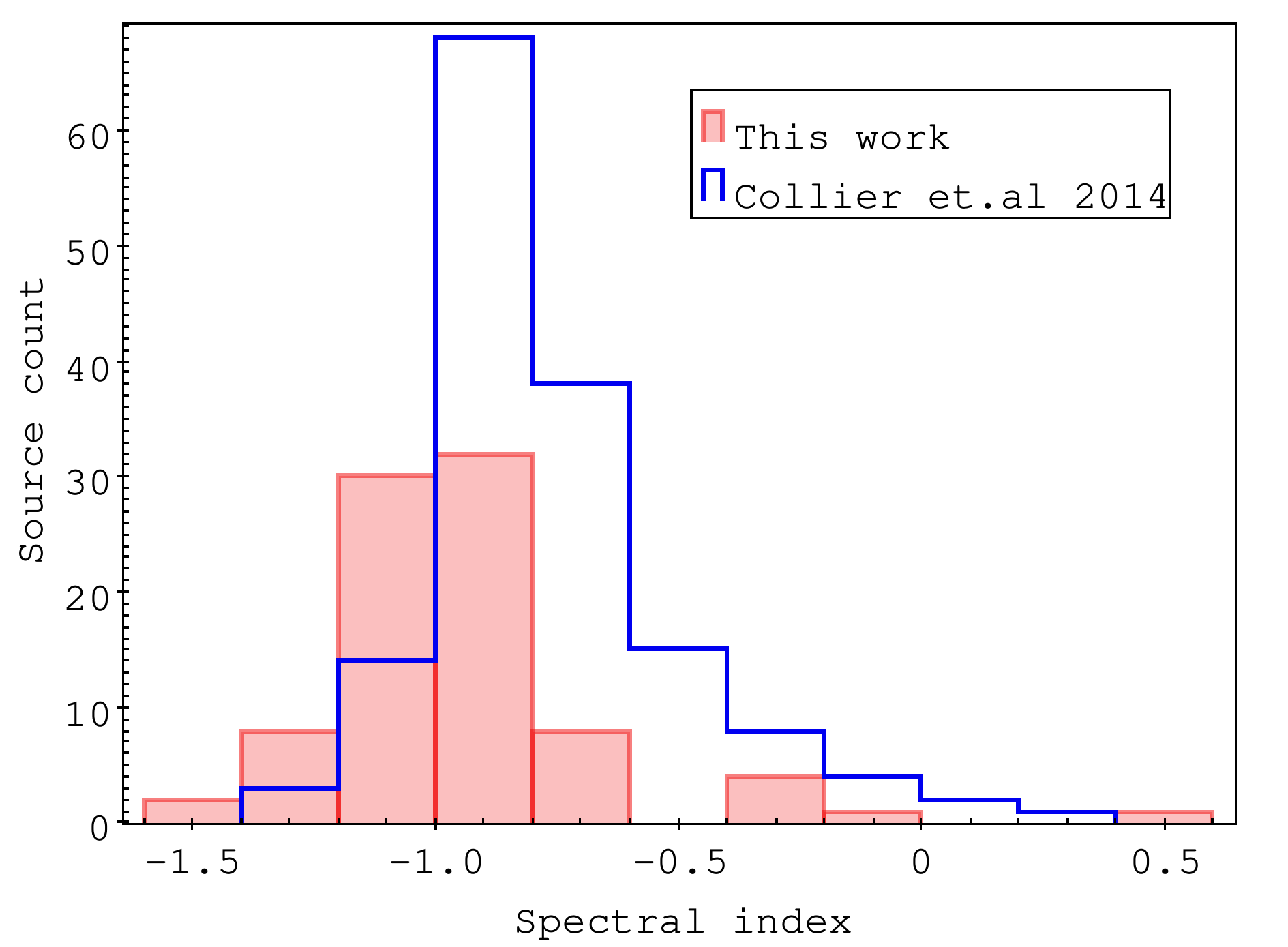}
   \caption{Radio spectral index distribution of our \ac{IFRSs} sample in the field of the \ac{LMC}, plotted as red solid boxes and \ac{IFRSs} from \citet[][]{2014MNRAS.439..545C}, plotted as blue dashed lines. Only sources with $\alpha_N\ge$3 are plotted where $\alpha_N$ is the number of data points used to measure spectral index. Spectral index measurements in \citet[][]{2014MNRAS.439..545C} are based on 92, 20 and 6~cm data}. 
  \label{fig:ifrs_SI}
\end{figure}

A total of 123 \ac{IFRSs} are identified in which 47 are found in SAGE, 71 in the CatWISE2020, and 5 in both. Out of 71 CatWISE2020 selected \ac{IFRSs} 64 have a radio-to-IR flux density as defined by \citet{2011A&A...531A..14Z} larger than 500 at both 20~cm \& 40~cm, 3 at 20~cm, and 11 at 40-cm only. Similarly, out of 47 SAGE-IFRSs 43 satisfied the Zinn ratio criteria at both 20~cm \& 40~cm, and the remaining 4 at 40~cm only. The five \ac{IFRSs} identified in both SAGE and CatWISE2020, all have ratios above 500 at both 20~cm \& 40~cm. All of our \ac{IFRSs} are found to have faint 3.4 or 3.6~$\mu$m counterparts which are reliably detected (S/N $>$ 5) in either of the CatWISE2020 or SAGE surveys. In Table~\ref{tab:IFRS}\footnote{The full table will be in the supplementary material of the journal.}, we present our \ac{IFRSs} sample. In Figure~\ref{fig:ifrs_SI}, we present the spectral index distribution of our \ac{IFRSs} sample, which follows a skewed distribution towards steep spectral indices with a median of \SpecIndexIFRS\ -- very similar to the distribution shown by \citet{2014MNRAS.439..545C}. 

So far, only 131 \ac{IFRSs} are known with spectroscopic redshifts lying in the range 1.63--4.4 \citep{orenstein2019redshift,2014MNRAS.439..545C}. The only \ac{IFRSs} in our sample with a measured redshift is J044537–685946 at $z = 1.714$ \citep[][]{2018ApJ...867..131Z}.

\input{IFRS/IFRS_table_sample}

To date, only two \ac{IFRSs} have been reported with an X-ray counterpart, suggesting that they are \ac{AGN}s of Type~I \citep{2014MNRAS.439..545C}. We analysed whether any of our \ac{IFRSs} from this study had been detected in the {\it XMM-Newton} survey \citep{2019svmc.confE..63H}, and three \ac{IFRSs} were found to have an X-ray counterpart within 15~arcsec radius. They are J051233--675037, J050634--675643, and J052350--704213, which we plan to follow up for further studies.

\subsection{Spectral Index for the Entire Sample}
 \label{sec:SpectralIndicies}
 
\ac{CSS}, \ac{GPS}, Quasars and distant \ac{AGN} cannot be distinguished easily by position in the sky in our catalogue as the angular resolutions of these existing radio surveys are, in most cases, insufficient. To remedy this, we have created a spectral index for each object to determine the dominant emission mechanism (see Table~\ref{tab:Catalogue}). 

We perform a least squares linear fit to model the source's spectral index through the use of a power law consistent with the decay of synchrotron emission (a curved spectrum that would model synchrotron losses). However, fitting this simple single power-law model is not sufficient to categorise the turnover of \ac{GPS} sources and other curved spectrum populations. \cite{1985MNRAS.217..601P} showed that this type of analysis is only viable when applied to a small frequency range that ignores the effects of a curved spectrum. 

\begin{figure*}
	\includegraphics[width=.495\textwidth]{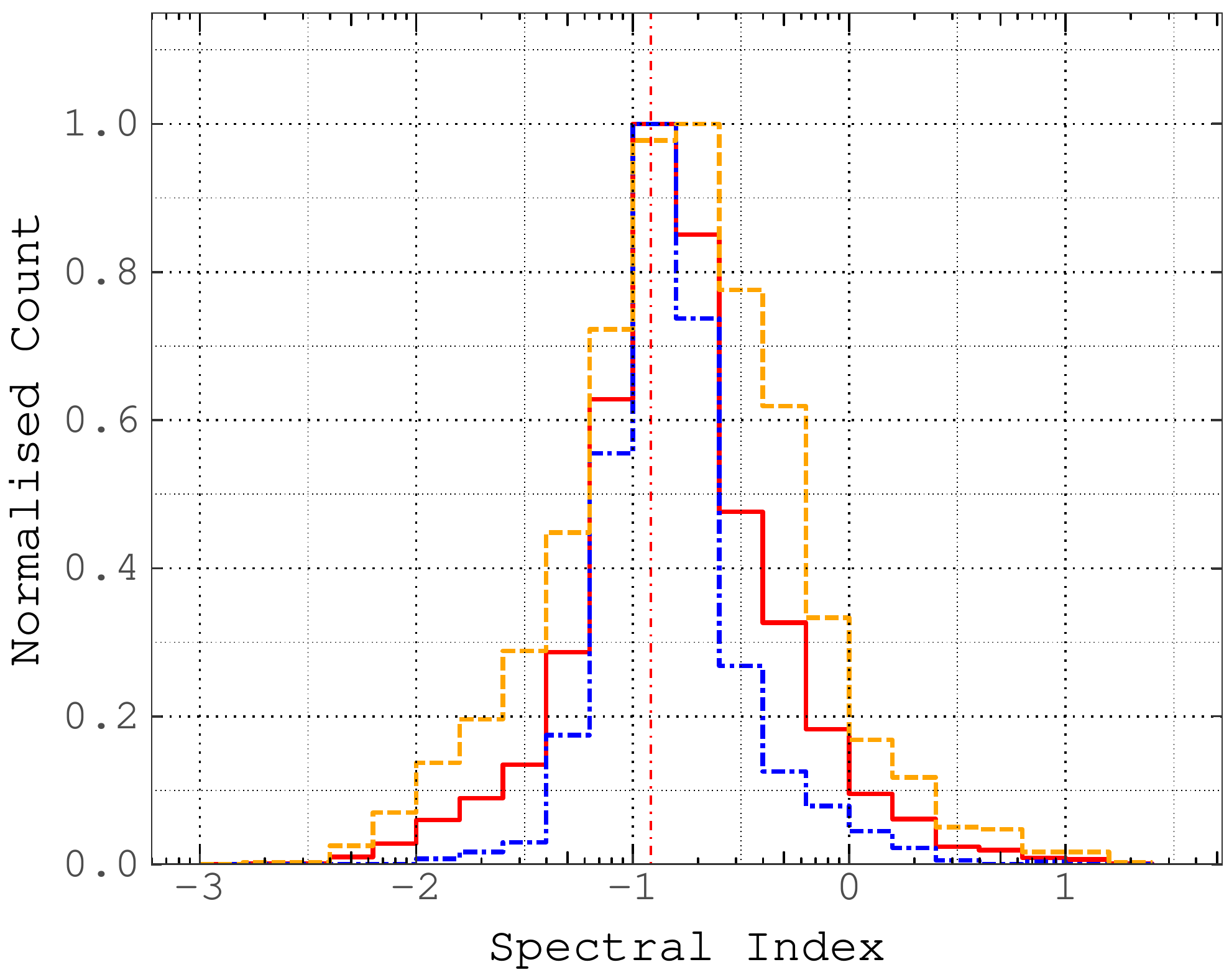}
	\includegraphics[width=.495\textwidth]{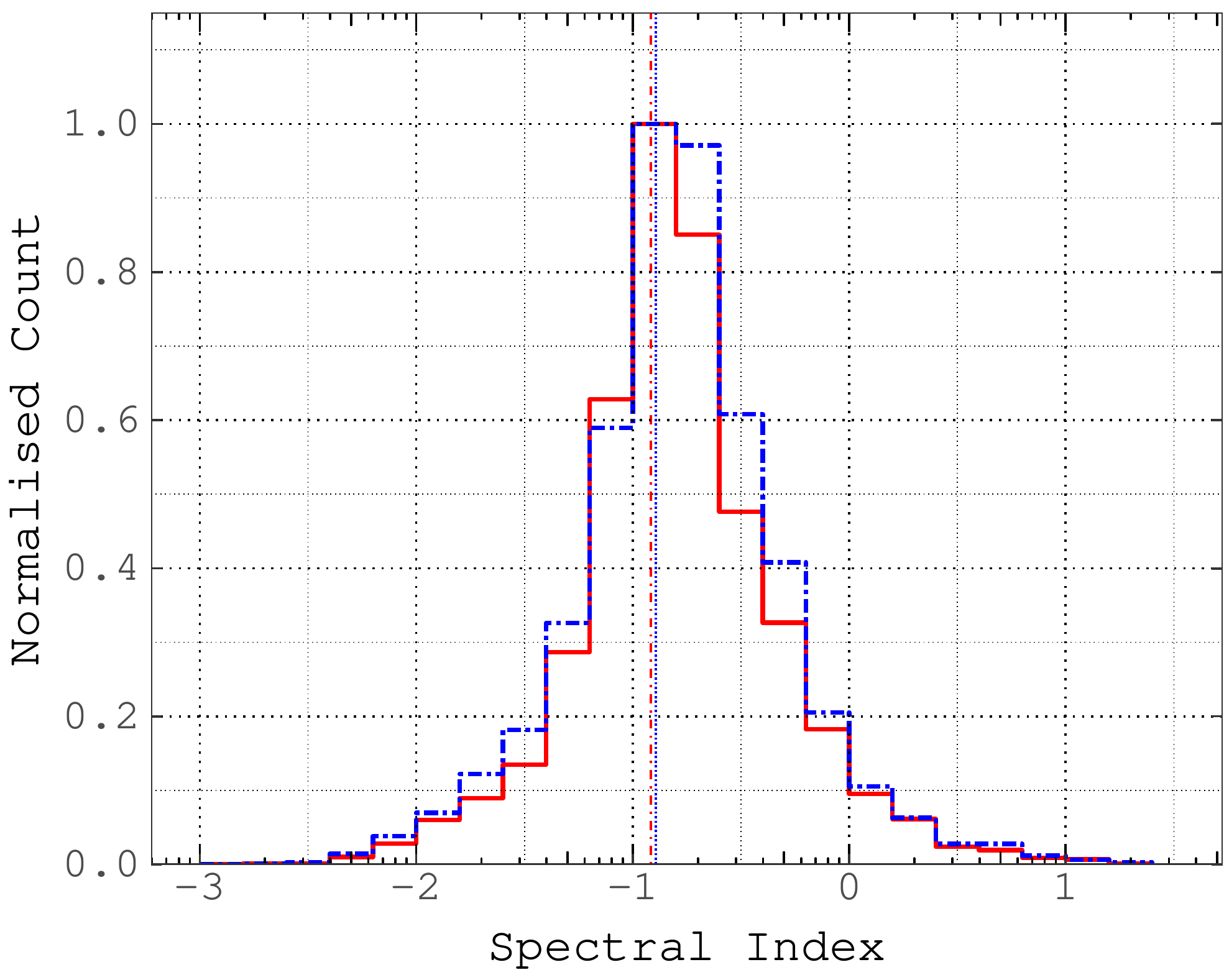}
\caption{{\bf Left:} Spectral index distribution for \SpecIndexSourceCount\ sources from all radio continuum catalogues. The red continuous line is the full catalogue, the orange dashed line are spectral indices estimated from sources with only two available data points and blue dot-dashed line are estimates where more than two data points were available. 
	The median $\alpha$ (vertical red dash-dotted line) for the full catalogue is \SpecIndex\ (Table~\ref{tab:SpecIndexSubset}).
{\bf Right:} A comparison between spectral index distributions for the full catalogue (red continuous line; same as in left figure) and for \SpecIndexLowestTwoFreqSourceCount\ spectral indices estimated from two data points between 0.843~GHz and 1.384~GHz (blue dot-dashed line). 
	The spectral index measure is restricted to these two frequencies as they provide similar angular resolutions. The median $\alpha$ for the 0.843~GHz to 1.384~GHz two-point spectral indices (vertical blue dotted line) is \SpecIndexLowestTwoFreq\ (Table~\ref{tab:SpecIndexSubset}).  
	Counts per bin are normalised to the maximum count per bin in each sample. } 
	\label{fig:SpecIndexDistributionnew}		
\end{figure*}

As mentioned earlier, significant care must be taken when calculating the spectral index of a source. For instance, the resolution of each observation should be similar to avoid contributions from other sources. Initially, using all catalogues, we find a mean spectral index of \SpecIndexMean\ $\pm$\SpecIndexSD\ (and median of \SpecIndex) from \SpecIndexSourceCount\ sources that are detected at two or more radio frequencies (Table~\ref{tab:SpecIndexSubset}). We subsequently refine this by presenting two spectral distributions; the first in Figure~\ref{fig:SpecIndexDistributionnew} (left), utilises all catalogues as presented in this paper. The second (and more accurate), Figure~\ref{fig:SpecIndexDistributionnew} (right) uses only the catalogues at 0.843~GHz and 1.384~GHz given their similar resolution as shown in Table~\ref{tab:Surveys}. Consequently, we find a median spectral index of \SpecIndexLowestTwoFreq\ and mean of \SpecIndexLowestTwoFreqMean\ $\pm$\SpecIndexLowestTwoFreqSD\ from \SpecIndexLowestTwoFreqSourceCount\ sources, summarised in Table~\ref{tab:SpecIndexSubset}. However, one should bear in mind that these spectral indices are measured between two close frequency values (0.843 and 1.384~GHz) which are affected by larger uncertainties. Indeed, some 50 sources (less than 1 per cent of the total population) in Table~\ref{tab:Catalogue} have questionable spectral index estimates of $\alpha<-2$. Where the spectral index values are extreme, we flag those sources to emphasis caution. The reason for such unrealistic spectral indices for these few sources is that the flux density measurements are made between only two nearby frequency bands where a small change (or error) in size or flux density leads to large changes and somewhat unrealistic estimates in spectral index. Finally, looking at the median values of $\alpha$ in the last 3 lines of Table~\ref{tab:SpecIndexSubset}, one can see that the more distant the sources are (or the fewer
optical/IR counterparts they have) the steeper their spectral index.

\begin{table}
    \centering
    \caption{Median and mean spectral index distributions for each subset in the catalogue for sources that are detected at more than one radio frequency.}
    \label {tab:SpecIndexSubset}	
    \begin{tabular}{ l c c c c }
    	\hline
   		Subset & Median & Mean & Standard & Number of \\
   		 & $\alpha$ & $\alpha$ & Deviation & Sources \\
    	\hline
    	All & \SpecIndex & \SpecIndexMean & \SpecIndexSD & \SpecIndexSourceCount \\
    	Resolution Matched & \SpecIndexLowestTwoFreq & \SpecIndexLowestTwoFreqMean & \SpecIndexLowestTwoFreqSD & \SpecIndexLowestTwoFreqSourceCount \\
    	\ac{RMS} Limited & \SpecIndexLowestTwoFreqLimited & \SpecIndexLowestTwoFreqLimitedMean & \SpecIndexLowestTwoFreqLimitedSD & \SpecIndexLowestTwoFreqLimitedSourceCount \\
    	\ac{AT20G} Detection & \SpecIndexATG & \SpecIndexATGMean & \SpecIndexATGSD & \SpecIndexATGSourceCount \\
    	\hline
    	No \ac{IR} or optical & \SpecIndexNoIR & \SpecIndexNoIRMean & \SpecIndexNoIRSD & \SpecIndexNoIRSourceCount \\
    	IR Counterpart & \SpecIndexRadioToIR & \SpecIndexRadioToIRMean & \SpecIndexRadioToIRSD & \SpecIndexRadioToIRSourceCount \\
    	Optical Counterpart & \SpecIndexOptical & \SpecIndexOpticalMean & \SpecIndexOpticalSD & \SpecIndexOpticalSourceCount \\
        \hline
    \end{tabular}   
\end{table}

From this analysis, we find \SourcesCSS\ ($\sim$25 per cent) sources with an $\alpha < -0.8$ that we classify as \ac{CSS} source candidates. Additionally we find \SourcesHFPpossible\ sources that have a highly inverted spectrum ($\alpha > 0.5$) that we classify as potential \ac{HFP} sources in which the turnover is unconstrained due to our limited frequency range, as shown in Table~\ref{tab:classification}. 

    \begin{table} 
        \caption{Source classification criteria for the radio continuum catalogue. $\dag$ is unique sources detected at one or more frequencies. }
        \label {tab:classification}	
        \begin{center}
    	    \begin{tabular}{ c c c c c }
        	\hline
       		Source         & Source & Classification & Catalogue \\
       		Classification & Count  & Criteria       & Flag \\
        	\hline
        	All$\dag$ & \SourcesTotalUnique &  & \\
        	Inverted $\alpha$ & \SourcesInverted & $\alpha > 0$ & \\
        	USS & \SourcesUSS & $\alpha < -1.3$ & \ac{USS}\\
    		CSS & \SourcesCSS & $-1.3 < \alpha < -0.8$ & \ac{CSS}\\
    		GPS & \SourcesGPS & 100~MHz $<$ turnover $<$ 5~GHz & \ac{GPS} \\
    		HFP & \SourcesHFP & SED peak above 5~GHz & HFP \\
    		\ac{IFRSs}        & \SourcesIFRSs              & various & IFRSs \\
            \hline
        \end{tabular}   
        \end{center}
    \end{table}

\begin{figure}
	\includegraphics[width=.47\textwidth]{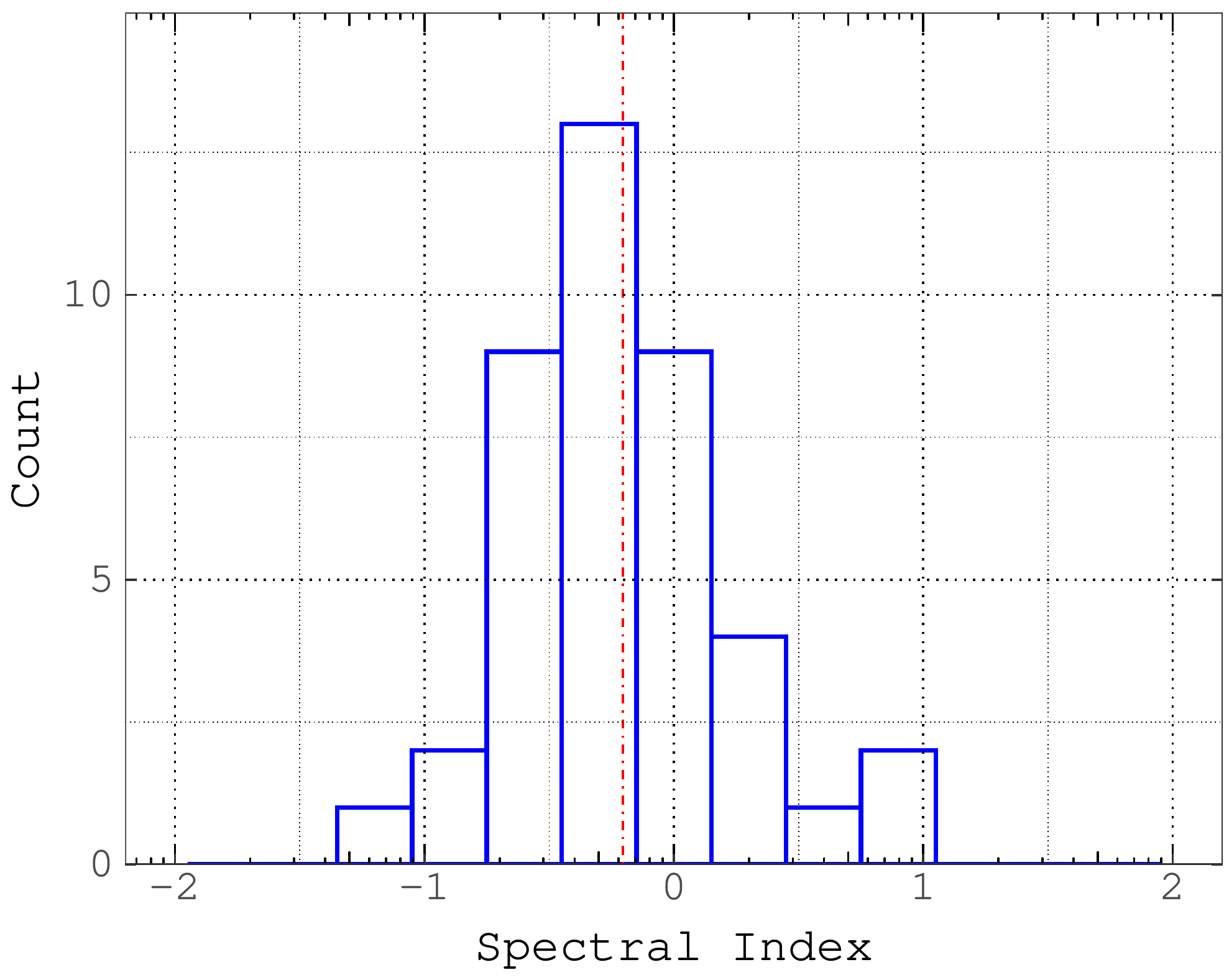}
	\caption{Spectral index distribution for \SpecIndexATGSourceCount\ sources that have detection in the \ac{AT20G} catalogue, measured using all radio continuum catalogues. The median $\alpha$ (vertical dash-dotted line) is \SpecIndexATG.} 
	\label{fig:SpecIndexDistributionAT20G}		
\end{figure}

We closely inspected the \SpecIndexATGSourceCount\ \ac{AT20G} sources in our catalogue (as shown in Figure~\ref{fig:SpecIndexDistributionAT20G}), and found that the spectral index distribution of these sources is significantly different from the distribution for the catalogue as a whole (shown in Figures~\ref{fig:SpecIndexDistributionnew}). This distribution has a mean $\alpha$ of \SpecIndexATGMean\ $\pm$\SpecIndexATGSD\ (Table~\ref{tab:SpecIndexSubset}). Among these \SpecIndexATGSourceCount\ sources, we find \ATGSourcesCSS\ \ac{CSS}, \ATGSourcesGPS\ \ac{GPS}, \ATGSourcesHFP\ \ac{HFP} and \ATGSourcesHFPpossible\ \ac{pHFP} sources which heavily skew the overall spectral index distribution towards $\alpha\sim$ 0. This is due to a selection bias as we are selecting only sources with a detection in a shallower high frequency survey which tends to select bright sources with flat or inverted spectra. However, in practice we are biased towards \ac{FSRQ} or \ac{BL}s.

\begin{figure} 
	\includegraphics[width=.47\textwidth]{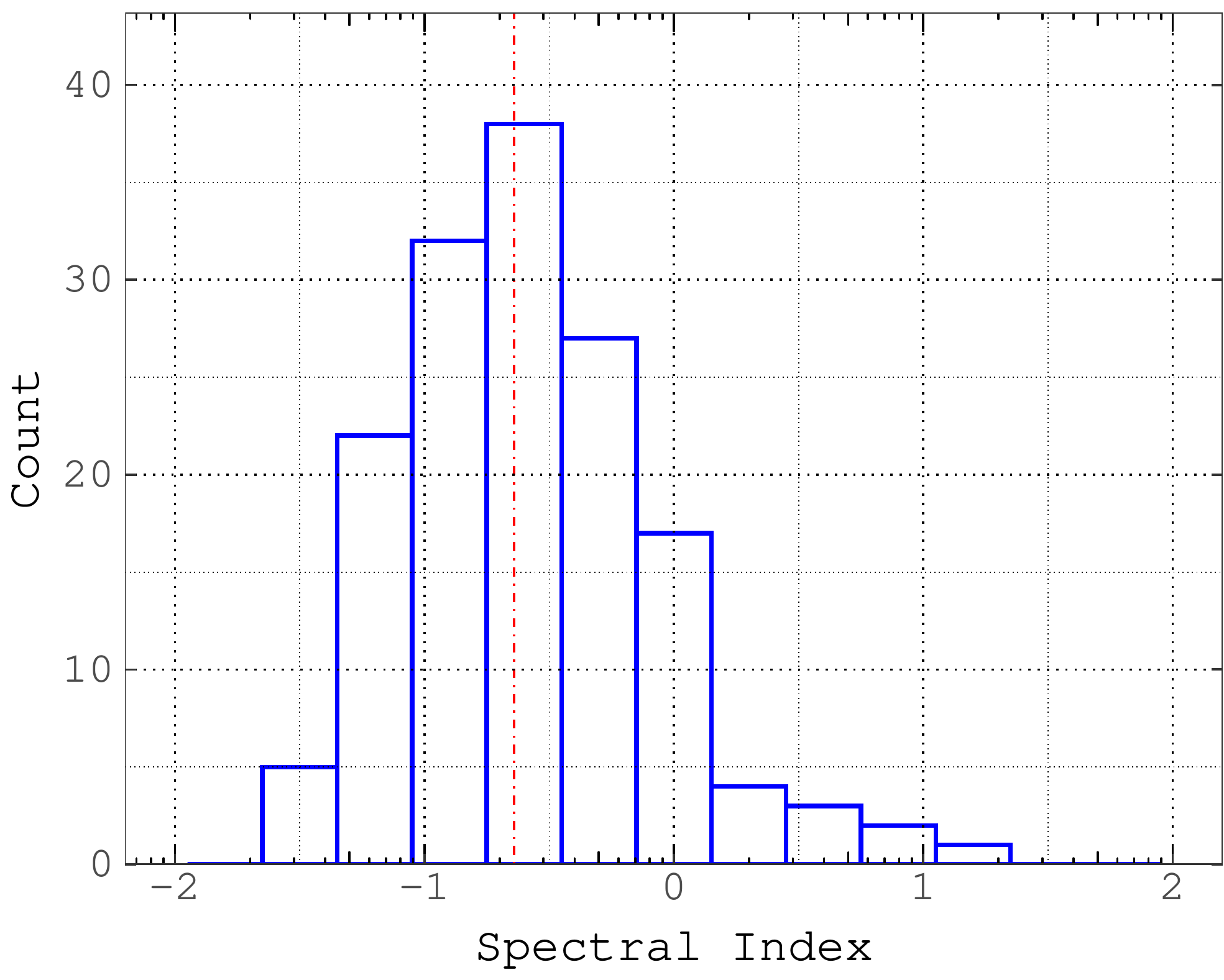} 
	\caption{Spectral index distribution for \SpecIndexOpticalSourceCount\ sources that have a detection in at least one optical catalogue, measured using all radio continuum catalogues. The median $\alpha$ (vertical dash-dotted line) is \SpecIndexOptical.} 
	\label{fig:SpecIndexDistributionOptical}		
\end{figure}

Interestingly, sources that have an optical counterpart tend to have a flatter spectral index as shown in Figure~\ref{fig:SpecIndexDistributionOptical} and summarised in Table~\ref{tab:SpecIndexSubset}. For example, we find the mean spectral index for this subsample to be \SpecIndexOpticalMean\ $\pm$\SpecIndexOpticalSD\ from \SpecIndexOpticalSourceCount\ sources (Table~\ref{tab:SpecIndexSubset}). This is also reflected in radio sources that have an \ac{IR} counterpart, but without an optical detection. Specifically, these sources have a mean spectral index of \SpecIndexRadioToIRMean\ $\pm$ \SpecIndexRadioToIRSD\ from \SpecIndexRadioToIRSourceCount\ sources. In contrast, sources that do not have an \ac{IR} or optical counterpart, have a far steeper mean spectral index at \SpecIndexNoIRMean\ $\pm$\SpecIndexNoIRSD\ from \SpecIndexNoIRSourceCount\ sources (Table~\ref{tab:SpecIndexSubset}). As for the \ac{AT20G} sample, this is probably a selection bias. These optical counterparts are mainly \ac{QSO}s or \ac{BL}s with well-established flat spectral index. Similarly, but to a lesser extent, IR detected sources are more likely to have a strong IR \ac{AGN} component, typical of a \ac{QSO} and/or Seyfert~1 \ac{AGN}, that tends to have flatter radio spectra. This interpretation seems supported by the fact that sources with no IR counterpart have steeper radio spectra on average, i.e. are more dominated by radio galaxies.

\subsubsection{Sensitivity Limitations}
 \label{sec:RMSEffects}
Almost all sources (\TwoLowestFreqCount\ sources, \TwoLowestFreqCountPerc\ per cent) in our main catalogue (Table~\ref{tab:Catalogue}) are detected at 0.843~GHz or 1.384~GHz. The remaining \TwoLowestFreqCountRemain\ sources are weak sources detected in either the 4.8 or 8.64~GHz catalogues. These \TwoLowestFreqCountRemain\ sources may represent either variable sources or \ac{HFP} (peaking their flux density above 5~GHz) as discussed in Sect.~\ref{sec:HFP}. 

For a 5$\sigma$ source to be included in the 1.384~GHz catalogue, an integrated flux density of at least 2.5~mJy is required (at the average \ac{RMS} noise of 0.5~mJy~beam$^{-1}$). A 2.5~mJy source in the \ac{ATCA} 1.384~GHz catalogue is expected to appear in the \ac{SUMSS} catalogue only if it has a 0.843~GHz flux of 6~mJy (5 times the average \ac{RMS}), which would imply a very extreme  spectral index of $-1.7$. For a more standard spectral index of $-0.7$ only sources with 1.384~GHz flux densities larger than 3.5~mJy can have a counterpart in the \ac{SUMSS} catalogue.

\begin{figure*}
	\centering
	\begin{tabular}{@{}cc@{}}
	        \includegraphics[width=.48\textwidth]{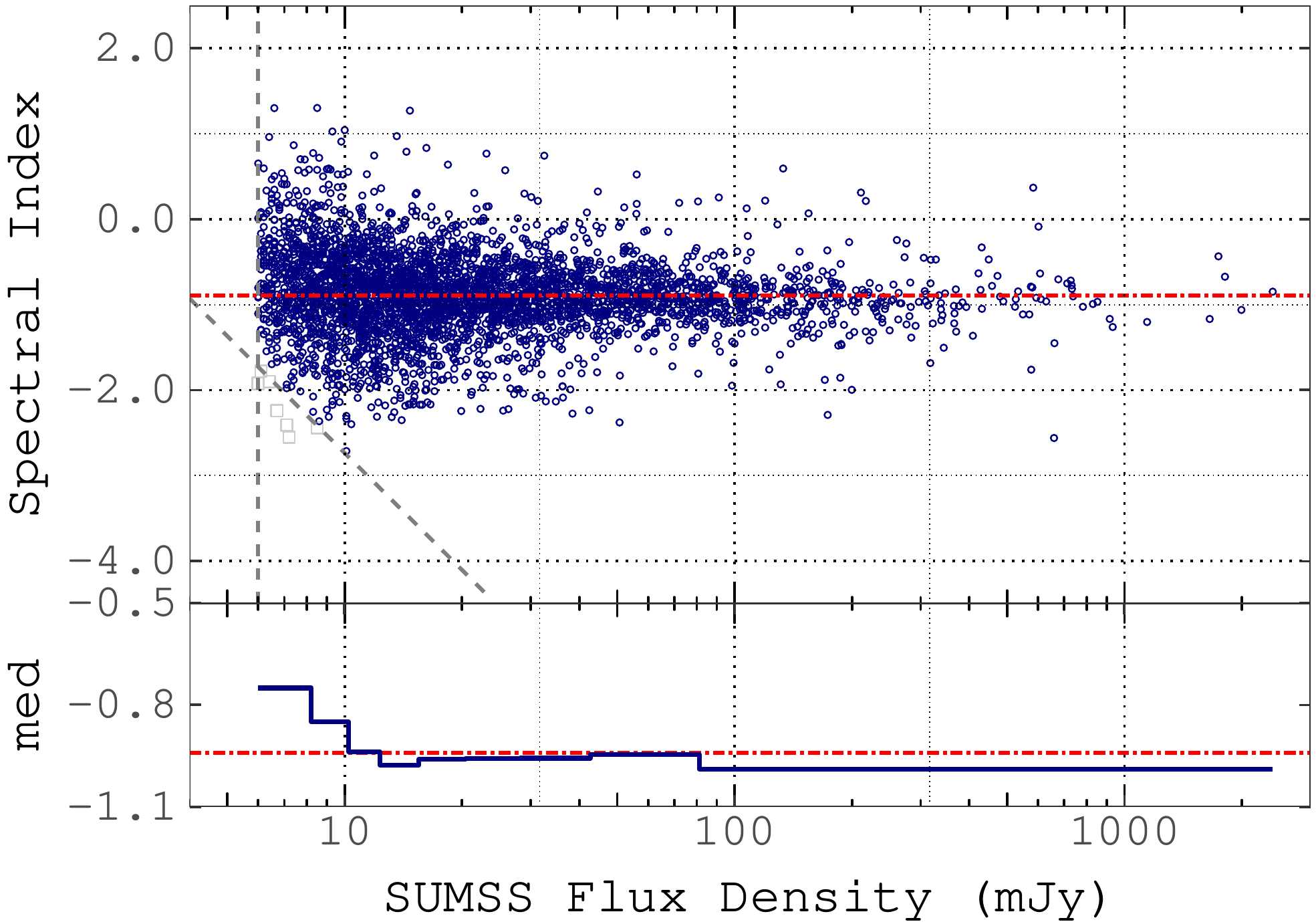} &
	        \includegraphics[width=.48\textwidth]{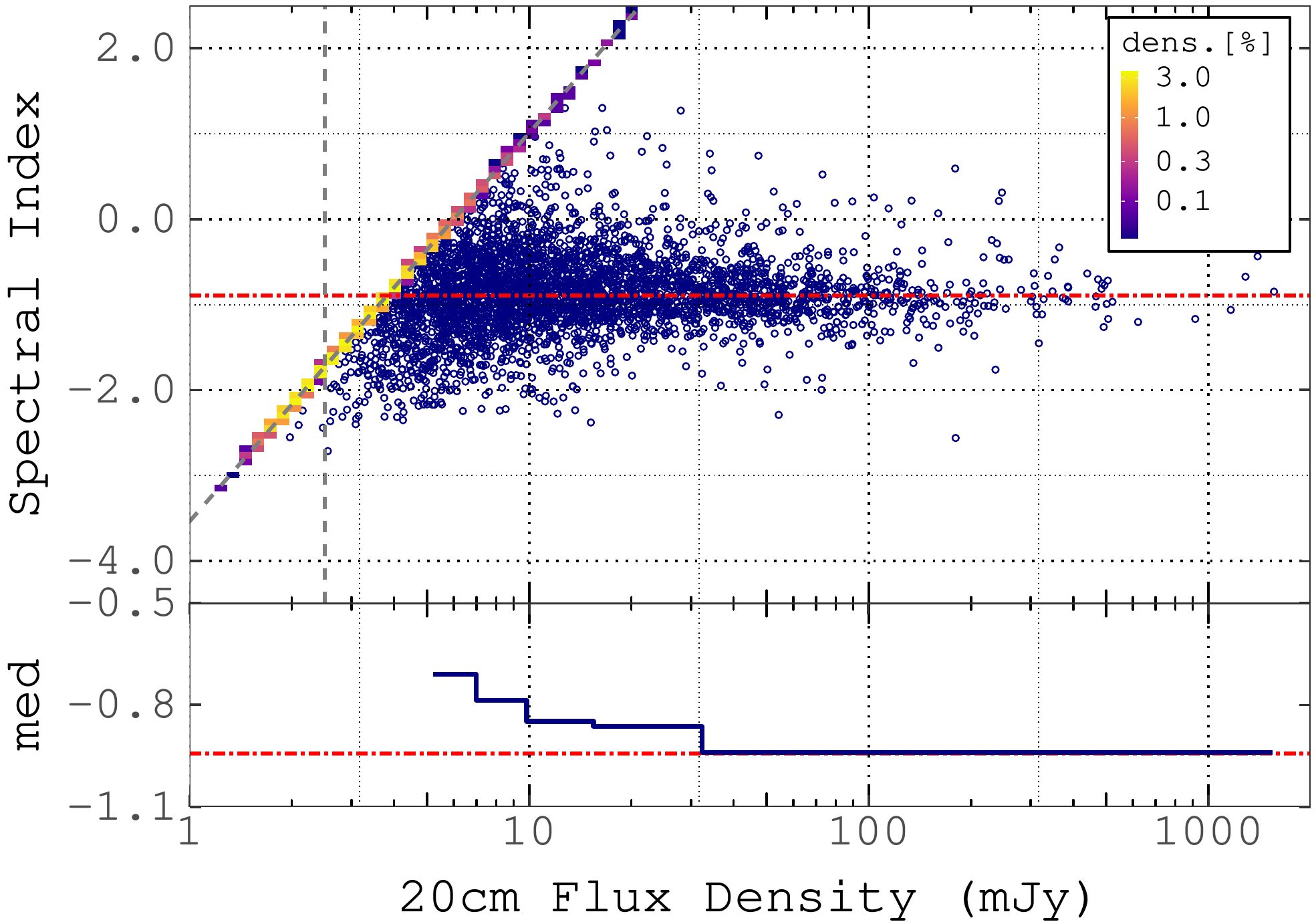} \\
	\end{tabular}
	\caption{Integrated flux density vs. spectral index for the two most complete survey lists: \ac{SUMSS} (left) and \ac{ATCA} at 1.384~GHz ({\bf top panels}) and the distribution of spectral index medians per flux density bin ({\bf bottom panels}). Vertical dashed lines in top panels indicate flux limits for each survey. Red horizontal dot-dashed lines in all panels represent the median spectral index  (\SpecIndexLowestTwoFreq) for the complete sample of all sources in common to \ac{SUMSS} and \ac{ATCA} surveys. For the \ac{ATCA} 1.384~GHz data points with no counterparts at 0.843 GHz we show the approximate source counts (in \%) of the limiting spectral indices as a binned density map (scale bar shown at the top right corner). For the median vs flux density plots data is split in 9 equally populated bins. Note that the bins with more points from limiting than calculated spectral indices are not shown.}
	\label{fig:IntegFluxSpecIndex_SUMSSLimited}
\end{figure*}

This clearly limits the study of the spectral index of the 1.384~GHz source catalogue at the low flux density end, since 5--10$\sigma$ 1.384~GHz sources might not be detected at 843~GHz due to the higher flux density limit of the \ac{SUMSS} catalogue. It also introduces a bias in the study of the spectral index of the faintest 1.384~GHz sources, as only very steep sources can be detected at 0.843~GHz. This bias against flat-spectrum sources at the low (3--10$\sigma$) 1.384~GHz flux densities can be seen in Figure~\ref{fig:IntegFluxSpecIndex_SUMSSLimited} (right panel), where we show the spectral index between 0.843~GHz and 1.384~GHz as a function of the integrated flux densities for the \ac{ATCA} 1.384~GHz sources. The diagonal dashed line shows the limit of the region of the spectral index -- flux parameter space where we can detect the source at both frequencies due to the limited sensitivity of the 0.843~GHz \ac{SUMSS} catalogue. For sources above this line only spectral index limits can be derived. A similar bias can be in principle introduced when studying the spectral index of the \ac{SUMSS} sources, this time against steep spectrum sources. Such a bias can however be neglected in our case, as the 1.384~GHz \ac{ATCA} catalogue is more sensitive, and the faintest \ac{SUMSS} sources (i.e. those detected at 5$\sigma=6$~mJy) can all have a counterpart in the 1.384~GHz \ac{ATCA} catalogue, except for very extreme (and rare) cases where the spectral index is steeper than $-1.7$. Indeed, Figure~\ref{fig:IntegFluxSpecIndex_SUMSSLimited} (left panel) shows that only a few \ac{SUMSS} sources are not detected at 1.384~GHz (upper limits are shown as open grey squares at the bottom left).

\begin{figure*}
	\centering
	\begin{tabular}{@{}cc@{}}
	        \includegraphics[width=.48\textwidth]{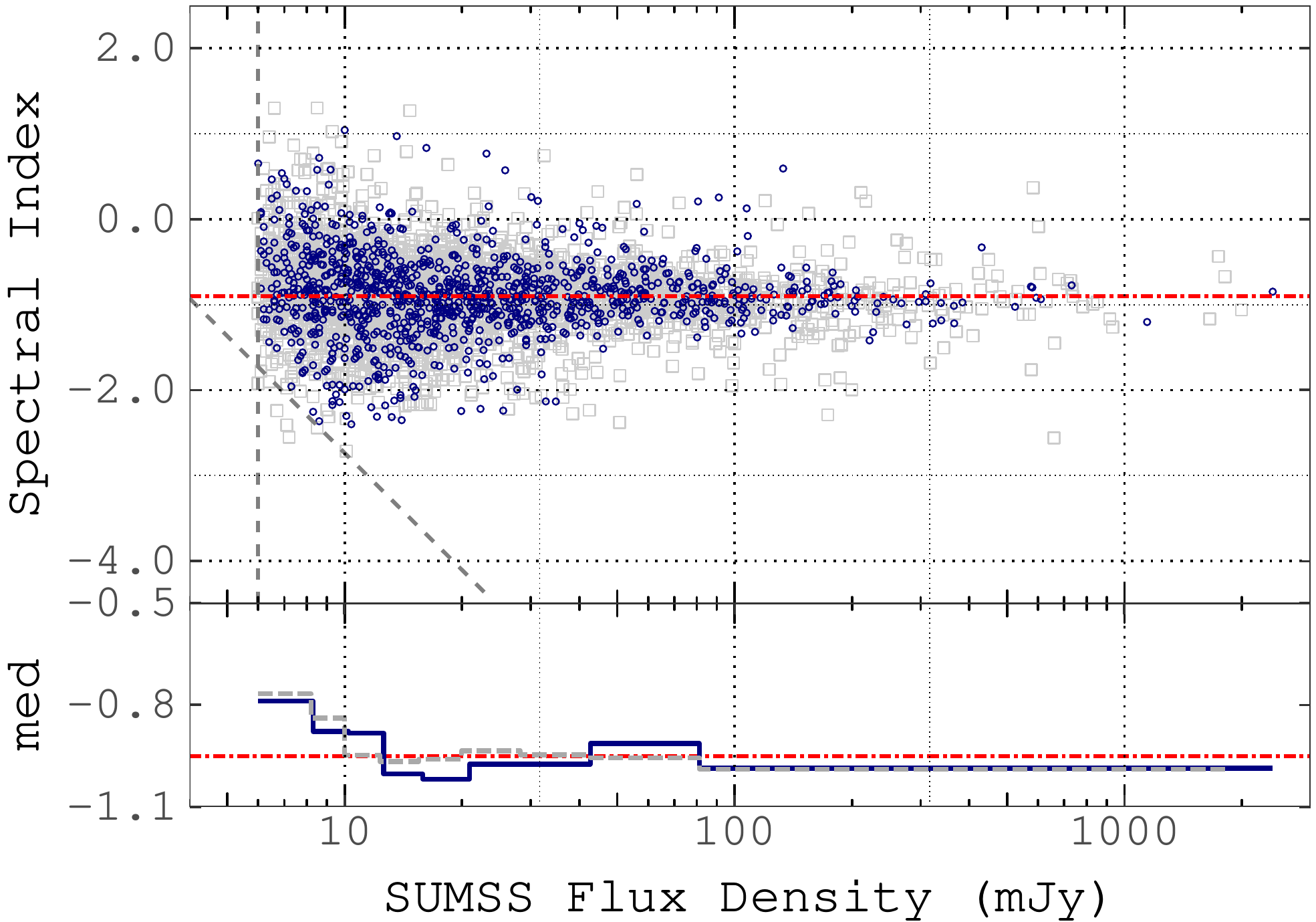} &
	        \includegraphics[width=.48\textwidth]{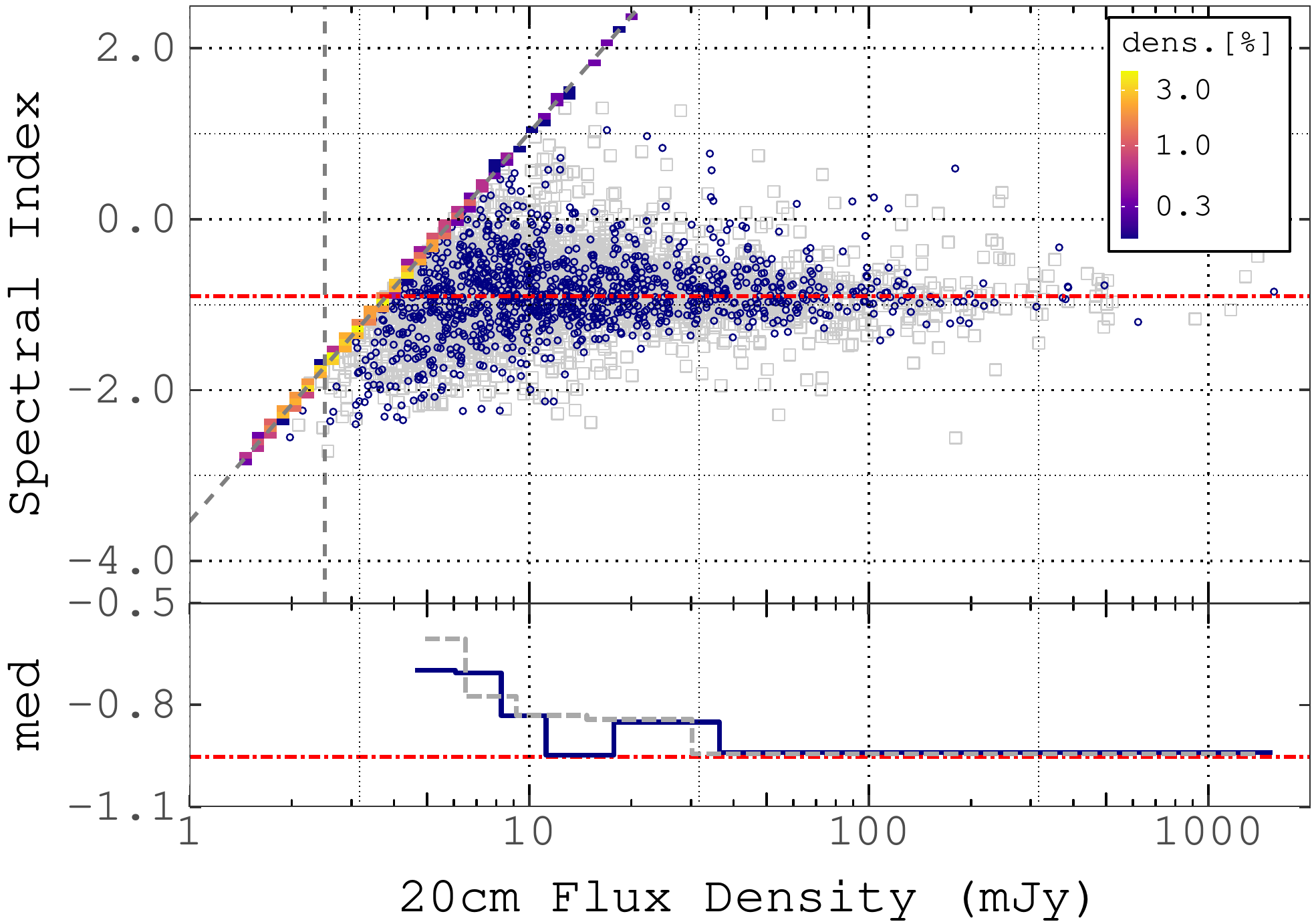} \\
	\end{tabular}
	\caption{As in Figure~\ref{fig:IntegFluxSpecIndex_SUMSSLimited} but splitting the sample into two groups: Open circles represent sources within  3.5\degr\ from the \ac{LMC} centre ({\it inner} subsample) while open squares are sources outside this radius ({\it outer} subsample). In the bottom panels we show the distribution of spectral index medians per, equally populated, flux density bin for the {\it inner} subsample (blue solid line) and {\it outer} subsample (gray dashed line) separately. No significant difference in source spectral index distribution is observed between the two groups, indicating that our catalogue is mainly comprised of background sources.}
	\label{fig:IntegFluxSpecIndex_SUMSSLimited3.5degr}
\end{figure*}


\subsubsection{Spectral Flattening}
 \label{sec:SpecFlat}
 
\input{Files/full_sumss_binned}
\input{Files/full_atca_binned}

Spectral flattening at low frequencies ($\le$ 5~GHz) has been suggested in the literature, from studies of deep radio surveys. It is conjectured that the faint radio source populations of a few mJy have a flatter mean spectral index than brighter source populations. For instance \cite{2006A&A...457..517P} studied the spectral index distribution of a sub-sample of sources extracted from the 1.4 and 5~GHz \ac{ATESP} surveys, finding a flatter mean (1.4~--~5~GHz) spectral index of --0.53$\pm$0.05 for sources with S$_{1.4~\mathrm{GHz}} \le 4$~mJy, when compared to --0.66$\pm$0.05 for sources above 4~mJy. These authors observed a similar behaviour for the 5~GHz selected sample, with mean spectral indices of --0.24$\pm$0.06 and --0.58$\pm$0.06 respectively below and above 4~mJy, in line with previous results based on smaller samples (\citealt{1987ApJ...321...94D,1991AJ....102.1258F}). The flatter values at 5~GHz were expected as higher frequency surveys tend to select source populations with flatter spectra. More recently \cite{2009AJ....137.4846O} pushed the spectral index analysis to lower flux densities (S$_{1.4~\mathrm{GHz}}>10$ $\mu$Jy) and lower frequencies (0.330~MHz), finding similar results: in the SWIRE deep field they measured median (0.33~--~1.4~GHz) spectral indices of \mbox{--0.54$\pm$0.06} in the flux interval S$_{1.4~\mathrm{GHz}} \sim 0.3-1$~mJy, and --0.66$\pm$0.11 at larger flux densities. Both \cite{2006A&A...457..517P} and \cite{2009AJ....137.4846O} found that this flattening is mostly associated with compact sources, likely core dominated \ac{AGN}. Interestingly \cite{2009AJ....137.4846O} find a significant re-steepening at S$_{1.4~\mathrm{GHz}} < 0.3$~mJy, which is interpreted as the result of star forming galaxies becoming the dominant population in the $\mu$Jy regime. 

It should be noted, however, that the tendency of radio source spectra to become flatter with decreasing flux density is not observed by all authors. For instance, \cite{2012MNRAS.421.1644R} and \cite{2012A&A...544A..38Z} found no statistical spectral flattening for mJy sources. This however could be a consequence of not well-matched sensitivities at the various frequencies, which might prevent the spectral index analysis to be applied to the faintest sources, as discussed in Sect.~\ref{sec:RMSEffects}. Indeed, the 1.4~GHz survey by \cite{2012A&A...544A..38Z} has a \ac{RMS} $\sim$3 times lower than that of their 2.3~GHz survey.

In order to provide further clues on this matter, we searched for spectral index -- flux density trends in our \ac{LMC} field catalogue. Tables~\ref{tab:atcaspindex} and \ref{tab:sumssspindex} show the mean and median values of the spectral index as a function of 1.384 and 0.843~GHz flux densities respectively. The median values are also shown in the bottom panels of Figure~\ref{fig:IntegFluxSpecIndex_SUMSSLimited}. As discussed in Sect.~\ref{sec:RMSEffects}, the limited sensitivity of the 0.843~GHz catalogue makes it unreliable for any spectral index analysis of the faintest 1.384~GHz selected sources, as only very steep-spectrum sources can be detected at both frequencies and spectral index limits become dominant. For this reason, we decided to restrict our analysis of the 1.384~GHz \ac{ATCA} sample to flux densities where spectral index limits represent less than 50~per~cent of the total number of sources in the flux bin under consideration, allowing us to get reliable median spectral index values. This happens at S$_{1.4~\mathrm{GHz}} >4.7$~mJy. For the shallower \ac{SUMSS} catalogue, we do not have significant biases and the analysis can be pushed down to the catalogue flux limit (i.e. 6~mJy).

We notice that the spectral flattening reported in previous works is observed at lower flux densities than those probed by the present study. Nevertheless, the larger statistics available allows us to split our sources in finer flux density bins. This in turn allows us to identify a small but clear trend towards flatter spectral indices already at S$<$10~mJy. We also notice that median spectral index values tend to be flatter for sources in the 1.384~GHz catalogue than for those in the 0.843~GHz catalogue, consistent with the results of \cite{2006A&A...457..517P}, where flatter spectral index values were found for the higher frequency selected sample.

Focusing on the 1.384~GHz catalogue, where spectral index values can be directly examined, we find that a median\footnote{The mean spectral index value in this flux bin is not well constrained due to the many spectral index limits. } spectral index value of --0.51 at S$\sim$5~mJy is consistent with the --0.53 found by \cite{2006A&A...457..517P} at S$<$4~mJy and the --0.54 found by \cite{2009AJ....137.4846O} at S$<$1~mJy, indicating we may be probing the same radio source population. Also, consistent with the mean values measured at higher fluxes: they vary from --0.64 to --0.69 in the flux range 6--17~mJy, against a value of --0.66 reported by \cite{2006A&A...457..517P} at S$>$4~mJy.

In order to test if there is any difference in spectral index distribution (including a possible contribution to the source spectral flattening) of sources belonging to the \ac{LMC}, we constructed the same graphs for the sources inside of the 3.5\degr\ radius from the approximate centroid of the \ac{LMC} at RA~=$~79.5$~deg and Dec~=~$-68.9$~deg (Figure~\ref{fig:IntegFluxSpecIndex_SUMSSLimited3.5degr}). We found no difference in source spectral index distribution within or outside the 3.5\degr\ radius (filled circles and open squares respectively), indicating that our catalogue is mainly comprised of background sources.

\subsubsection{Radio Two-Colour Diagram}
\label{sec:RadioTwoColour}

\begin{figure*}
	\includegraphics[width=0.495\textwidth]{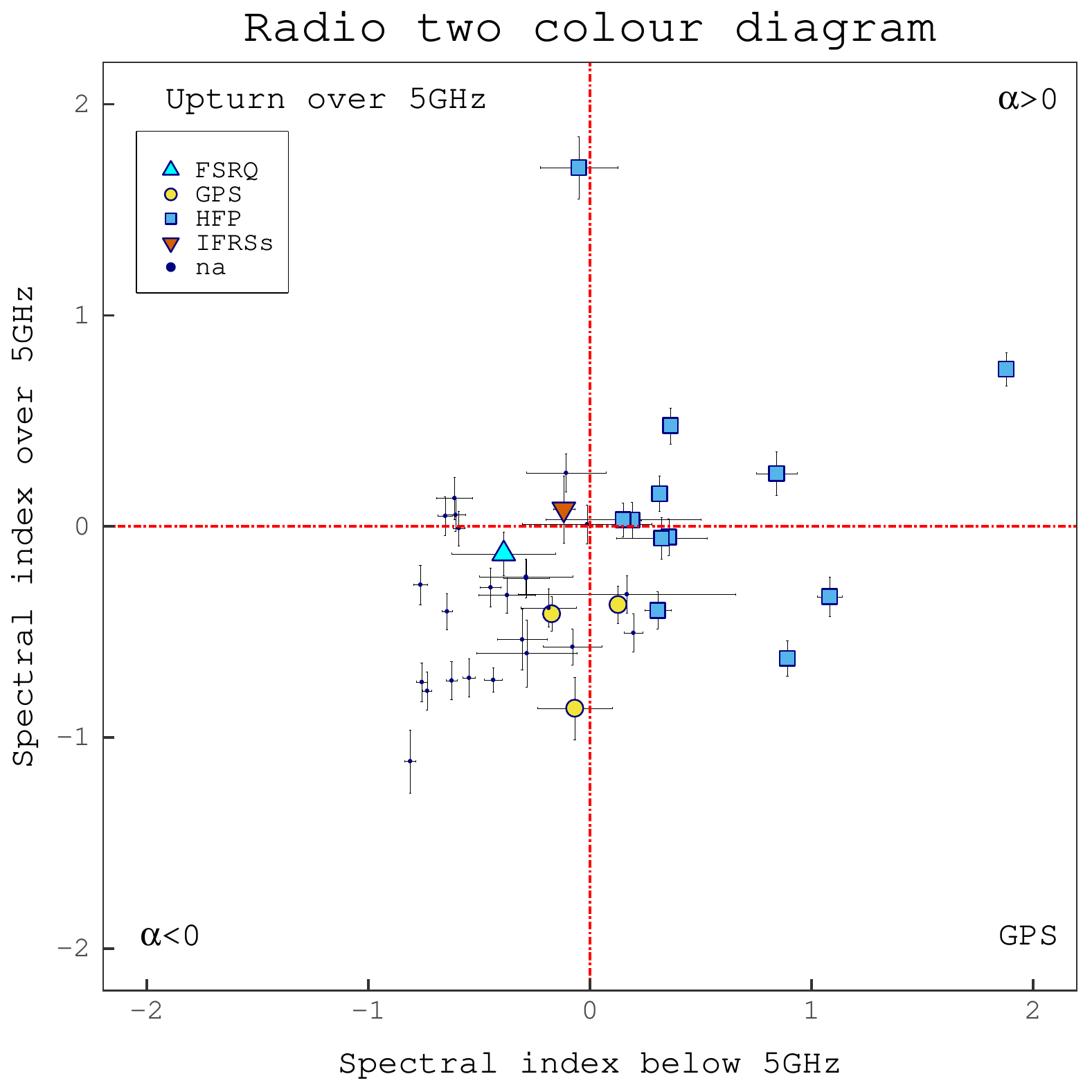} 
	\includegraphics[width=0.495\textwidth]{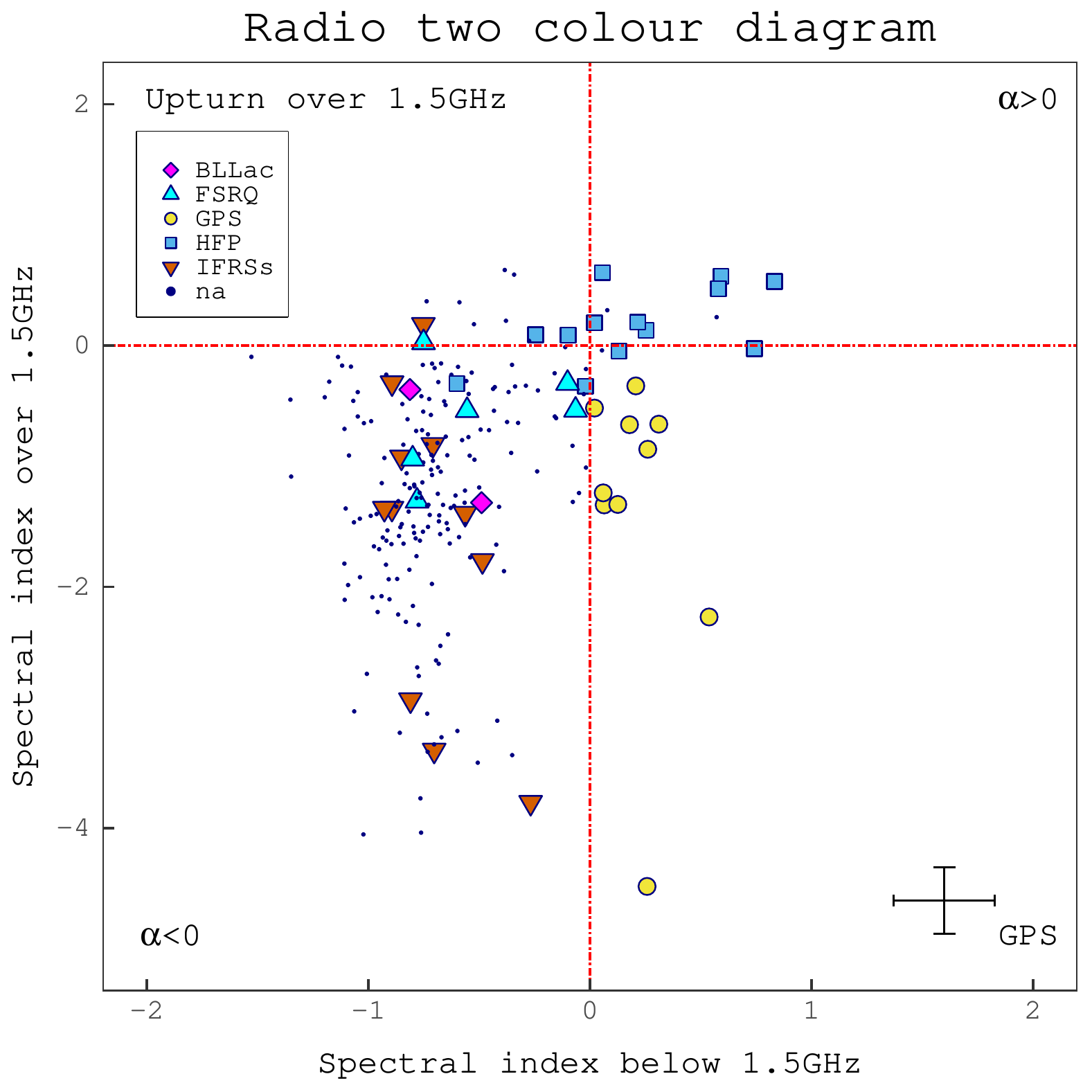}
        \caption[Radio \lq Two-Colour\rq\ Diagram]{5~GHz radio ``two-colour'' diagram (left) for \SpecIndexBreakHigh\ sources, contrasting two spectral index measurements for each source divided at 5~GHz. The median $\alpha$ for the low frequency measurement is \SpecIndexBreakHighleftMed\ and the mean $\alpha$ for the high frequency measurement is \SpecIndexBreakHighrightMed. 1.5~GHz radio two-colour diagram (right) for \SpecIndexBreakLow\ sources, contrasting two spectral index measurements for each source divided at 1.5~GHz. The median $\alpha$ for the low frequency measurement is \SpecIndexBreakLowleftMed\ and the mean $\alpha$ for the high frequency measurement is \SpecIndexBreakLowrightMed. In the lower right corner we show median errors in estimated spectral indices. In the upper left corner (both plots) we show the colour code of the points for different classes presented here.}
	\label{fig:RadioTwoColourLowHigh}
\end{figure*}

We adopt the radio ``two-colour'' diagram as shown by \cite{2006MNRAS.371..898S} in order to allow the identification of sources with curved spectra (Figure~\ref{fig:RadioTwoColourLowHigh}). This approach treats the high and low frequency components of each source separately, by calculating two spectral indices. Here, we define our high frequency sample as $>$1.5~GHz and our low frequency sample as the inverse case (i.e. $<$1.5~GHz). This analysis should be treated with caution as the spectral index was calculated based on flux densities from observations with different angular resolutions.

\begin{figure}
	\includegraphics[width=0.45\textwidth]{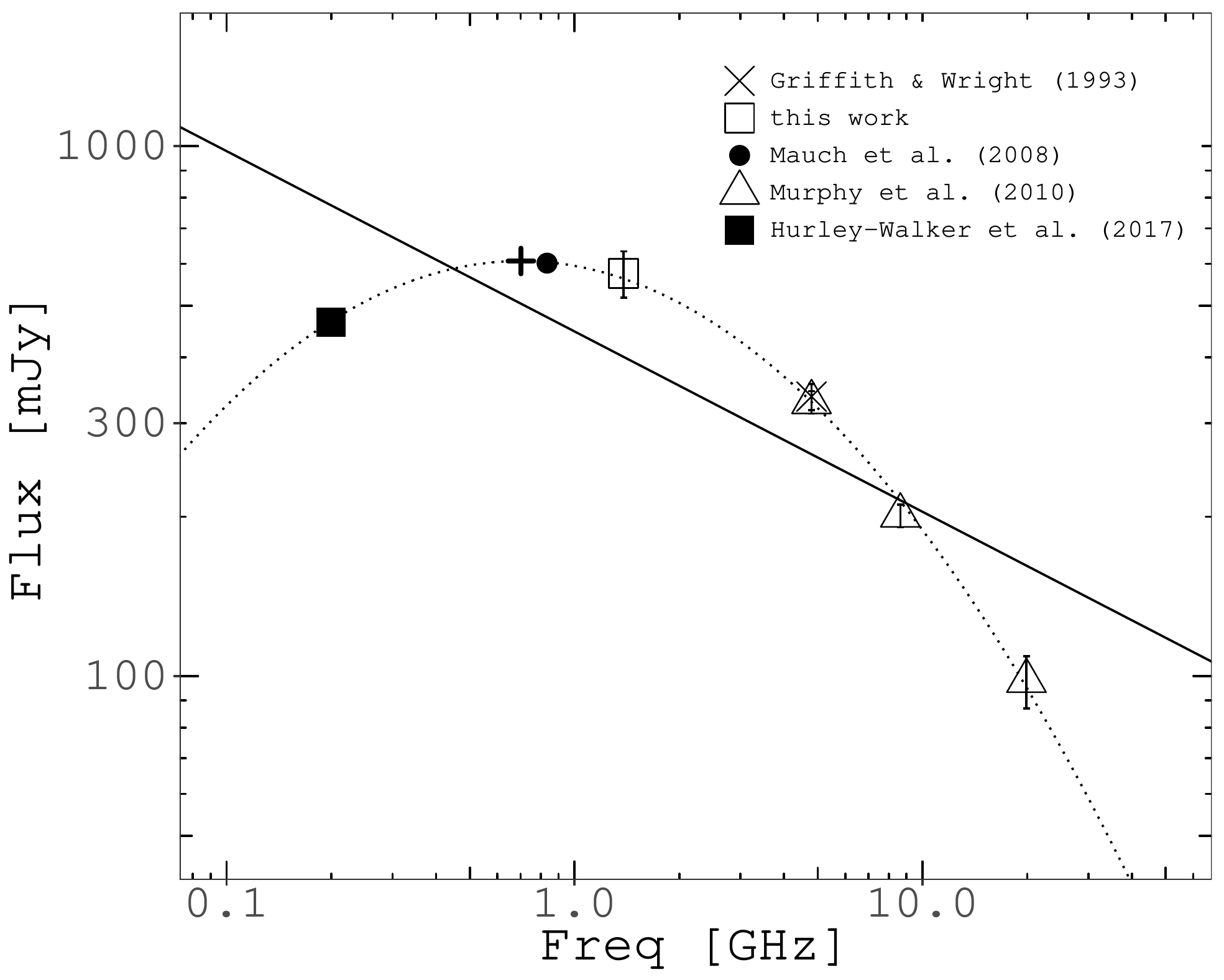} 
        \caption[Polynomial fit example]{Example of the radio spectra of one of the \ac{GPS} candidates (J061011--724814) fitted with a $2^{\rm nd}$ order polynomial (dotted line) and power-law function (continuous line). A cross (+) represents the location of the estimated turnover frequency.}
	\label{fig:polyfit}
\end{figure}

Using the radio two-colour diagram, we find \SourcesGPS\ sources with a spectral index of $\alpha_{1.5~\mathrm{GHz}}^{20~\mathrm{GHz}}\le0\le\alpha_{0.2~\mathrm{GHz}}^{1.5~\mathrm{GHz}}$. This limited sample is primarily due to the lower source density at 20~GHz. 

We used a $2^{\rm nd}$ order polynomial to determine the turnover frequency for GPS candidates. In Figure~\ref{fig:polyfit} we show an example of radio spectra of one of the objects from this sample fitted with $2^{\rm nd}$ order polynomial.

With these estimated turnover frequencies, we use Equation~\ref{eq:turnover} to determine the \ac{LLS} for each \ac{GPS} candidate, as shown in Table~\ref{tab:turnover}. The average turnover frequency for this sample of \ac{GPS} sources is 1.73~GHz with a corresponding average \ac{LLS} of 320~pc. These results are consistent with findings of \citet{2016MNRAS.458.3786J}.

\begin{table} 
    \caption{Turnover frequency and linear size relation (using equation \ref{eq:turnover}) for our sample of \ac{GPS} candidate sources.}
    \label {tab:turnover}	
    \begin{center}
	    \begin{tabular}{ l c c c c }
    	\hline
Source Name         & Turnover $\nu$ & \ac{LLS}  & Candidate  \\
                    & (GHz)          & (kpc)& Type    \\
    	\hline
  045358-713817 & 2.05 & 0.16 & GPS \\ 
  045516-705446 & 2.71 & 0.10 & GPS \\ 
  050000-632237 & 1.00 & 0.47 & GPS \\ 
  050139-662525 & 1.05 & 0.44 & GPS \\ 
  052310-720624 & 1.76 & 0.20 & GPS \\ 
  052505-664052 & 2.29 & 0.13 & GPS \\ 
  052517-672246 & 1.62 & 0.23 & GPS \\ 
  054317-662655 & 3.14 & 0.08 & GPS \\ 
  054736-680905 & 0.94 & 0.52 & GPS \\ 
  061011-724813 & 0.70 & 0.82 & GPS \\ 
        \hline
    \end{tabular}   
    \end{center}
\end{table}


\label{sec:HFP}

We find \SourcesHFP\  sources that have a convex spectrum with a peaked (turnover) integrated flux density above 5~GHz and, as such, we follow \cite{2000A&A...363..887D} classification and label them as \ac{HFP} sources.


\section{Summary and Conclusion}

We have compiled a master catalogue based on several new and existing radio continuum lists of sources behind the \ac{LMC}, to a 5$\sigma$ detection level from $\sim$200~mJy down to 0.2~mJy, depending on observing frequency (see Table \ref{tab:Surveys}). From this, a total of \SourcesTotalUnique\ sources have been identified, with multi-frequency detections of \SpecIndexSourceCount\ sources. We find the median spectral index of $\alpha$~=~\SpecIndexLowestTwoFreq\ for \SpecIndexLowestTwoFreqSourceCount\ sources which are detected at two frequencies (0.843 and 1.384~GHz) with similar resolution (\ac{FWHM} $\sim$40--45~arcsec). We have cross-matched these \SourcesTotalUnique\ sources with and five optical surveys (see Table \ref{tab:OpticalSurveys}). As a result, we find optical counterparts for \OpticalMatchesUnique, of which \OpticalMatchesUniqueWithZ\ have a redshift measurement with a mean redshift of 0.39. 

Based on the \ac{SUMSS} and 1.384~GHz \ac{ATCA} source lists, we present evidence that radio sources below 10~mJy exhibit flatter radio spectra than brighter ones. To take into account the biases introduced by the limited sensitivity of the 0.843~GHz \ac{SUMSS} catalogue, we restricted our analysis to 1.384~GHz flux densities S$>5$~mJy. As a result, we obtain median (0.843--1.384~GHz) spectral indices flattening from values of $<-0.8$ for S$_{1.4~\mathrm{GHz}} >10$ mJy to --0.51 for S$_{1.4~\mathrm{GHz}} <5$~mJy. This is in general agreement with findings from previous deep surveys (\citealt{2006A&A...457..517P}; \citealt{2009AJ....137.4846O}).

Additionally, \SourcesCSS\ unresolved sources have been detected with a spectral index of $< -0.8$ between 408~MHz and 20~GHz (Sect.~\ref{sec:gps_css}). This is a strong indicator for young \ac{AGN} that are either \ac{CSS} or \ac{GPS} objects. Furthermore we find evidence for \SourcesGPS\ of these sources to be \ac{GPS} candidates, that have $\alpha_{5~\mathrm{GHz}}^{20~\mathrm{GHz}} \le 0 \le \alpha_{843~\mathrm{MHz}}^{5~\mathrm{GHz}}$. We classify \SourcesHFP\ sources as \ac{HFP} and \SourcesHFPpossible\ as possible \ac{HFP} sources (see Table \ref{tab:classification}) that have steep inverted spectra where we do not observe a turnover (i.e. the turnover frequency is $>$ 20~GHz).

\section{Data Availability}

The catalogues underlying this article will be made available on the CDS\footnote{https://cds.u-strasbg.fr/} website when the paper is published.


\section*{Acknowledgements}

The Australian SKA Pathfinder is part of the Australia Telescope National Facility which is managed by \ac{CSIRO}. Operation of \ac{ASKAP} is funded by the Australian Government with support from the National Collaborative Research Infrastructure Strategy. \ac{ASKAP} uses the resources of the Pawsey Supercomputing Centre. 
Establishment of \ac{ASKAP}, the Murchison Radio-astronomy Observatory and the Pawsey Supercomputing Centre are initiatives of the Australian Government, with support from the Government of Western Australia and the Science and Industry Endowment Fund. 
We acknowledge the Wajarri Yamatji people as the traditional owners of the Observatory site. 
This work was supported by resources provided by the Pawsey Supercomputing Centre with funding from the Australian Government and the Government of Western Australia. 
This paper includes archived data obtained through the Australia Telescope Online Archive (\href{http://atoa.atnf.csiro.au}{http://atoa.atnf.csiro.au}).
The Australia Telescope Compact Array is part of the Australia Telescope National Facility which is funded by the Australian Government for operation as a National Facility managed by \ac{CSIRO}. 
Parts of this research were conducted with the support of Australian Research Council Centre of Excellence for AllSky Astrophysics in 3 Dimensions (ASTRO~3D), through project number CE170100013. 
IP acknowledges support from INAF under the SKA/CTA PRIN ``FORECaST'' and the PRIN MAIN STREAM ``SAuROS'' projects. 
H.A. has benefited from grant CIIC 174/2021 of Universidad de Guanajuato, Mexico.
D.U. acknowledges Ministry of Education, Science and Technological Development of the Republic of Serbia through the contract No. 4451-03-9/2021-14/200104.






\bibliographystyle{mnras}
\bibliography{references} 




\bsp	
\label{lastpage}
\end{document}

%% file: Files/catalogue_pars.tex
\newcommand{\SourcesTotalUnique}{6\,434}

\newcommand{\SpecIndex}{$-0.92$}  
\newcommand{\SpecIndexMean}{$-0.89$}  
\newcommand{\SpecIndexSD}{0.47}
\newcommand{\SpecIndexSourceCount}{3\,789}

\newcommand{\SpecIndexLowestTwoFreq}{$-0.89$}
\newcommand{\SpecIndexLowestTwoFreqMean}{$-0.88$}
\newcommand{\SpecIndexLowestTwoFreqSD}{0.48}
\newcommand{\SpecIndexLowestTwoFreqSourceCount}{3\,636}

\newcommand{\SpecIndexIFRS}{$-0.96$}

\newcommand{\SpecIndexBreakHigh}{41}
\newcommand{\SpecIndexBreakLow}{234}

\newcommand{\SpecIndexBreakHighleftMed}{$-0.17$}
\newcommand{\SpecIndexBreakHighrightMed}{$-0.29$}
\newcommand{\SpecIndexBreakLowleftMed}{$-0.71$}
\newcommand{\SpecIndexBreakLowrightMed}{$-1.04$}

\newcommand{\SpecIndexLowestTwoFreqLimited}{$-0.85$}
\newcommand{\SpecIndexLowestTwoFreqLimitedMean}{$-0.81$}
\newcommand{\SpecIndexLowestTwoFreqLimitedSD}{0.45}
\newcommand{\SpecIndexLowestTwoFreqLimitedSourceCount}{2\,984}

\newcommand{\SpecIndexATG}{$-0.2$}
\newcommand{\SpecIndexATGMean}{$-0.21$}
\newcommand{\SpecIndexATGSD}{0.44}
\newcommand{\SpecIndexATGSourceCount}{41}

\newcommand{\SpecIndexOptical}{$-0.71$}
\newcommand{\SpecIndexOpticalMean}{$-0.66$}
\newcommand{\SpecIndexOpticalSD}{0.49}
\newcommand{\SpecIndexOpticalSourceCount}{209}

\newcommand{\SpecIndexRadioToIR}{$-0.84$}
\newcommand{\SpecIndexRadioToIRMean}{$-0.78$}
\newcommand{\SpecIndexRadioToIRSD}{0.47}
\newcommand{\SpecIndexRadioToIRSourceCount}{917}

\newcommand{\SpecIndexNoIR}{$-0.95$}
\newcommand{\SpecIndexNoIRMean}{$-0.93$}
\newcommand{\SpecIndexNoIRSD}{0.46}  
\newcommand{\SpecIndexNoIRSourceCount}{2\,872}

\newcommand{\SourcesMOST}{2\,084} 
\newcommand{\SourcesTwentyCm}{6\,434}   
\newcommand{\SourcesSixCm}{570}         
\newcommand{\SourcesThreeCm}{279}    
\newcommand{\TwoLowestFreqCount}{6\,434}
\newcommand{\TwoLowestFreqCountPerc}{100}
\newcommand{\TwoLowestFreqCountRemain}{0}

\newcommand{\SourcesMWAsingle}{1\,277}
\newcommand{\SourcesMWAmultiple}{255}

\newcommand{\SourcesAT}{38} 
\newcommand{\SourcesPMN}{191}         
\newcommand{\SourcesSUMSS}{3\,636}    
\newcommand{\SourcesMRC}{40}    
\newcommand{\SourcesASKAPBETA}{1\,995}    

\newcommand{\OpticalSourcesTotalUnique}{7\,883}

\newcommand{\OpticalMatchesUnique}{343}

\newcommand{\OpticalMatchesKozOne}{122}
\newcommand{\OpticalMatchesKozTwo}{10}
\newcommand{\OpticalMatchesKozThree}{20}
\newcommand{\OpticalMatchesSixDf}{111}
\newcommand{\OpticalMatchesMACHO}{31}

\newcommand{\OpticalMatchesUniqueWithZ}{128}

\newcommand{\OpticalMatchesWithZKozOne}{3}
\newcommand{\OpticalMatchesWithZKozTwo}{0}
\newcommand{\OpticalMatchesWithZKozThree}{20}
\newcommand{\OpticalMatchesWithZMACHO}{27}
\newcommand{\OpticalMatchesWithZSixDf}{84}

\newcommand{\SourcesUSS}{546}
\newcommand{\SourcesCSS}{1\,866}
\newcommand{\SourcesGPS}{10}
\newcommand{\SourcesHFP}{14}
\newcommand{\SourcesHFPpossible}{0}

\newcommand{\SourcesInverted}{146}
\newcommand{\SourcesIFRSs}{67}

\newcommand{\ATGSourcesCSS}{1}
\newcommand{\ATGSourcesGPS}{0}
\newcommand{\ATGSourcesHFP}{14}
\newcommand{\ATGSourcesHFPpossible}{0}

\newcommand{\OffsetAtcaTwcmVSaskapRA}{4.23} 
\newcommand{\OffsetAtcaTwcmVSaskapDEC}{0.36} 
\newcommand{\OffsetMostVSaskapRA}{2.79} 
\newcommand{\OffsetMostVSaskapDEC}{0.72} 

\newcommand{\OffsetSumssVSaskapRA}{3.18} 
\newcommand{\OffsetSumssVSaskapDEC}{-0.36} 
\newcommand{\OffsetMostVSsumssRA}{-0.11} 
\newcommand{\OffsetMostVSsumssDEC}{1.44} 

\newcommand{\CommonMostVsSumss}{1638} 
 
\newcommand{\ExtraMostSources}{446}

%% file: Files/main_fluxes.tex
\begin{table*} 
    \caption{Excerpt of the point source catalogue of \SourcesTotalUnique\ objects in the directions of the \ac{LMC} with its positions, integrated flux densities with associated uncertainty and spectral index. The columns provided are as follows: (1) Source name derived from the highest frequency source component; (2) position of J2000 right ascension from the highest frequency detection; (3) position of J2000 declination from the highest frequency detection; (4) flux density flag for the GLEAM survey; (5~\&~6) GLEAM MWA 200\,MHz survey integrated flux density; (7~\&~8) SUMSS 843\,MHz survey integrated flux density; (9) flux density flag for the 20~cm ATCA 1.384\,GHz survey; (10~\&11) 20~cm ATCA 1.384\,GHz survey integrated flux density; (12~\&~13) 6~cm ATCA 4.85\,GHz survey integrated flux density; (14~\&~15) 3~cm ATCA 8.64\,GHz survey integrated flux density; (16~\&~17) MRC 408\,MHz survey integrated flux density; (18~\&~19) PMN 4.8\,GHz survey integrated flux density; (20~\&~21) AT20G survey integrated flux density at (20~\&~21) 4.8\,GHz; (22~\&~23) 8.6\,GHz; (24~\&~25) 19.904\,GHz;  (26~\&~27) spectral index fit to all available source components with the associated error; (28) number of flux densities used in the spectral index fit and (29) source classification. Upper limits and uncertain values in flux densities flag columns are designated with preceding $<$ and colon (:), respectively. The columns not presented in this table but available in the full version are: flux densities from MRC, PMN and AT20G surveys. The full catalogue is provided through the \textsc{VizieR} service.}
    \begin{tabular}{ccccccccccc}
    \hline
    (1) & (2) & (3) & (4, 5~\&~6) & (7~\&~8) & (9, 10~\&~11) & (12~\&~13) & (14~\&~15) & (...) & (26~\&~27) & (29) \\
    Name J & RA & DEC & 
    $S_{\mathrm{MWA}}$ & 
    $S_{\mathrm{SUMSS}}$ & 
    $S_{\mathrm{20cm ATCA}}$ & 
    $S_{\mathrm{6cm ATCA}}$ & 
    $S_{\mathrm{3cm ATCA}}$ & 
    ... &
    $\alpha \pm \Delta\alpha$ &
    Class\\
    & (J2000) & (J2000) &  
    0.200\,GHz & 
    0.843\,GHz & 
    1.384\,GHz & 
    4.85\,GHz & 
    8.64\,GHz & ...& &  \\
    & &  & (mJy) & (mJy)  & (mJy) & (mJy) & (mJy) & ...& & \\
    \hline
  ... & ... & ... & ... & ... & ... & ... & ... & ... & ... \\ 
  053145--721857 & 05:31:45.9 & --72:18:57 & 165 $\pm$ 17 & 41.2  $\pm$ 1.6 & 24.9 $\pm$ 2.5 & ... & ... & ... & --0.98 $\pm$ 0.07 & ... \\ 
  051637--723707 & 05:16:37.8 & --72:37:08 & 319 $\pm$ 16 & 261   $\pm$ 11  & 230 $\pm$ 23   & ... & ... & ... & --0.03 $\pm$ 0.02 & HFP \\ 
  054427--715526 & 05:44:27.2 & --71:55:27 & 982 $\pm$ 18 & 334   $\pm$ 10  & 213 $\pm$ 21   & 55.1 $\pm$ 5.5 & 21.7 $\pm$ 2.2 & ... & --0.99 $\pm$ 0.02 & ... \\ 
  060531--711229 & 06:05:31.3 & --71:12:29 & 230 $\pm$ 19 & 93.9  $\pm$ 3.1 & 75.9 $\pm$ 7.6 & ... & ... & ... & --0.32 $\pm$ 0.02 & ... \\ 
  055632--712906 & 05:56:32.2 & --71:29:06 & 558 $\pm$ 21 & 189.7 $\pm$ 5.8 & 115 $\pm$ 12   & 33.0 $\pm$ 3.3 & 14.9 $\pm$ 1.7 & ... & --0.93 $\pm$ 0.03 & ... \\ 
  
   ... & ... & ... & ... & ... & ... & ... & ... & ... & ... \\ 
  054749--635200 & 05:47:49.1 & --63:52:01 & 80 $\pm$ 22 & 19.5 $\pm$ 1.3 & 9.4 $\pm$ 1.0 & ... & ... & ... & --1.08 $\pm$ 0.16 & ... \\ 
  054831--654816 & 05:48:31.0 & --65:48:17 & 146 $\pm$ 25 & 39.0 $\pm$ 1.4 & 22.2 $\pm$ 2.2 & 6.6 $\pm$ 1.1 & ... & ... & --0.98 $\pm$ 0.07 & ... \\ 
  054845--655349 & 05:48:45.1 & --65:53:50 & 584 $\pm$ 78 & 116.6 $\pm$ 3.6 & 70.9 $\pm$ 7.1 & 17.2 $\pm$ 1.8 & 7.6 $\pm$ 1.0 & ... & --1.14 $\pm$ 0.04 & ... \\ 
  054845--665725 & 05:48:45.6 & --66:57:25 & 173 $\pm$ 28 & 60.0 $\pm$ 1.9 & 38.7 $\pm$ 3.9 & 16.2 $\pm$ 1.7 & 7.1 $\pm$ 1.1 & ... & --0.82 $\pm$ 0.05 & ... \\ 
  054850--625443 & 05:48:50.2 & --62:54:43 & $<$51 & 18.6 $\pm$ 1.2 & :10.5 $\pm$ 1.2 & ... & ... & ... & --1.13 $\pm$ 0.23 & ... \\ 
  054919--643738 & 05:49:19.5 & --64:37:38 & 110 $\pm$ 21 & 32.9 $\pm$ 1.3 & 19.1 $\pm$ 1.9 & ... & ...& ... & --0.89 $\pm$ 0.11 & ... \\ 
  
  ... & ... & ... & ... & ... & ... & ... & ... & ... & ... \\ 
    \hline
    \end{tabular}   
    \label{tab:Catalogue}
\end{table*}

%% file: Files/askap_fluxes.tex

\begin{table*}
\centering
\caption{Excerpt for the source list (total of \SourcesASKAPBETA\ sources) at 0.843~GHz derived from \ac{ASKAP}-Beta survey. Note that the flux densities shown are flux corrected (see text for more details). The uncertainties are only from the fitting method i.e. flux calibration errors have not been included in the presented uncertainty.} 
\label{tab:askaptable}

\begin{tabular}{lccccc}
  \hline
No & RA & DEC & Peak & Int. Flux & Local rms \\
     & (J2000)   &  (J2000)  & (mJy beam$^{-1}$)& (mJy)& (mJy beam$^{-1}$)\\ 
  \hline
  1431 & 05:58:22.4 & --65:07:23 & 34.7 $\pm$ 0.8 & 40.2 $\pm$ 0.9 & 0.78\\
  1432 & 05:58:23.2 & --67:55:28 & 9.1 $\pm$ 0.6 & 9.6 $\pm$ 0.7 & 0.65\\
  1433 & 05:58:23.9 & --67:18:23 & 5.2 $\pm$ 0.6 & 7.0 $\pm$ 0.8 & 0.59\\
  1434 & 05:58:25.3 & --66:05:16 & 22.2 $\pm$ 0.7 & 27.5 $\pm$ 0.8 & 0.69\\
  1435 & 05:58:28.4 & --68:24:37 & 22.5 $\pm$ 0.7 & 30.0 $\pm$ 1.0 & 0.76\\
   \hline
\end{tabular}
\end{table*}

%% file: Files/most_fluxes.tex

\begin{table*}
\centering
\caption{Excerpt for the source list (total of \SourcesMOST\ sources) at 0.843~GHz derived from \ac{MOST}. The uncertainties are only from the fitting method i.e. flux calibration errors have not been included in the presented uncertainty.} 
\label{tab:mosttable}

\begin{tabular}{lccccc}
  \hline
No & RA        & DEC       & Peak & Int. Flux & Local rms \\ 
     & (J2000)   &  (J2000)  & (mJy beam$^{-1}$)& (mJy)& (mJy beam$^{-1}$)\\ 
  \hline
  1116 & 05:24:32.4 & --66:42:36 & 26.0 & 29.5 $\pm$ 0.3 & 0.7\\
  1117 & 05:24:41.8 & --71:01:33 & 14.0 & 14.0 $\pm$ 1 & 0.75\\
  1118 & 05:24:41.7 & --67:18:07 & 19.0 $\pm$ 2 & 19.2 $\pm$ 0.2 & 1.17\\
  1119 & 05:24:46.7 & --70:25:19 & 4.0 & 4.0 & 0.59\\
  1120 & 05:24:48.5 & --65:15:57 & 13.0 $\pm$ 1 & 13.2 $\pm$ 0.1 & 0.61\\
  
   \hline
\end{tabular}
\end{table*}

%% file: Files/optical_detections.tex

\begin{table}
\centering
\caption{Excerpt of the list (total of \OpticalMatchesUnique\ sources) of sources from our base catalogue cross-matched with six optical surveys used in this study (see Sect.~\ref{sec:SupplementarySurveys}). We provide the name of the source as defined in Table~\ref{tab:Catalogue}, catalogued redshift value and the associated reference (columns {\it z} and {\it z}~Ref. and the references to optical detection (Det. Ref.)). Number-coded references are: (1) 6dF: \citet{2009MNRAS.399..683J}, (2) MACHO: \citet{2012ApJ...747..107K}, MQS~I: \citet{2009ApJ...701..508K}, (4) MQS~II: \citet{2012ApJ...746...27K}, (5) MQS~III: \citet{2013ApJ...775...92K} and (6) MCELS: \citet{1998PASA...15..163S}. 
} 
\label{tab:optdet}

\begin{tabular}{lccc}
  \hline
Name J & $z$        & $z$ Ref.       & Det. ref. \\ 
  \hline
... & ... & ... & ...\\ 
  043259-713059 & 0.04383 & 1 & 1\\
  043359-695840 & 0.06808 & 1 & 1\\
  043429-694910 & ... & ... & 3\\
  043436-693615 & ... & ... & 6\\
  043436-705252 & ... & ... & 6\\
  043442-691448 & ... & ... & 6\\
  043521-722117 & ... & ... & 6\\
  043523-690349 & 0.061 & 5 & 3;5\\
  043555-722838 & 0.04391 & 1 & 1\\
  043642-664254 & 0.02948 & 1 & 1\\
    ... & ... & ... & ...\\ 
  
   \hline
\end{tabular}
\end{table}

%% file: IFRS/IFRS_table_sample.tex
\begin{table}
\scriptsize 
   \caption{Excerpts of the table that list properties of 123 \ac{IFRSs} studied here. Column 2 is source IR name or IDs from SAGE and/or CatWISE2020 catalogue; Column 5 represent 3.4$-$3.6\,$\mu$m flux density. SAGE is 3.6\,$\mu$m and CatWISE2020 3.4\,$\mu$m}
\begin{tabular}{c c c c c c}
   \hline
ATCA Name  & IR Source ID  & S\textsubscript{1.4\,GHz}  & S\textsubscript{843\,MHz} & S\textsubscript{3.6\,$\mu$m}  &  $\alpha$ \\
    J      &     CWISE         & (mJy)                      & (mJy)                     & ($\mu$Jy)                           &         \\
    \hline 
054349-652228 & J054349.72-652228.2 & 51.6 & 94.7   & 3.2 & -1.0        \\
 055750-630159 & J055750.40-630158.9 & 44.9 & 65.1  & 3.9  & -0.63       \\
 051859-640238 & J051859.33-640238.7 & 49.1 & 80.5  & 4.6 & -0.7        \\
 051415-645354 & J051415.33-645354.0 & 9.0  & 19.4  & 4.9  & -1.15       \\
 061501-684234 & J061501.18-684234.0 & 32.3 & 50.9  & 5.5  & -0.88       \\
 ...          & ...                 & ... & ... & ... & ...  \\ 
\hline
    \end{tabular}
    \label{tab:IFRS}
\end{table}

%% file: Files/full_sumss_binned.tex
\begin{table}
\centering

\caption{Spectral index statistics for the 1.384~GHz selected catalogue (ATCA), as derived in different flux density bins. } 
\label{tab:atcaspindex}

\begin{tabular}{cccccc}
  \hline
Lower & Upper & N & Median     & Mean       & $\sigma_{\alpha}$ \\ 
(mJy) & (mJy) &   & ($\alpha$) & ($\alpha$) &  \\ 
  \hline
  4.69 & 5.96 & 642 & --0.51 & --0.71 & 0.53 \\ 
  5.96 & 7.94 & 642 & --0.67 & --0.64 & 0.61 \\ 
  7.94 & 10.96 & 643 & --0.77 & --0.65 & 0.64 \\ 
  10.96 & 17.38 & 642 & --0.81 & --0.69 & 0.67 \\ 
  17.38 & 35.48 & 642 & --0.83 & --0.73 & 0.59 \\ 
  35.48 & 1548.82 & 643 & --0.89 & --0.79 & 0.75 \\ 
   \hline
\end{tabular}
\end{table}

%% file: Files/full_atca_binned.tex
\begin{table}
\centering

\caption{Spectral index statistics for the 0.843-GHz selected catalogue (SUMSS), as derived in different flux density bins. } 
\label{tab:sumssspindex}

\begin{tabular}{cccccc}
  \hline
Lower & Upper & N & Median     & Mean       & $\sigma_{\alpha}$ \\ 
(mJy) & (mJy) &   & ($\alpha$) & ($\alpha$) &  \\ 
  \hline
6.00 & 8.00 & 373 & --.68 & --0.67 & 0.57 \\ 
  8.00 & 9.79 & 369 & --0.79 & --0.79 & 0.61 \\ 
  9.79 & 11.75 & 359 & --0.89 & --0.90 & 0.53 \\ 
  11.75 & 14.45 & 362 & --0.89 & --0.94 & 0.53 \\ 
  14.45 & 17.78 & 357 & --0.95 & --0.97 & 0.49 \\ 
  17.78 & 22.91 & 360 & --0.89 & --0.89 & 0.43 \\ 
  22.91 & 31.62 & 363 & --0.92 & --0.92 & 0.42 \\ 
  31.62 & 47.86 & 361 & --0.91 & --0.92 & 0.38 \\ 
  47.86 & 89.13 & 361 & --0.90 & --0.87 & 0.31 \\ 
  89.13 & 2398.83 & 363 & --0.95 & --0.95 & 0.35 \\ 
   \hline
\end{tabular}
\end{table}